\documentclass[reprint,showpacs,preprintnumbers,amsmath,amssymb]{revtex4}

\usepackage{graphicx,psfrag,color}
\usepackage{amsmath}
\usepackage{amsfonts}
\usepackage{mathrsfs}
\usepackage{amssymb}
\usepackage{bm}
\usepackage{url}
\usepackage{color}
\usepackage{natbib}
\usepackage{mdwlist}
\usepackage{dcolumn}% Align table columns on decimal point

\usepackage[export]{adjustbox}
\usepackage{tikz}
\usepackage{pgfplots}

\usepackage{epstopdf}
\usepackage{float}

\newcommand{\bnab}{\mbox{\boldmath$\nabla$}}
\newcommand{\eps}{\epsilon}

\def\half {{\textstyle{1 \over 2}}}

\newcommand{\bx}{\mathbf{x}}
\newcommand{\bhx}{\mathbf{\hat{x}}}
\newcommand{\bhy}{\mathbf{\hat{y}}}
\newcommand{\bhz}{\mathbf{\hat{z}}}
\newcommand{\bu}{\mathbf{\tilde{u}}}
\newcommand{\rh}{\tilde{\rho}}
\newcommand{\bnu}{\mbox{\boldmath$\nu$}}

\newcommand{\G}{{\cal G}}
\newcommand{\La}{{\cal L}}
\newcommand{\beq}{\begin{equation}}
\newcommand{\eeq}{\end{equation}}

\renewcommand{\tabcolsep}{9pt}

\begin{document}

\title{Optimising energy growth as a tool for finding exact coherent structures}

\author{D. Olvera}
\email{do12542@bris.ac.uk}
\affiliation{School of Mathematics, Bristol University, Bristol, BS8 1TW, UK}
\author{R. R. Kerswell}
\email{R.R.Kerswell@bris.ac.uk}
\affiliation{School of Mathematics, Bristol University, Bristol, BS8 1TW, UK}

\date{\today}

\begin{abstract}

We discuss how searching for finite amplitude disturbances of a given energy which maximise their subsequent energy growth after a certain later time $T$ can be used to probe phase space around a reference state and ultimately to find other nearby solutions. The procedure relies on the fact that of all the  initial disturbances on a constant-energy hypersphere, the optimisation procedure will naturally select the one which lies nearest to the stable manifold of a nearby solution in phase space if $T$ is large enough.  Then, when in its subsequent evolution, the optimal disturbance transiently approaches the new solution, a flow state at this point can be used  as an initial guess  to converge the solution to machine precision. We illustrate this approach in  plane Couette flow by: a) rediscovering the spanwise-localised `snake' solutions of Schneider et al. (2010b); b) probing phase space at very low Reynolds numbers ($< 127.7$) where the constant linear-shear solution is believed to be the global attractor; and finally c) examining how  the edge between laminar and turbulent flow evolves when stable stratification kills the turbulent attractor. We also show that the steady snake solution smoothly delocalises as unstable stratification is gradually turned on until it connects (via an intermediary global 3D solution) to  2D Rayleigh-Benard roll solutions.

\end{abstract}
%

%\pacs{}

\maketitle

\section{Introduction}

% Linear
Optimisation has proved a powerful tool to extract information from the Navier-Stokes equations.  In the shear flow transition problem,  optimising over all possible infinitesimal disturbances to find the one which maximises the subsequent energy growth after some pre-selected time $T$ has proven invaluable in  exposing the generic energy amplification mechanisms present. Called variously `transient growth' \cite{Reddy93, Henningson94}, `nonmodal instability' \cite{Schmid07} or `optimal perturbation theory' \cite{Butler92} (see the reviews \cite{Grossmann00, Schmid07} and book \cite{SchmidHenningson01}),  the approach reveals key aspects of the linearised dynamics around the reference state which has helped to interpret finite-time flow phenomena and pick apart what causes transition.  The approach owes its popularity to its linearity which means that there are multiple ways to extract the optimals and the mathematics in each case is well understood (e.g. \cite{Trefethen93, TrefethenEmbree05, Schmid07, SchmidBrandt14}). The downside of the approach is that it can say nothing about {\em finite} amplitude disturbances or, in other words, what can happen a finite distance away from the reference state in phase space \cite{Waleffe95, Dauchot97}
 
 % Nonlinear
Conceptually, the remedy to this is simple: let competing disturbances seeking to maximise the energy growth after time $T$ all have the same initial finite energy $E_0$ and use the fully nonlinear Navier-Stokes equations as a constraint \cite{Pringle10, Cherubini10, Kerswell14}. This, however, doubles the number of parameters ($E_0$ joins $T$) over which the results must be interpreted {\em and} leads to a fully nonlinear, non-convex optimisation problem where much less is known about its possibly multiple solutions (local and well as global maxima) or how to find them. So far, the solution technique has necessarily been iterative and this has revealed a number of interesting new insights in the transition problem \cite{Pringle10, Cherubini10, Cherubini11, Monokrousos11, Pringle12, Rabin12, Cherubini12, Cherubini13, Duguet13, Pringle15} (see the review \cite{Kerswell14}). For example, one can ask what is the smallest (most `dangerous') energy disturbance which can trigger transition by some time $T$, with the answer in the large $T$ limit labelled the {\em minimal seed} for transition \cite{Pringle12, Rabin12, Pringle15}.  The minimal seeds which emerge from this procedure are fully-localised and are therefore realistic targets for experimental investigations (e.g. \cite{Pringle15}).

% Here 
The optimisation approach works by naturally selecting disturbances on the energy hypersphere if they lie outside the basin of attraction of the reference state since then the energy remains finite for $T \rightarrow \infty$ (all other disturbances have to decay eventually).  The new state to which these disturbances are drawn needn't be a turbulent attractor and so the minimal disturbance to reach another simple stable state can also be calculated (see \S6.2 in \cite{Rabin13}).  What is not so clear is whether the approach can find {\em unstable} solutions although a similar line of reasoning seems to hold. The  optimisation procedure would be expected to select disturbances from the energy hypersphere which lie nearest or on the stable manifold of a nearby solution in phase space if $T$ is large enough as this is the best way to avoid energy decay. The difference now, however, is since the nearby solution is unstable, $T$ cannot be too large otherwise  even these  disturbances will have  decayed away (realistically it is improbable to stay on the stable manifold to converge in to the unstable state).  Once such an optimal disturbance has been found,  its temporal evolution will show evidence of a  transient approach to the new solution. A sufficiently close visit should yield flow states which can then be converged to the new solution. The main purpose of this paper is to demonstrate that this approach can work.

We illustrate this  in the context of  plane Couette flow (pCf) by rediscovering the spanwise-localised `snake' solutions of Schneider et al. \cite{Schneider10b} building upon the prior exploratory work of Rabin (see \S6.3 in \cite{Rabin13}) who identified the key role played by the choice of $E_0$.  In a wide geometry, snake solutions coexist with repeated copies of Nagata's well known solution \cite{Nagata90} in a narrow geometry and, not surprisingly, the stable manifolds of the (lower energy) snake solutions pass closer (in energy norm) to the simple shear solution in phase space than those of Nagata's solutions. However, the latter offer the possibility of greater energy growth as they lead to a global flow state and hence are preferred by the optimisation algorithm if they pass close to the energy hypersphere. As a result, Rabin found a threshold  initial energy below which the optimal disturbance appears to approach a snake solution and above which Nagata's solution is approached (\S 6.3 \cite{Rabin13}).  We complete this calculation here by  recomputing these optimal disturbances and converging out both snake solutions.

Armed with this success, we then probe phase space of pCf at very low Reynolds numbers ($< 127.7$ \cite{Waleffe03}) looking for new solutions where the (basic) constant shear solution is believed to be the global attractor (a proof only exists for $Re<20.7$ \cite{Joseph66}). We find evidence of  solutions but these turn out only to be the ghosts of known solutions at higher Reynolds numbers. Finally, as another example of how the optimisation approach can be utilised, we examine how  the edge between laminar and turbulent flow evolves when stable stratification suppresses the turbulence. With the turbulent attractor present, a `bursting' phenomenon forms a distinctive initial  feature  of the transition process for disturbances `above' the edge. This bursting is  found to change little  when the turbulent attractor vanishes under increasing stratification but disappears when the solution acting as the edge state ceases to exist. This then indicates that the bursting is directly related to the presence of the unstable manifold of the edge state directed away from the uniform shear solution in phase space.

The paper starts with the  formulation of the stratified plane Couette flow problem in \S\ref{pCf} which introduces the three non-dimensional parameters that fully specify the problem once the computational box is chosen: the Reynolds number $Re$, the bulk Richardson number $Ri_b$ and the Prandtl number $Pr$ with $Pr=1$ throughout. The optimisation approach used and the iterative solution technique adopted  are then described in \S\ref{meth}. The results section \S\ref{res} is divided into 3 parts: a description of the wide domain computations to find the snake solutions is given first in \S\ref{wide}; followed by a discussion of efforts to probe pCf at very low $Re$ in  \S\ref{low}; and then  the calculations examining the bursting phenomenon are presented in \S\ref{burst}. A final discussion in \S\ref{Disc} recaps the various results, provides some prospectives and then looks forward to future work.

\section{Formulation}

%--------------------------------------------------------------
\subsection{Stratified plane Couette flow \label{pCf}}
%--------------------------------------------------------------

The usual plane Couette flow set-up is considered in this paper of two (horizontal) parallel plates separated by a distance $2h$ with the top plate
moving at $U \mathbf{\hat{x}}$ and the bottom plate moving at $-U \mathbf{\hat{x}}$. Stable stratification is added by imposing that the fluid density is $\rho_0 -\Delta_\rho$ at the top plate and  $\rho_0+\Delta_\rho$ at the bottom plate (gravity $g$ is normal to the plates and directed downwards from the top plate to the bottom plate). Using the Boussinesq approximation ($\Delta_\rho \ll \rho_0$), the governing equations can be non-dimensionalised using $U$, $h$ and $\Delta_\rho$ to give
\begin{align}
\frac{\partial\mathbf{ u} }{\partial t} +\mathbf{u} 
\cdot \nabla \mathbf{u}  = -\nabla p -{Ri}_b \; \rho  \; \mathbf{\hat{y}}
   + \frac{1}{Re} \nabla^2 \mathbf{u},                                     \label{NS_1}
\end{align}
\begin{align}
\nabla \cdot \mathbf{u}  = 0,                                              \label{NS_2}
\end{align}
\begin{align}
\frac{\partial \rho }{\partial t} +\mathbf{u}  \cdot \nabla \rho =  \frac{1}{Re \; Pr} \nabla^2 \rho \label{NS_3}
\end{align}
where the bulk Richardson number $Ri_{\textrm{b}}$, Reynolds number $Re$, and the Prandtl number $Pr$ (always set to 1 in this study) are respectively 
defined as:
\begin{align}
Ri_{\textrm{b}} := \frac{\Delta_\rho \: g \, h}{\rho_0 \, U^2}, \;\;\;\;\;\;\;\;\;\;\;\;\;\;\;\;
  Re :=\frac{U \, h}{\nu}, \;\;\;\;\;\;\;\;\;\;\;\; \;\;\;\;  Pr := \frac{\nu }{\kappa }.
\label{numbers} 
\end{align}
Here $\mathbf{u}=(u,v,w)$ is the velocity field, $\kappa$ the thermal diffusivity, the total dimensional density is $\rho_0+\rho \Delta_\rho$, $p$ is the pressure  and $\nu$ is the kinematic viscosity. The boundary conditions are then
\begin{equation}
 u(x,\pm 1,z,t))=\pm 1 \quad \& \quad \rho(x,\pm 1,z,t)= \mp 1.
\end{equation}
which admit the steady 1D solution
\begin{equation}
 \mathbf{u}= y \, \mathbf{\hat{x}} \quad \& \quad \rho =  -y. 
 \label{basic2}
 \end{equation}
The (possibly large) disturbance fields away from this basic state,  
\begin{equation}
 \bu(x,y,z,t)=\mathbf{u}-y\, \mathbf{\hat{x}}, \qquad
 \rh(x,y,z,t)=\rho +y, 
\label{notation}
 \end{equation}
 conveniently satisfy homogeneous boundary conditions at $y=\pm 1$. Periodic boundary conditions
 are used in both the  ($x$) streamwise and ($z$) spanwise directions over wavelengths $L_x h$ and $L_z h$ so that the (non-dimensionalised) computational domain is $L_x \times 2 \times L_z$. The total `energy' of the disturbance  is taken as 
\begin{align}
 E :=   \langle  \,\half \bu^2+ \half Ri_b \rh^2 \,\rangle 
          \;   = \; \langle \, \half (\mathbf{u} - y \, \mathbf{\hat{x}}  )^2 + \half Ri_b(\rho +y )^2 \,\rangle .
 \end{align}
where 
$ \langle \,(\cdot)\,\rangle:= \frac{1}{V} \iiint \,(\cdot) \,dV $ is a volume average.

%---------------------------------------------
%
\subsection{Methods \label{meth}}
%
%---------------------------------------------

To find the  largest  energy growth that a (finite-amplitude)  perturbation $(\bu,\rh)$ can experience over a fixed time interval $[0,T]$ requires seeking the global maximum of the constrained Lagrangian
\begin{align}
\La \,:=\,\,& \langle \, \half \bu(\bx,T)^2+\half Ri_b \,\rh (\bx,T)^2 \, \rangle 
+ \lambda \biggl[ \,\langle \, \half \bu(\bx,0)^2+\half Ri_b \,\rh (\bx,0)^2 \, \rangle
-E_0 \,\biggr]+ \int^T_0 \langle \, \pi(\bx,t)\bnab \cdot \bu\, \rangle \\
&+\int^T_0 \langle \, \bnu(\bx,t) \cdot  \biggl[ \frac{\partial \bu}{\partial t}+y \frac{\partial \bu}{\partial x}+\tilde{v} \bhx+\bu \cdot \bnab \bu+\bnab \tilde{p}+Ri_b\, \rh \bhy-\frac{1}{Re} \nabla^2 \bu \biggr]\, \rangle \, dt \\
&+\int^T_0 \langle \, \tau(\bx,t) \cdot \biggl[ \frac{\partial \rh}{\partial t}+y \frac{\partial \rh}{\partial x}
-\tilde{v}+\bu \cdot \bnab \rh-\frac{1}{RePr} \nabla^2\rh   \biggr]\, \rangle \, dt
\end{align}
where $\lambda$, $\pi$, $\bnu$ and $\tau$ are the Lagrange multiplier fields imposing the constraints that the initial perturbation 
energy is $E_0$, the perturbation is incompressible,  and both the perturbation  Navier-Stokes equation and the density equation are satisfied respectively. Taking variations with respect  to all the degrees of freedom leads to the  Euler-Lagrange equations which, beyond the aforementioned constraints, comprise of the `dual' evolution equations for the fields, $\bnu=\nu_1 \bhx+\nu_2 \bhy+\nu_3 \bhz$ and $\tau$, 
\begin{align}
\frac{\partial \bnu}{\partial t}+y \frac{\partial \bnu}{\partial x}-\nu_1 \bhy-\bnu \cdot (\bnab \bu)^T+\bu \cdot \bnab \bnu+\bnab \pi+\frac{1}{Re} \nabla^2 \bnu &=\tau \bnab \rh -\tau \bhy,  \label{dual1}\\
\frac{\partial \tau}{\partial t}+y \frac{\partial \tau}{\partial x}+\bu \cdot \bnab \tau+\frac{1}{Re Pr} \nabla^2 \tau &= Ri_b \,\nu_2, \label{dual2}
\end{align}
the temporal end conditions
\beq
\bu(\bx,T)+\bnu(\bx,T)={\bf 0}, \qquad Ri_b \,\rh(\bx,T)+\tau(\bx,T) =0,  \label{final}
\eeq
and the initial conditions
\beq
\frac{\delta  \La}{\delta \bu(\bx,0)}:=\lambda \bu(\bx,0)-\bnu(\bx,0)={\bf 0},  \qquad \frac{\delta \La}{\delta \rh(\bx,0)}:=\lambda Ri_b \,\rh(\bx,0)-\tau(\bx,0) =0. \label{initial}
\eeq
To eliminate spatial boundary terms, $\bnu$ and $\tau$ are taken to obey the same homogeneous boundary conditions as $\bu$ and $\rh$, and $\bnu$ is further assumed   incompressible to automatically satisfy the Euler-Lagrange equation with respect to $\tilde{p}$. The solution strategy to find the global maximum of ${\cal L}$ is iterative, starting with a `guess' for the initial perturbation  $(\bu,\rh)$   which is then time-stepped across the time interval $[0,T]$ via the Navier-Stokes equation. The final values of $\bu$ and $\rh$ `initiate'  $\bnu$ and $\tau$ (via conditions (\ref{final})\,) for the time integration of the dual equations (\ref{dual1}) and (\ref{dual2}) {\em backwards} to $t=0$ where the fact that the equations (\ref{initial}) are generally not satisfied is used to update the form of the initial perturbation (subject to it staying of total energy $E_0$) in the direction of increasing ${\cal L}$. We use a simple steepest ascent method
where
\beq
(\, \bu^{(m+1)}(\bx,0),\rh^{(m+1)}(\bx,0) \,) \,=\, (\, \bu^{(m)}(\bx,0),\rh^{(m)}(\bx,0)\,)+ \frac{\eps}{\lambda} 
\biggl( 
\frac{\delta  \La}{\delta \bu(\bx,0)^{(m)}},     \frac{1}{Ri_b}    \frac{\delta \La}{\delta \rh(\bx,0)^{(m)}}
\biggr)
\eeq
% {\color{red} [CHECK THIS IS WHAT DANIEL DOES ? ]} Yes. just a 1/Ri factor was missing 
where, for example, $\bu(\bx,t)^{(m)}$ is the $m$th iterate, and $\lambda$ is subsequently chosen to ensure the new $(m+1)^{th}$ iterate has energy $E_0$ as discussed in \cite{Pringle10, Pringle12, Rabin12, Kerswell14}, but other approaches are possible (e.g. \cite{Monokrousos11, Duguet13, Cherubini10,Cherubini11,Cherubini13}). This direct and adjoint looping method is now well used (e.g. see the reviews \cite{Luchini14, Kerswell14}) but of course there is no  guarantee that the global maximum always emerges for this fully nonlinear problem.  The hoped output of the procedure is the optimal initial condition - the optimal disturbance - which experiences the largest growth over a time horizon $[0,T]$ of all initial conditions with the same initial total energy $E_0$ and so is a function of $T$ and $E_0$.

%
% Fig 1
%
\begin{figure}%  
    \begin{center}   
\includegraphics[angle=0,height=4.5cm, trim=2.1cm 0.28cm 2.1cm 0.3cm]{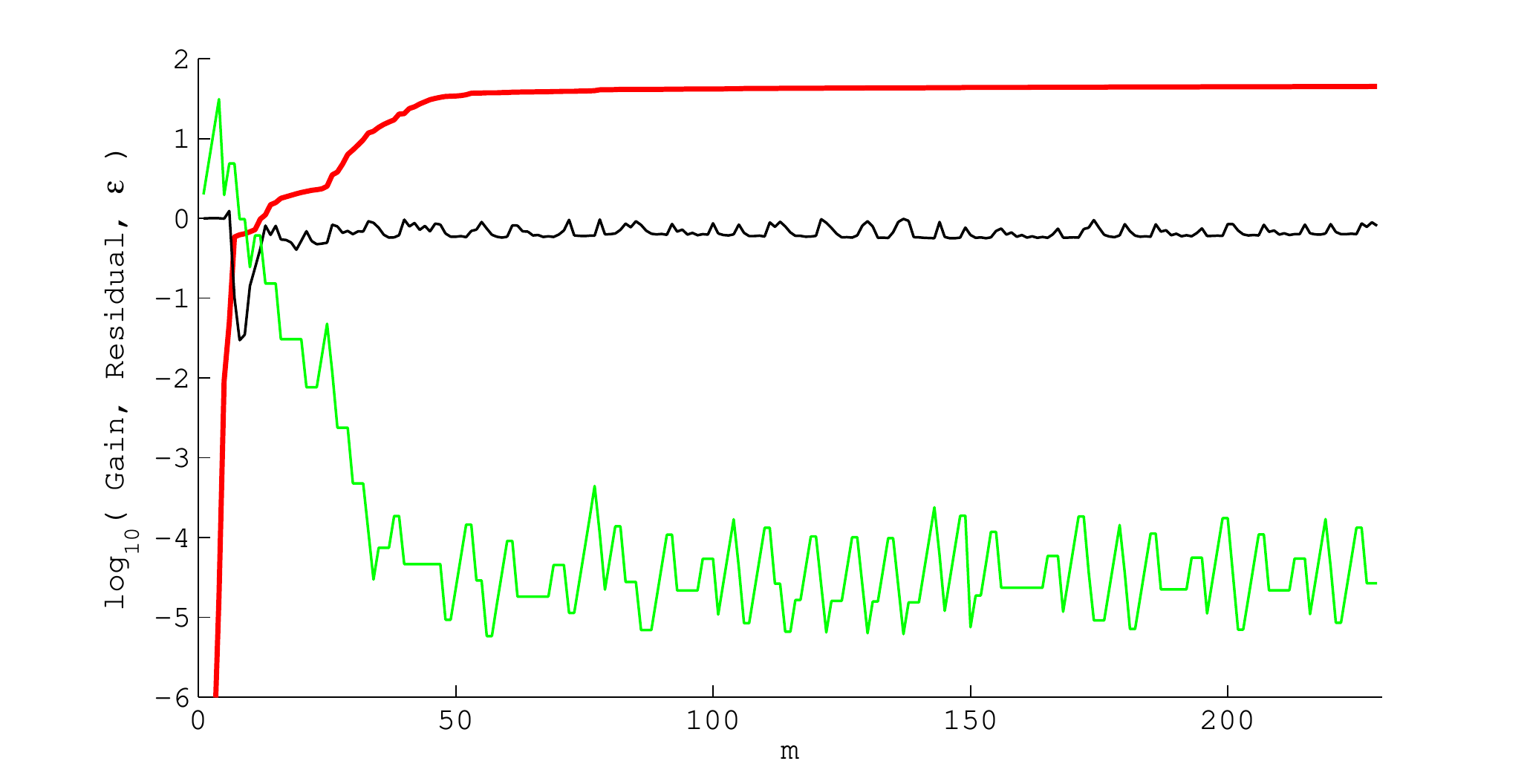}
\includegraphics[angle=0,height=4.5cm, trim=2.1cm 0.28cm 2.1cm 0.3cm]{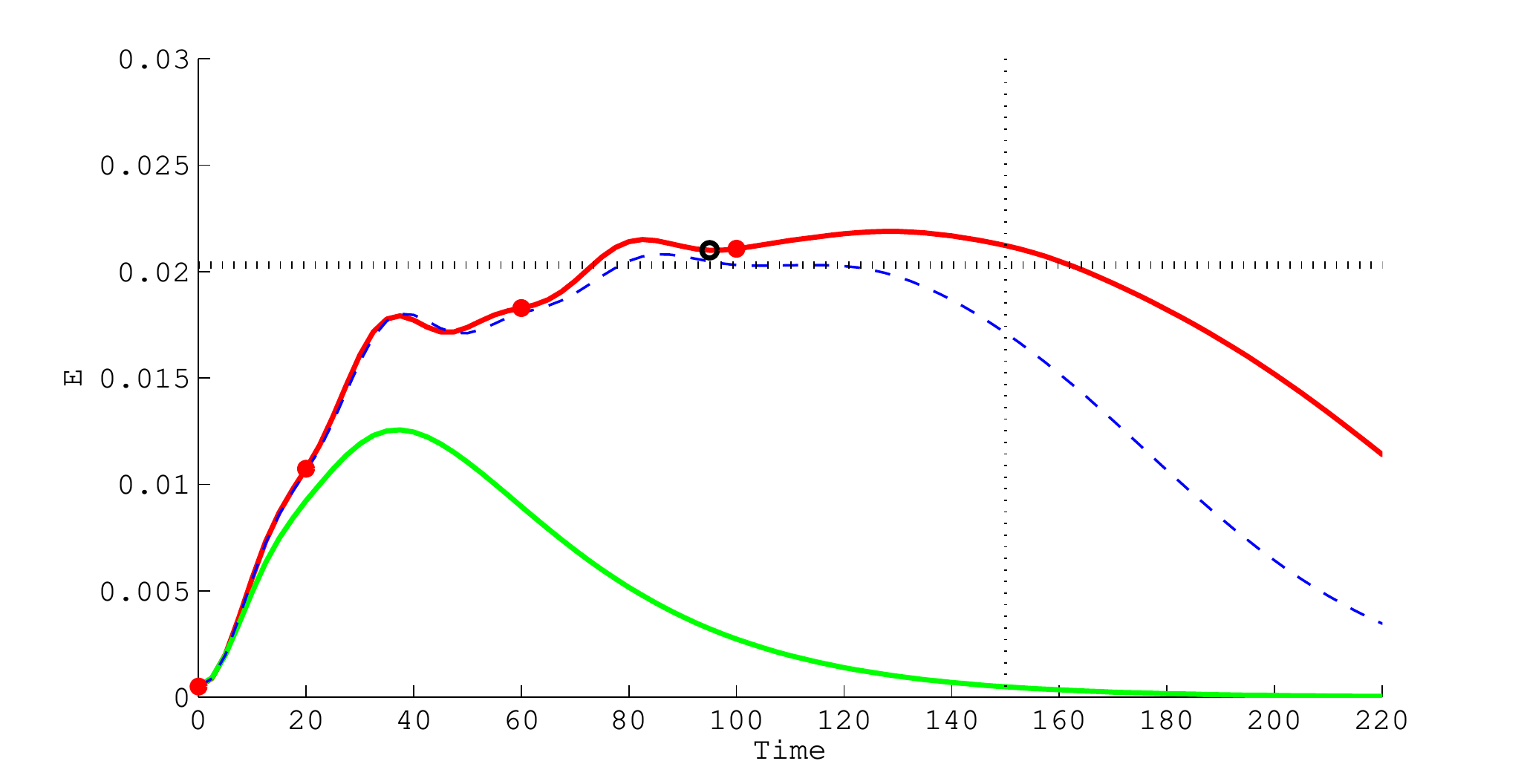}
%
%  -- NEW FIGURE-- evolution of perturbation used as guess for converging Snake solution (EQ)
%     t=0
%\adjincludegraphics[angle=0,height=3.5cm,width=12cm,trim={{.075\width} {.025\height} {.075\width}  {.06\height}},bb=-108   275   704   566]{./FiguresWide/A_GUESS_XZ_Fig1-eps-converted-to.pdf}  % t=0
\includegraphics[angle=0,height=3.5cm,width=12cm,clip, trim=2.1cm 0.28cm 2.1cm 0.3cm]{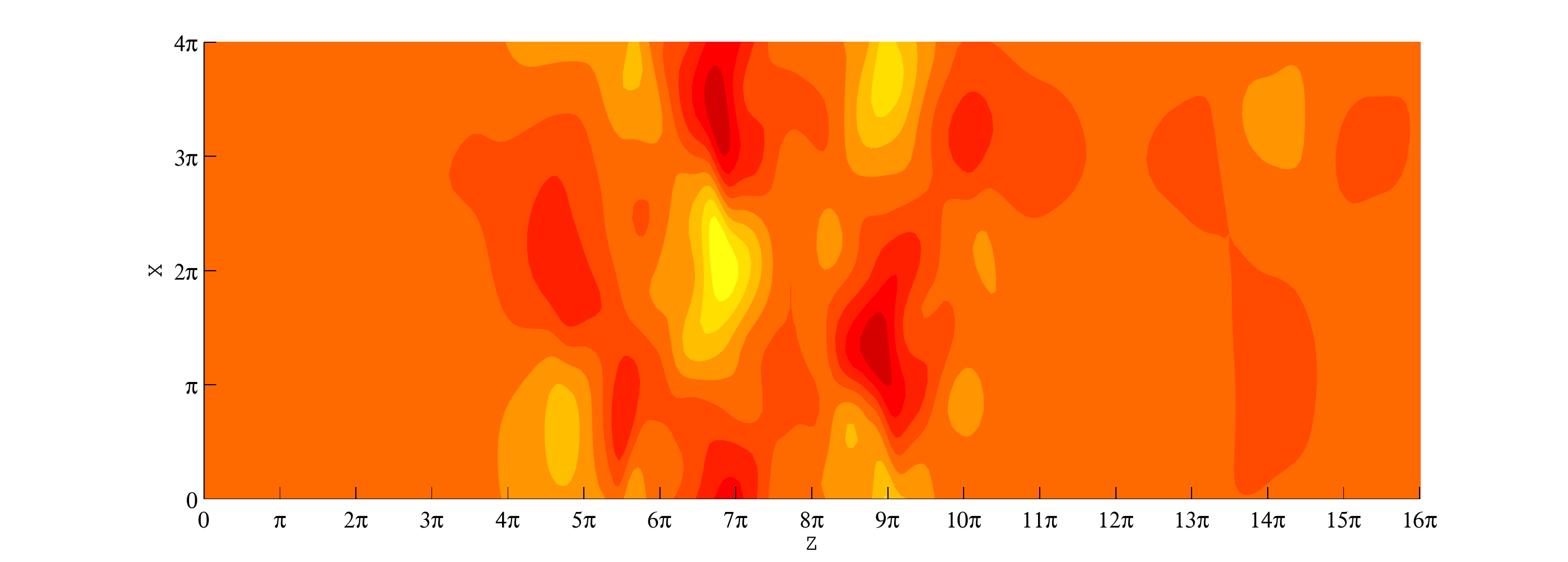}  % t=0
%\adjincludegraphics[scale=1, angle=0,height=4cm ,trim={{.06\width} {.029\height} {.03\width}  {.06\height}},clip]{./FiguresWide/A_GUESS_ISOS_Fig1-eps-converted-to.pdf} \\  % GUESS_T150_16pi_XZ_t0_ISOS
\includegraphics[scale=1, angle=0,height=4cm, trim=1.1cm 1.5cm 1.1cm 0.5cm]{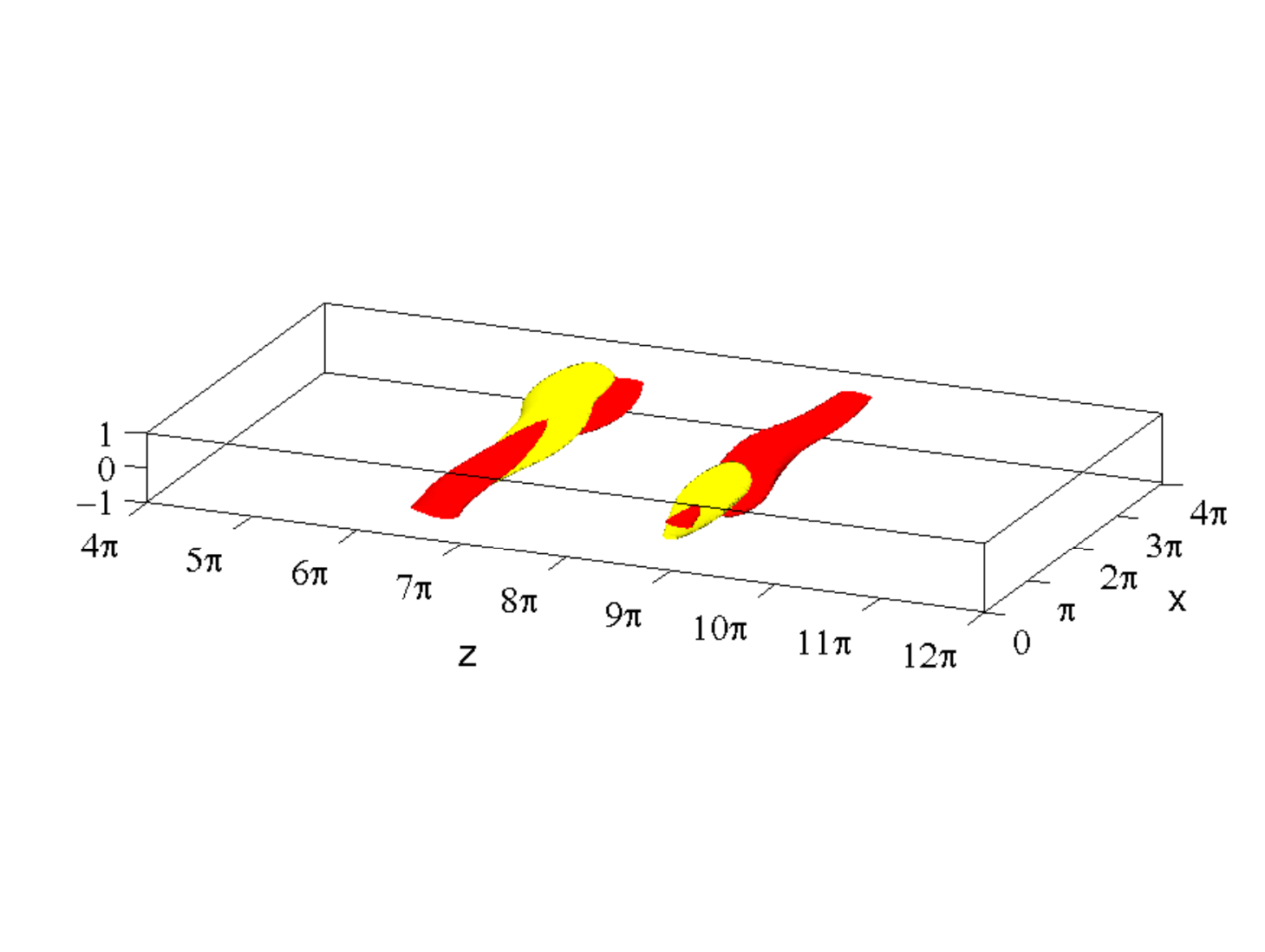} \\  
%  t=20
%\adjincludegraphics[angle=0,height=3.5cm,width=12cm    ,trim={{.075\width} {.025\height} {.075\width}  {.06\height}},clip]{./FiguresWide/B_GUESS_XZ_Fig1-eps-converted-to.pdf}  % t=20   % GUESS_T150_16pi_XZ_t20
\includegraphics[angle=0,height=3.5cm,width=12cm,trim=2.1cm 0.28cm 2.1cm 0.3cm]{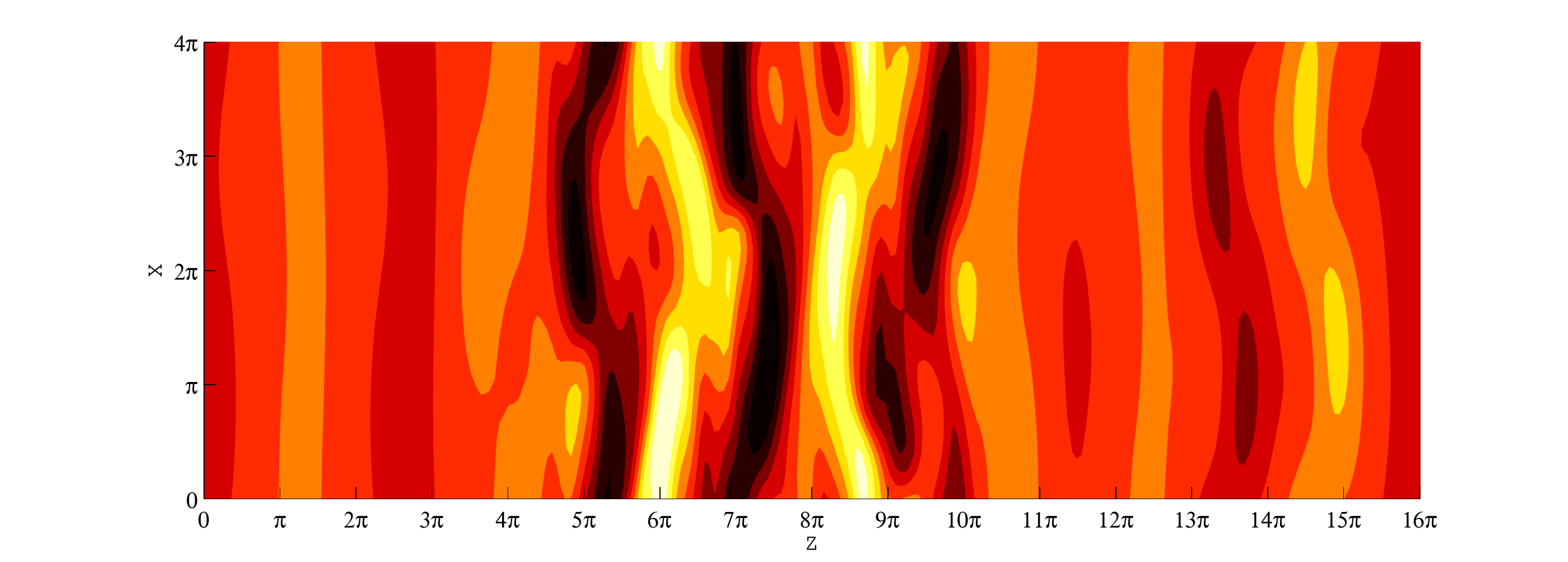}  % t=20   % GUESS_T150_16pi_XZ_t20
\includegraphics[angle=0,height=4cm]{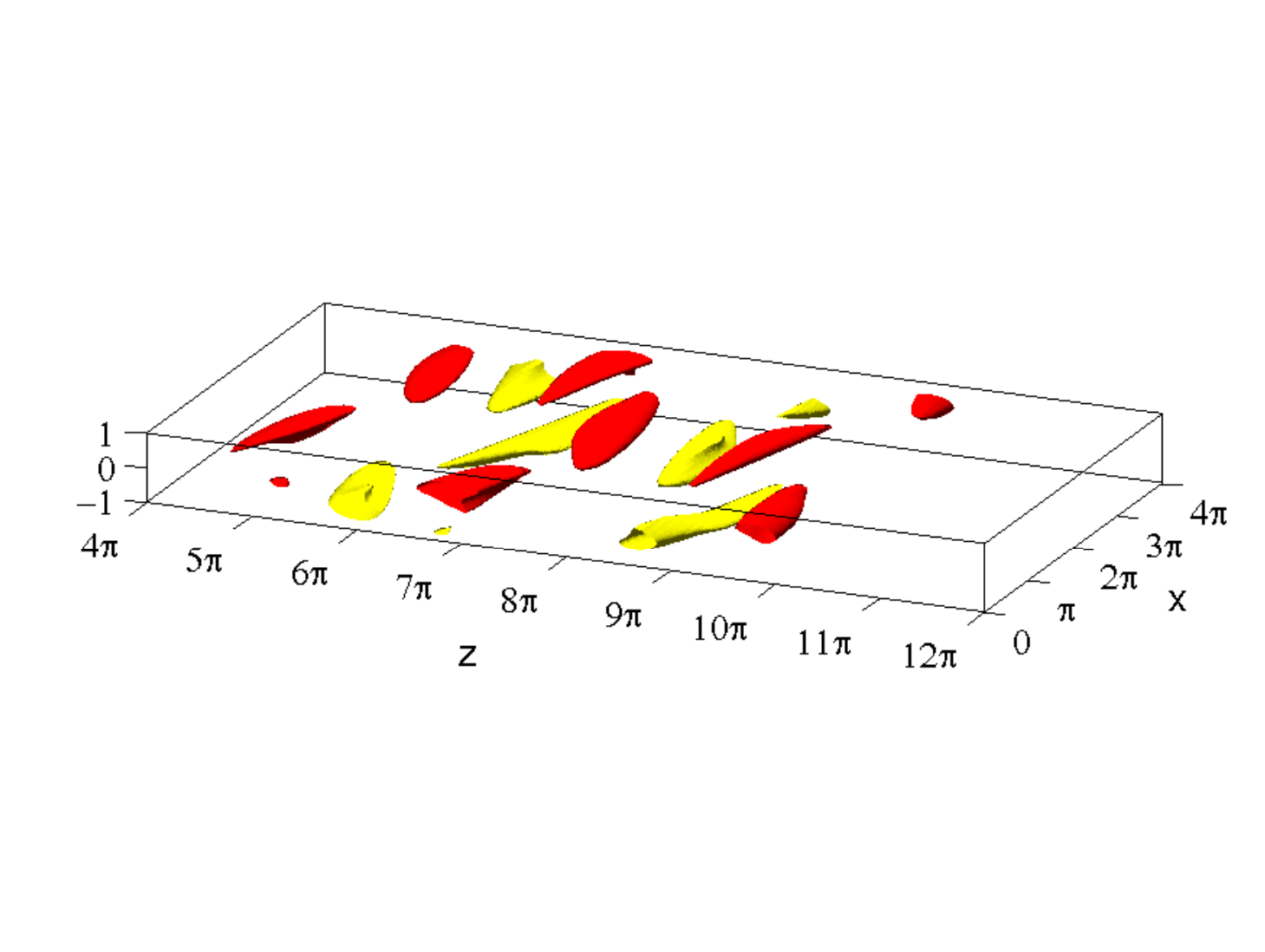} \\     % GUESS_T150_16pi_XZ_t20_ISOS
%
%  t=60
\includegraphics[angle=0,height=3.5cm,width=12cm,trim=2.1cm 0.28cm 2.1cm 0.3cm]{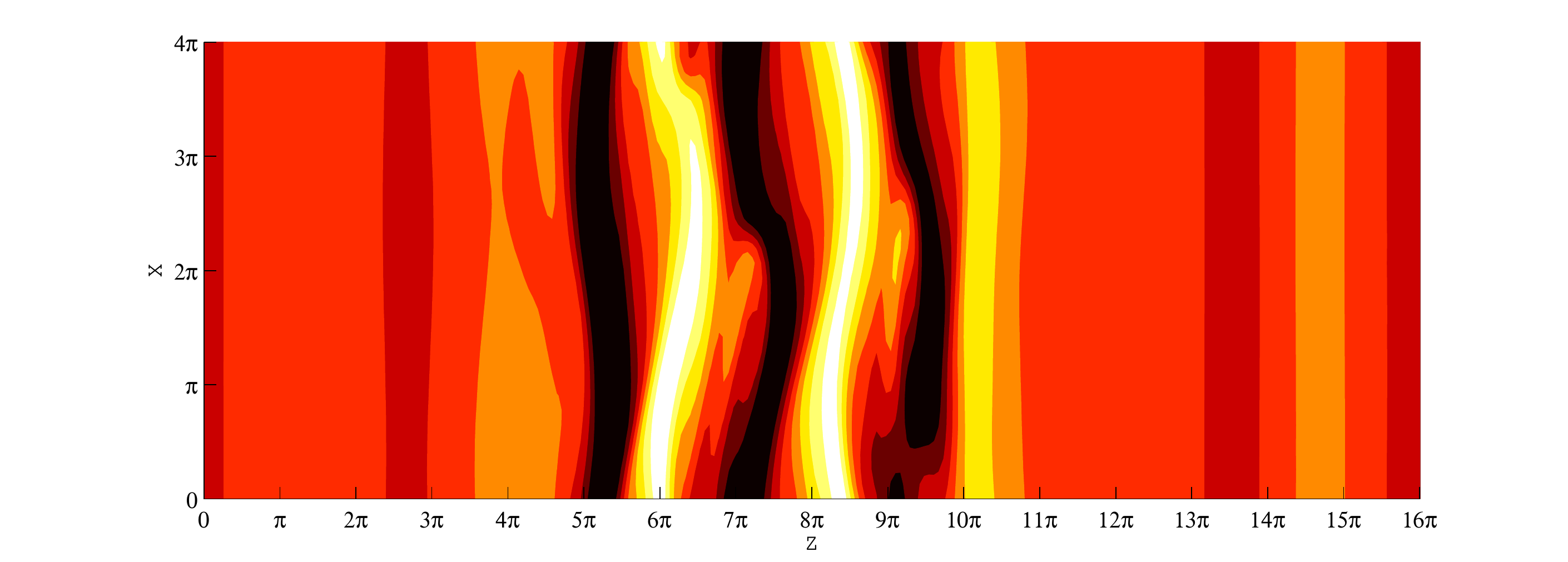}  % t=60 GUESS_T150_16pi_XZ_t60
\includegraphics[angle=0,height=4cm,trim=2.1cm 0.28cm 2.1cm 0.3cm]{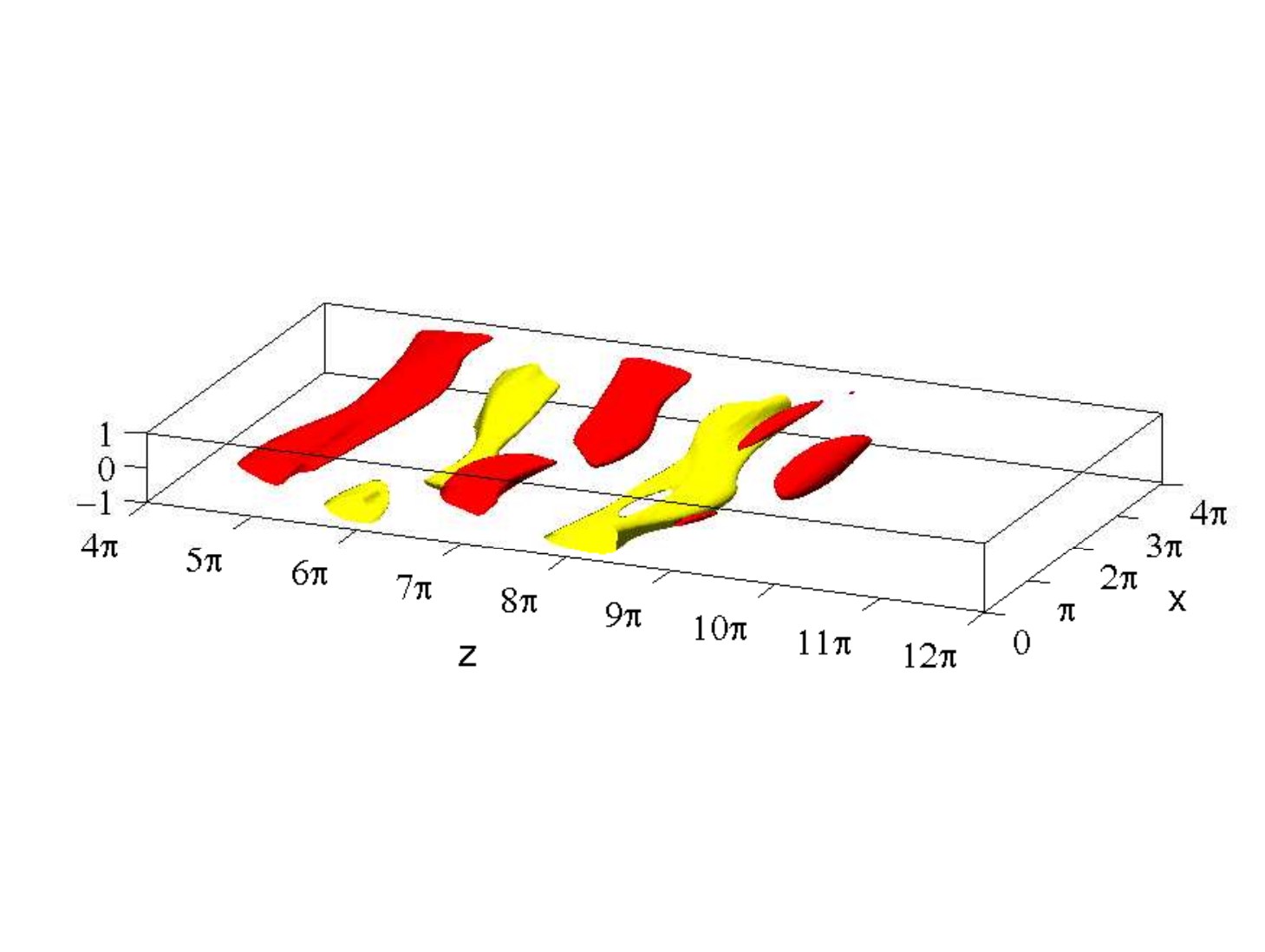} \\ 
%
%  t=100
\includegraphics[angle=0,height=3.5cm,width=12cm, trim=2.1cm 0.28cm 2.1cm 0.3cm]{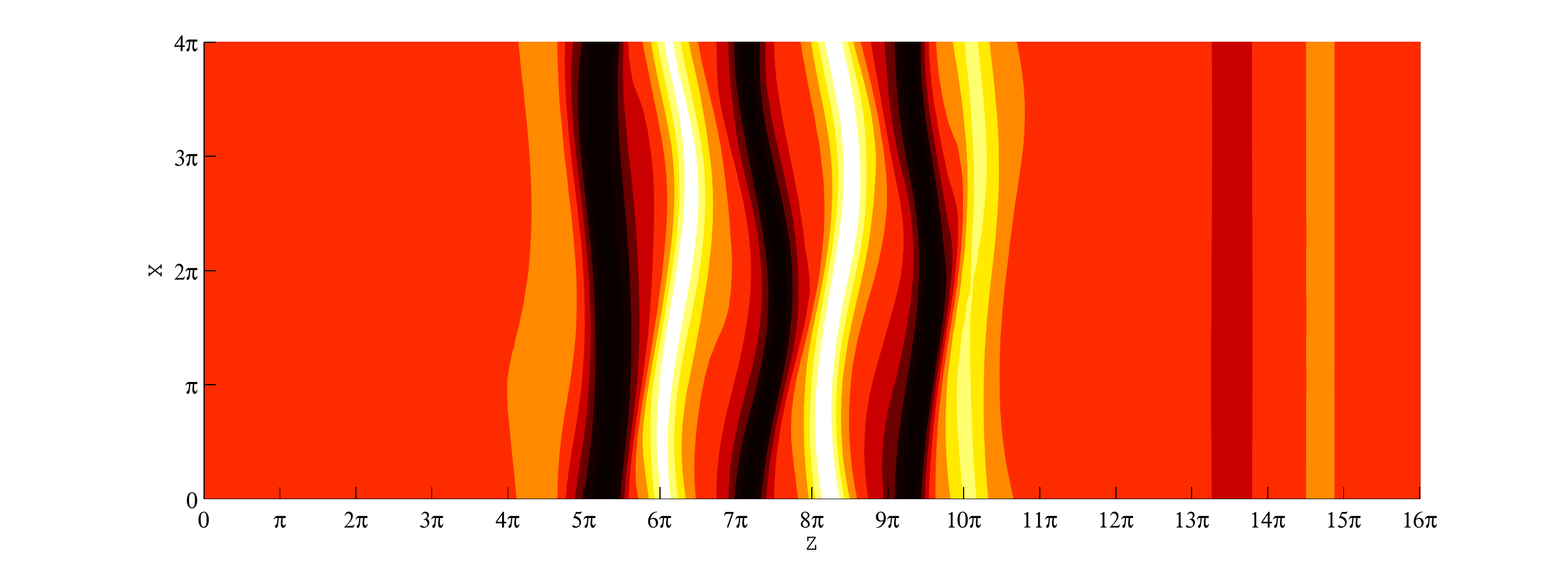}  % t=100    GUESS_T150_16pi_XZ_t100
\includegraphics[angle=0,height=4cm,trim=2.1cm 0.28cm 2.1cm 0.3cm]{./FiguresWide/C_GUESS_ISOS_Fig1-eps-converted-to.pdf} \\ 
\caption[]{
Top left: Gain (bold red), residual (black) and $\epsilon$ (green) for variational computations with 
initial energy $E_0=5 \times 10^{-4}$ at $Re=180$, $Ri_b=0$  and  $T=150$ in a $4 \pi \times 2 \times 16\pi$ domain (the computation is started with random noise so the gain is $O(10^{-9})$ after 1 step of the algorithm\,). Top right: time evolution of the  initial perturbation for iteration $m=15$, $m=50$  and $m=200$. The black circle indicates the state at $t=95$ from the initial condition at $n=200$ which is used as initial guess for the Newton-GMRES method. 
A horizontal dotted line shows the level of kinetic energy of the subsequently  converged solution shown in figure \ref{Guess_converged_16pi}. The red dots indicate the evolution of the perturbation shown below  at times $t=0$, $20$, $60$, $100$ (top to bottom). 
Left column, $xz$ cross-section at $y=0$ using 8 contour levels (between -0.58 and 0.059). Right column, isocontours showing $\pm 60\%$ of maximum streamwise perturbation velocity.
}
\label{variational_16pi}
\end{center}
\end{figure}

%
% Fig 2
%
\begin{figure}
    \begin{center}   
\begin{tabular}{c}
\adjincludegraphics[angle=0,height=3.0cm,width=12cm    ,trim={{.075\width} {.025\height} {.075\width}  {.06\height}},clip]{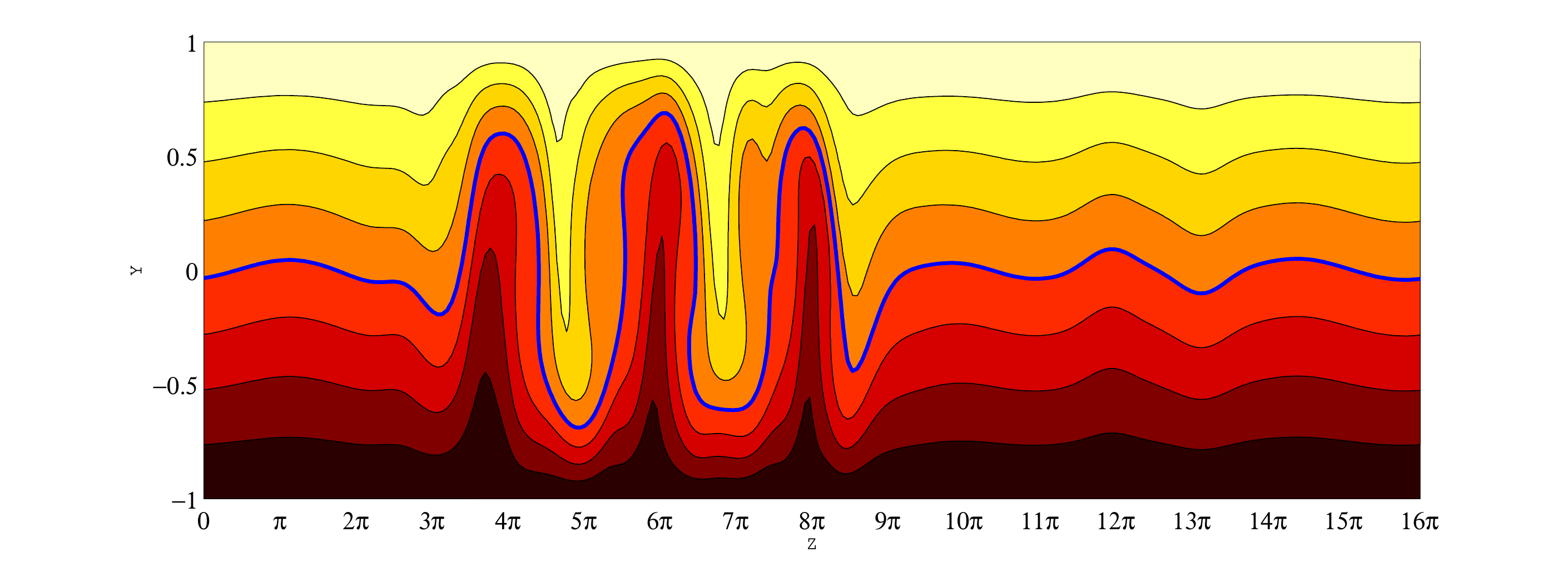}  % GUESS_WIDE_16pi_isoline0
\end{tabular}
\begin{tabular}{c}
\adjincludegraphics[angle=0,height=3.0cm,width=12cm    ,trim={{.075\width} {.025\height} {.075\width}  {.06\height}},clip]{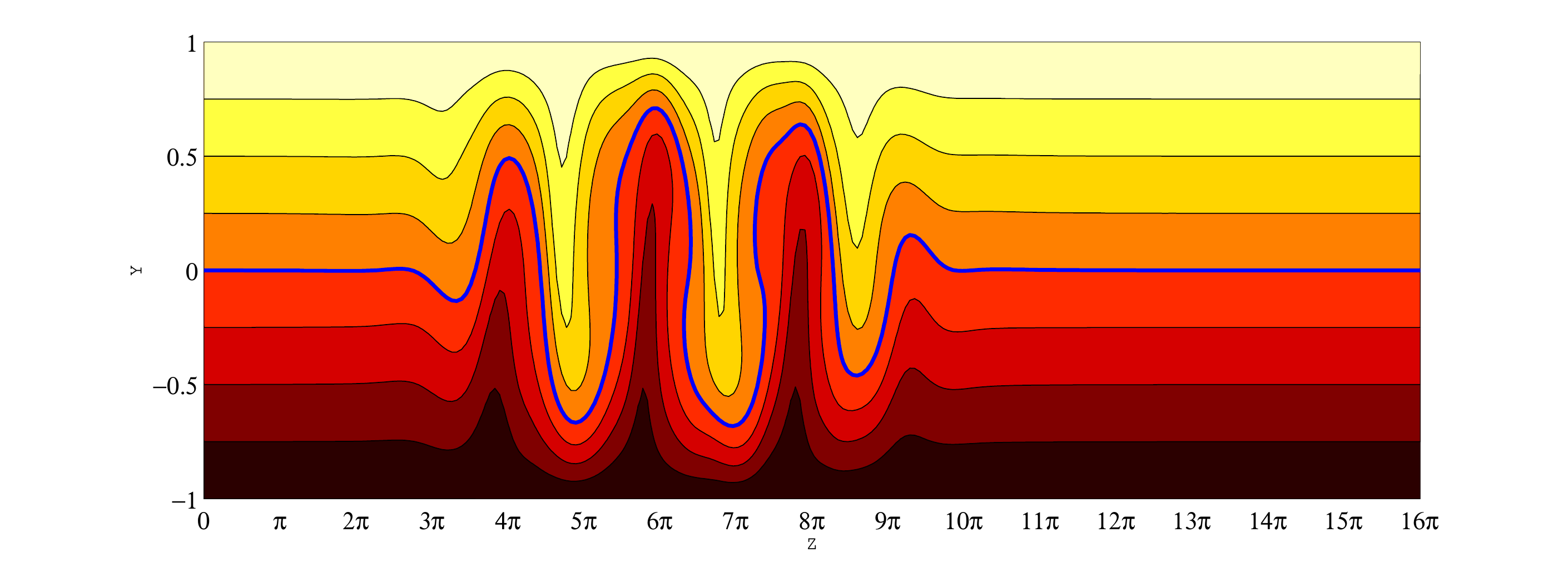}  % converged_WIDE_16pi_isoline0
\end{tabular}
 \caption[]{Top: the flow state at $t=95$ used as initial guess indicated by the black circle  in figure \ref{variational_16pi}.
Bottom: the  converged state after  26 Newton steps. There are 7 contour levels going from -1 to 1 with a  blue isoline indicating $u=0$.
}
 \label{Guess_converged_16pi}
\end{center}
\end{figure}

The status of the iterative procedure is monitored by computing the residual
%
%\beq
%{\cal R}(m)\,:=\, \biggl\langle \left| \frac{\delta \La}{\delta \bu(\bx,0)^{(m)}}\right|^2  \biggr\rangle/\langle \left| \bnu(\bx,0)^{(m)} \right|^2 \rangle         {\color{red} \,\,[\,STRATIFIED \,\,\, VERSION?!]}  
%\eeq
%
%  STRATIFIED VERSION  OF RESIDUAL, where ||.||^2 = <a,a>
\beq
{\cal R}(m)\,:=\,  \frac{  \,\langle \delta \La/\delta \bu(\bx,0)^{(m)}  \, \rangle  \,+\,
               \frac{1}{Ri_b} \langle \,  \delta \La/\delta \rh(\bx,0)^{(m)}  \, \rangle         }
{\langle\,  \bnu(\bx,0)^{(m)} \rangle \,+\,
\frac{1}{Ri_b}  \langle \tau(\bx,0)^{(m)} \rangle}
\eeq
which should approach 0 for convergence.
The time integrations of the Boussinesq equations forward in time and the dual equations backwards in time were carried out using an adapted version of the parallelized DNS code `Diablo' (\cite{Taylor08} and http://www.damtp.cam.ac.uk/user/jrt51/files.html)  which uses a third-order mixed Runge-Kutta-Wray/Crank-Nicolson timestepper. The horizontal directions are periodic and treated pseudospectrally, while a second-order finite-difference discretization is used in the cross-stream direction. The resolution used was typically 64 Fourier modes per $2 \pi$ in $x$ and $z$  and 128 finite difference points in $y$. If needed, this resolution was doubled to ensure numerical accuracy. Diablo was coupled it to a Newton-Raphson-GMRES algorithm \citep{Viswanath07} which allowed ECS to be converged from a good guess and continued around in parameter space (most notably by varying $Ri_b$). In what follows, it is usually more convenient to plot the gain $\G$ of iterates which is defined as
\beq
\G(E_0,T)\,:= \frac{E(T)}{E(0)}
\eeq
where $E_0=E(0)$ is the a priori-fixed initial perturbation energy: maximizing this, of course, is equivalent to maximizing the final total energy over all perturbations with given initial energy $E_0$.

%-------------------------------------------------------------------------------------
%
\section{Results \label{res}}
%
%-------------------------------------------------------------------------------------

%
%   Fig 3 0ld
%
%\begin{figure} 
%   \begin{center}   
%\adjincludegraphics[angle=0,height=3.5cm ,trim={{.02\width} {.02\height} {.08\width} {.058\height}},clip]{./FiguresWide/GMRES_OUTPUT_snake.eps}
%\adjincludegraphics[angle=0,height=3.5cm ,trim={{.02\width} {.02\height} {.08\width} {.058\height}},clip]{./FiguresWide/GMRES_OUTPUT_SAMVariational.eps}\\%
% \caption[]{Left: Initial guess at $t=95$ in $4\pi \times 2 \times 16 \pi$ domain converges in 26  Newton steps. The residual is defined in the usual way as the $L^2$-norm of the error from solving the Navier-Stokes equation.
%Right: 
%  }
%\label{GMRES_output}
%\end{center}
%\end{figure}

%-------------------------------------------------------------------------------------
\subsection{Wide domain pCf: Rediscovering Snakes \label{wide}}
%-------------------------------------------------------------------------------------

Optimal energy growth calculations were performed at $Re=180$  with stratification turned off ($Ri_b=0$) in the two wide domains, $4\pi \times 2 \times 8 \pi$  and $4\pi \times 2 \times 16 \pi$ where the snake solutions are known to exist  \cite{Schneider10a,Schneider10b}. The calculations were initiated with random initial conditions (energy scattered in the lowest modes) normalised so that the total kinetic energy was $E_0$. If $E_0$ is too small, only the immediate neigbourhood of  the constant shear solution is explored with a nonlinear version of the 2D linear optimal -  a global set of streamwise rolls -  emerging as the optimal.  If $E_0$ is too large, the optimal perturbation leads to another  global state resembling multiple copies of Nagata's solution (e.g. see figure 6.13 in \cite{Rabin13}). Flows states from this optimal evolution could, presumably, be used to converge Nagata's solution  but this was not pursued (it would be more numerically efficient to treat a narrower domain which supports just one spanwise wavelength if that was an objective). At intermediary $E_0$, the optimal perturbation is more spanwise localised and stays spanwise localised as it evolves into a state  suggestive of a snake solution (see figure  \ref{variational_16pi}).

Figure \ref{variational_16pi} (top left) shows the convergence features of the optimal growth calculation at this intermediary initial energy using $T=150$ in the $16\pi$ wide box. The iterative algorithm is clearly struggling to converge - the residual remains $O(1)$ - yet the gain has levelled off and most importantly, a plateau has emerged in the time evolution of the optimal iterates (see top right figure). This signals a  close approach to a constant-energy saddle (either an equilibrium or travelling wave) and it is from here that we take a flow snapshot (specifically at $t=95$)  which is spanwise localised:  see figure \ref{Guess_converged_16pi}(upper). This state converged in 26 Newton steps to a very similar looking steady solution - see figure \ref{Guess_converged_16pi}(lower) - which is the equilibrium snake solution  (hereafter referred to as EQ)  of \cite{Schneider10a}.

\begin{figure} 
    \begin{center}   
\adjincludegraphics[angle=0,height=4.5cm ,trim={{.08\width} {.02\height} {.08\width} {.058\height}},clip]{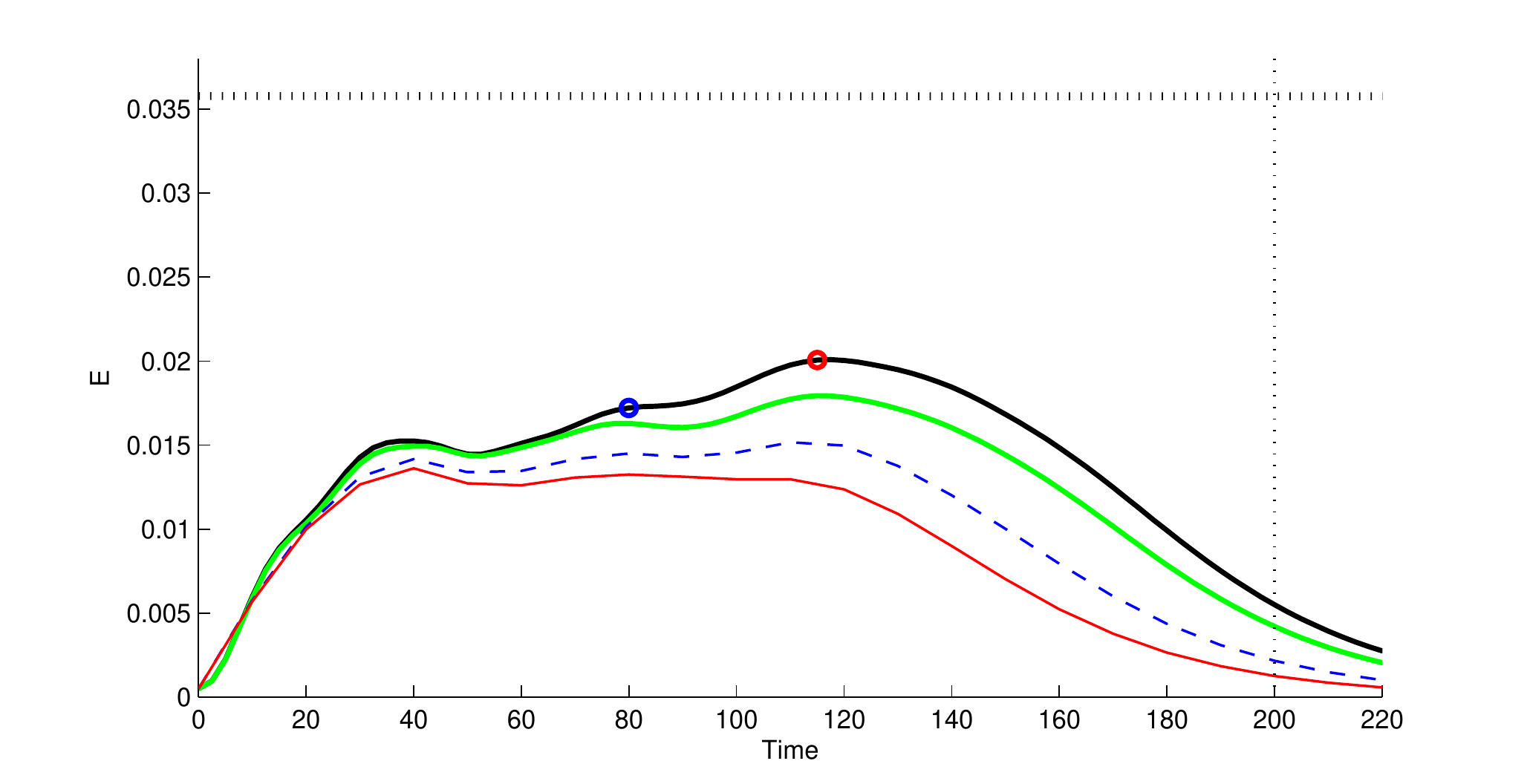}
\adjincludegraphics[angle=0,height=4.5cm ,trim={{.02\width} {.02\height} {.08\width} {.058\height}},clip]{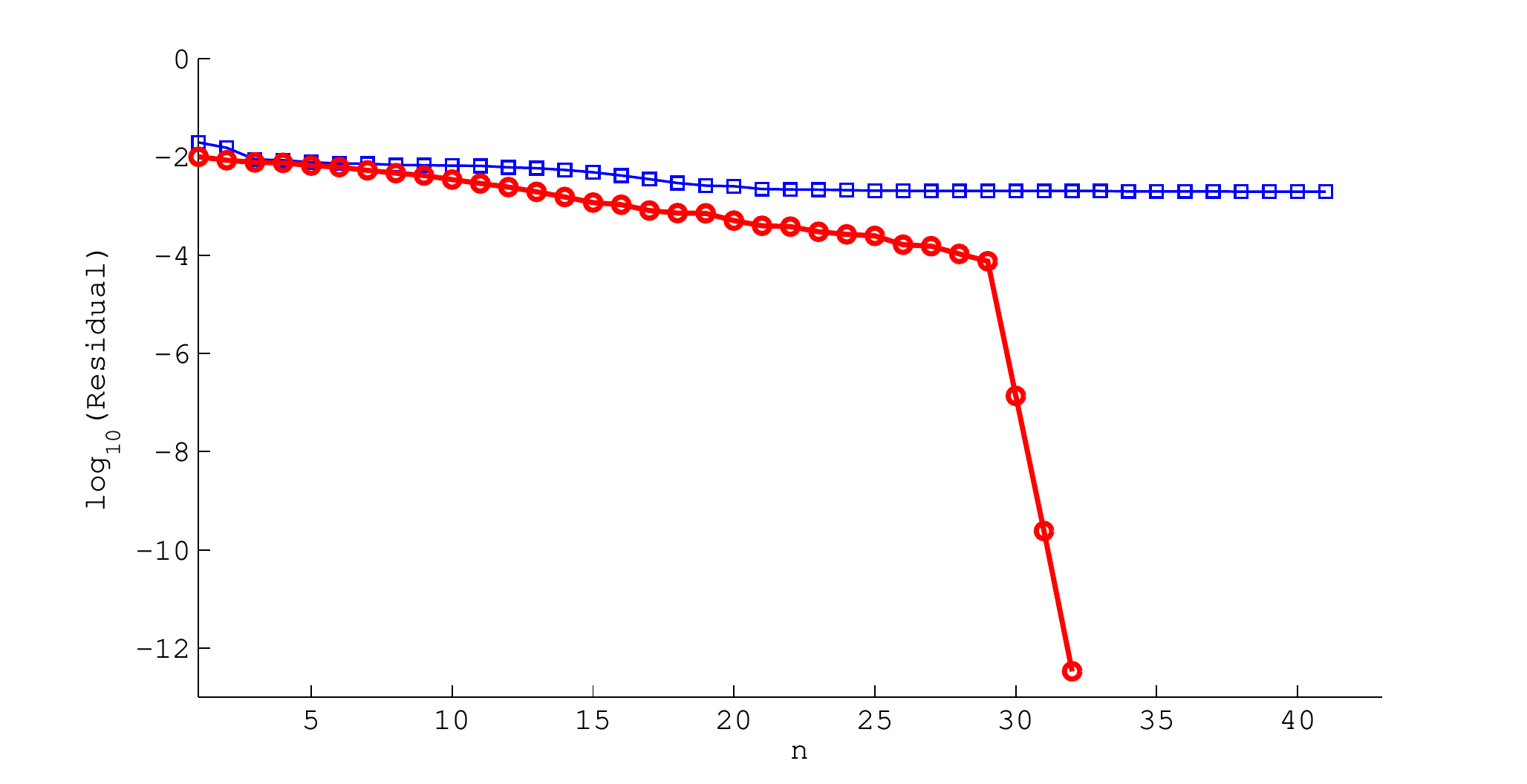}
 \caption[]{Left: Time evolution of  initial perturbations at various iterative stages  ($m=200$ red; $m=485 $ blue dashed, $m=2280 $ green and $m= 3000$, black) in a domain $4 \pi \times 2 \times 8 \pi$ with $T=200$. Dots on the $m=3000$ iterate indicate states used as initial guess for the Newton-GMRES method. The horizontal dotted line shows the level of kinetic energy $E$ of the solution  converged from the $t=115$ flow (bottom right, in figure \ref{Guess_converged_8pi}). Right:  The results of applying  Newton-GMRES to states taken at $t=80$ and $115$. 
Squares indicate the unsuccessful attempt starting from the state at  $t=80$ whereas the circles show successful convergence after 32 Newton steps starting with the  state at  $t=115$.}
\label{variational_8pi}
\end{center}
\end{figure}

\begin{figure}
    \begin{center}   
\begin{tabular}{c}
\adjincludegraphics[angle=0,height=3cm,width=7cm    ,trim={{.075\width} {.025\height} {.075\width}  {.06\height}},clip]{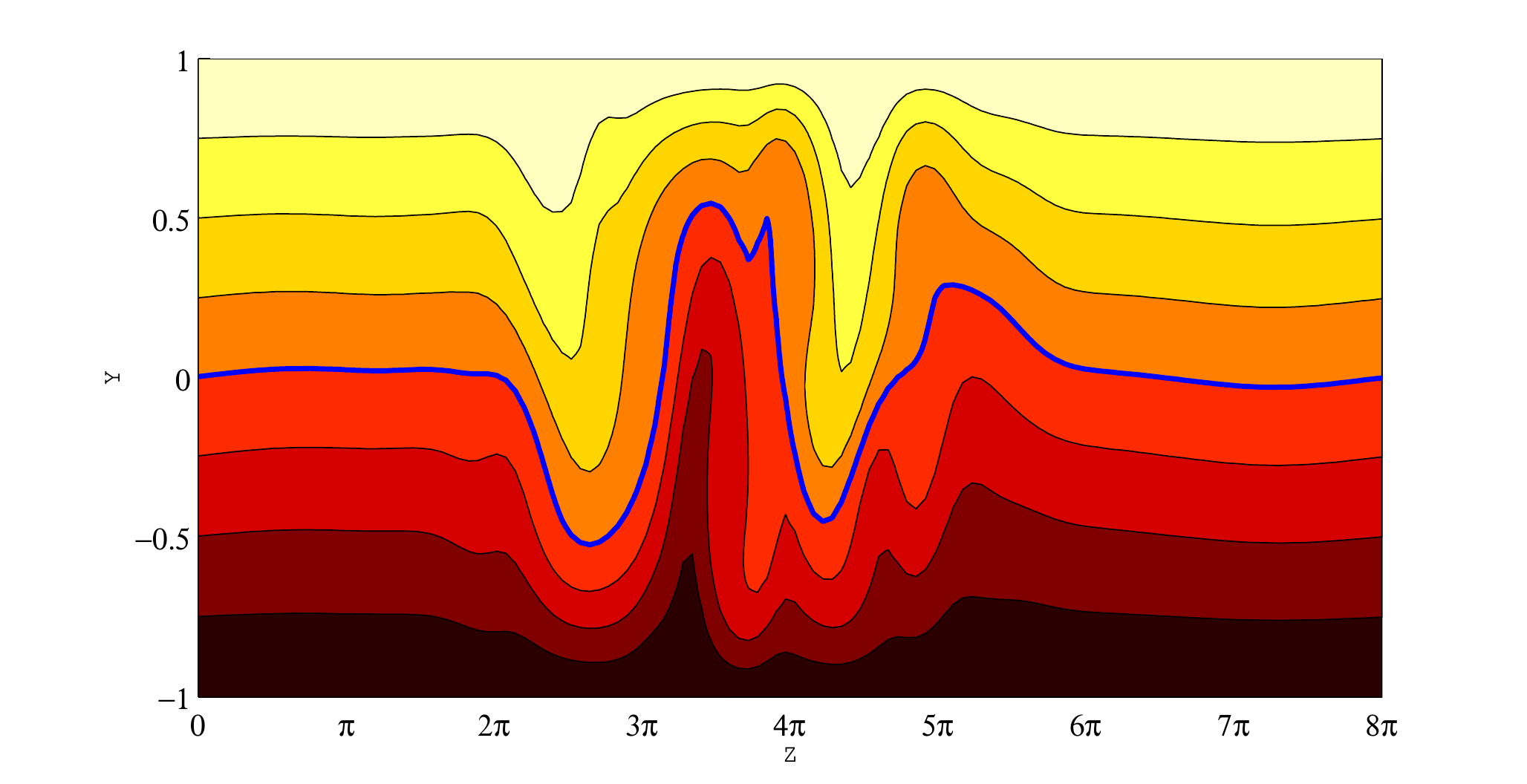}% SAM_BOX_T200_t80
\adjincludegraphics[angle=0,height=3cm,width=7cm    ,trim={{.075\width} {.025\height} {.075\width}  {.06\height}},clip]{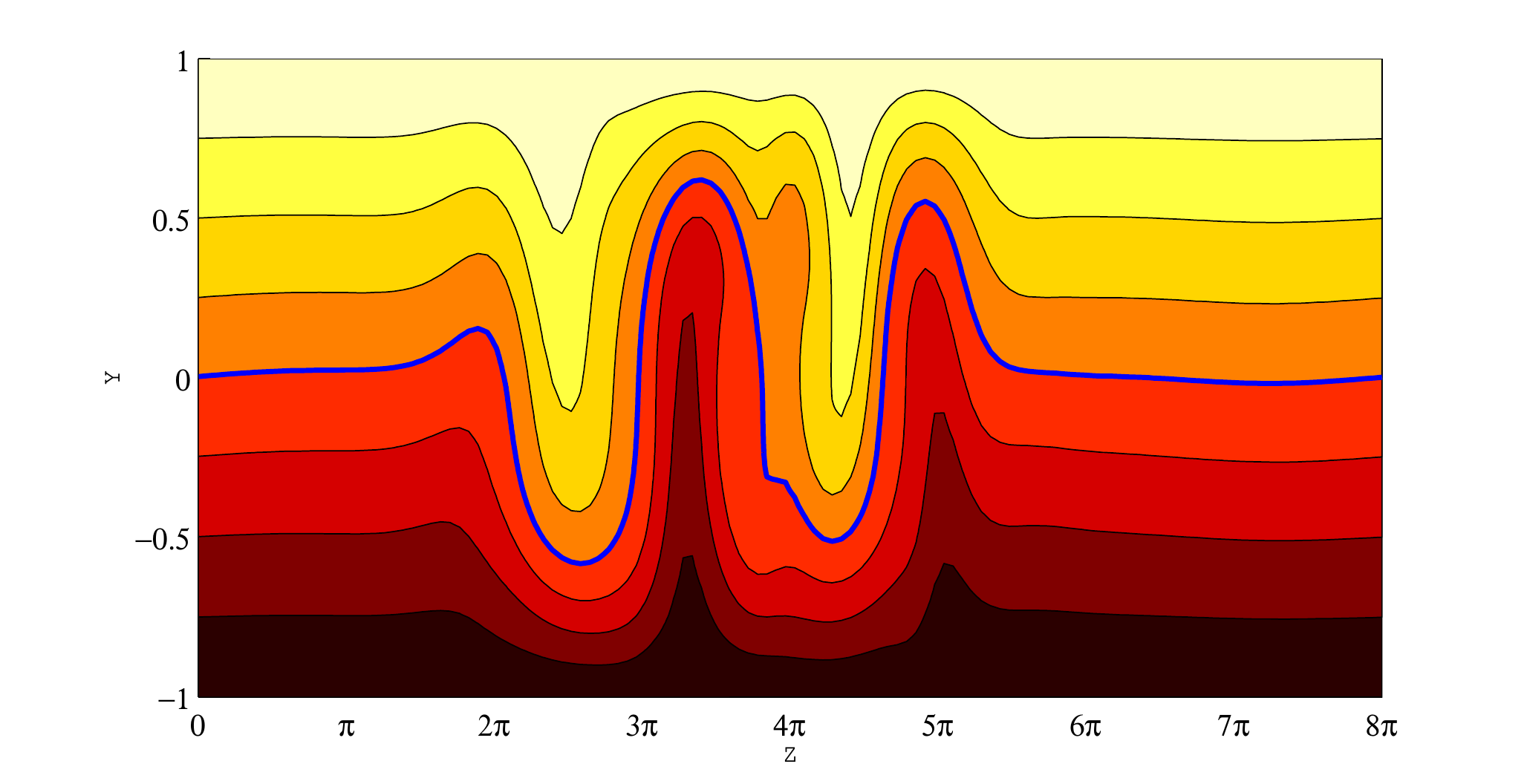}\\% SAM_BOX_T200_t115
\end{tabular}
\begin{tabular}{c}
\adjincludegraphics[angle=0,height=3.cm,width=7cm    ,trim={{.075\width} {.025\height} {.075\width}  {.06\height}},clip]{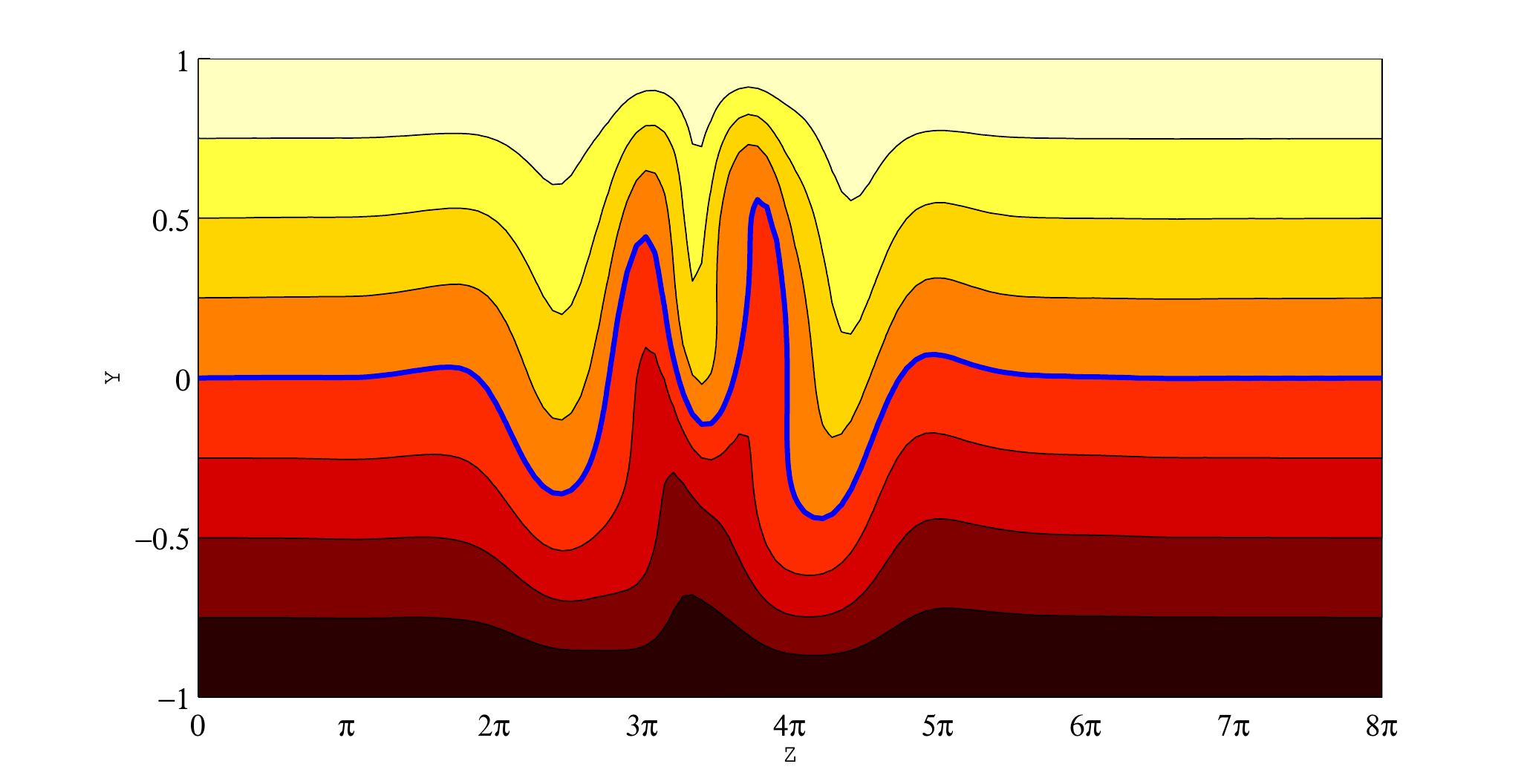}%  SAM_BOX_T200_t80_STALLED
\adjincludegraphics[angle=0,height=3.cm,width=7cm    ,trim={{.075\width} {.025\height} {.075\width}  {.06\height}},clip]{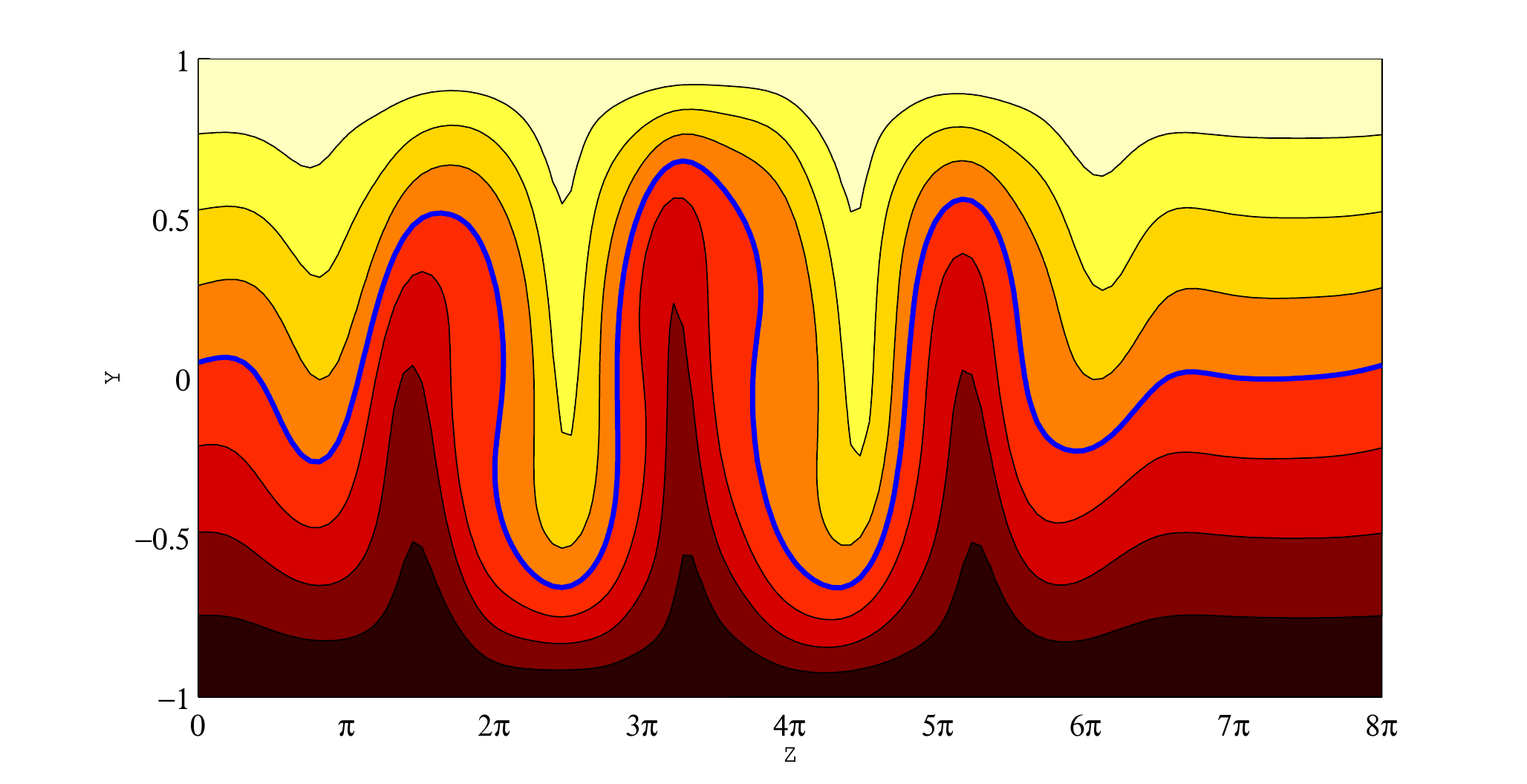}\\% SAM_BOX_T200_CONVERGED_TW
\end{tabular}
 \caption[]{
Top left, state at $t=80$ (blue circle in figure \ref{variational_8pi}) with the unconverged result after  $n=41$ iterations in bottom left.
Top right, state at $t=115$ (red circle in figure \ref{variational_8pi}) and the converged solution achieved after $n=32$ iterations in bottom right.
The  domain is $4 \pi \times 2 \times 8\pi$ and there are 7 contour levels going from -1 to 1 with a blue isoline indicating $u=0$. }
\label{Guess_converged_8pi}
\end{center}
\end{figure}

The optimisation procedure proceeded much more slowly in the narrower domain $4\pi \times 2 \times 8\pi$ for reasons which are unclear but again a plateau is eventually established in the optimal evolution: see figure \ref{variational_8pi}(left). Two flow states were extracted from this - see figure \ref{Guess_converged_8pi}(upper) - with one converging and one apparently not - see  figure \ref{Guess_converged_8pi}(lower) and the convergence behaviour in figure \ref{variational_8pi}(right). The converged state this time was the travelling wave snake solution (hereafter referred to as TW) of \cite{Schneider10a}.
% (TW was also found by edge tracking at Re=800 in the $4 \pi \times 2 \times 16 \pi$ geometry whereas edgetracking in $4 \pi \times 2 \times 8\pi$  at $Re=800$ produced EQ7-1 of \cite{Gibson14}).

%
% Fig 3
%

%
% Fig 4
%

%
% Fig 5
% 
\begin{figure}
\adjincludegraphics[angle=0,width=7cm,trim={{.05\width} {.025\height} {.05\width}  {.049\height}},clip]{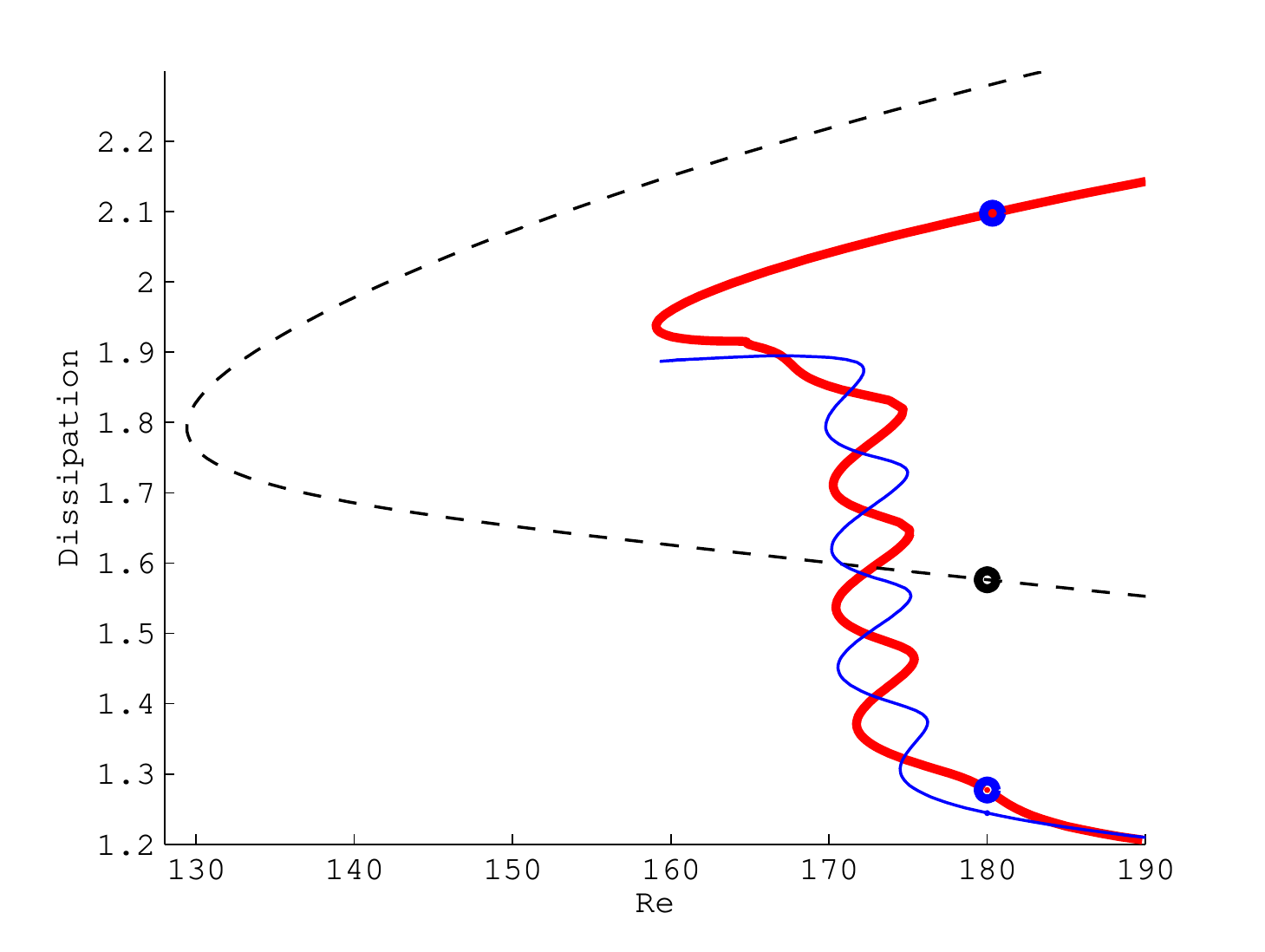}
\adjincludegraphics[angle=0,width=7cm,trim={{.05\width} {.025\height} {.05\width}  {.049\height}},clip]{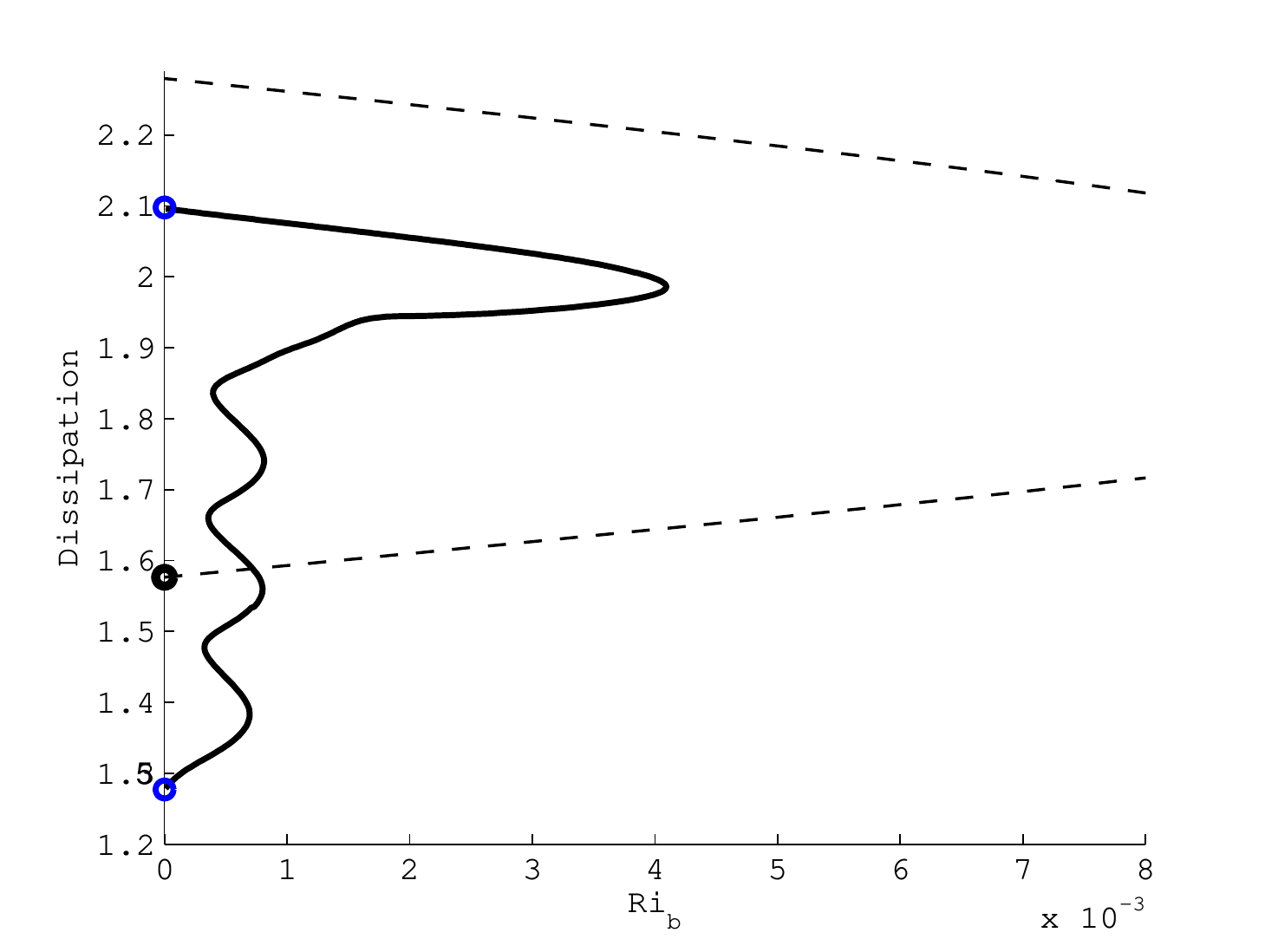}
\caption{Left: continuation in Reynolds number of the solutions converged in the domain $4\pi \times 2 \times 16\pi$ (the TW snake  is the thin  blue line and the  EQ snake is the thick red line): compare with figure 2 from \cite{Schneider10b}. Circles indicates the states
where continuation in $Ri_b$ was started as shown in the right figure which shows continuation in $Ri_b$ at $Re=180$ for the Nagata solution (spanwise wavenumber $k=1$)
and the EQ snake. }
\label{SNAKES_1_2}
\end{figure}

%
% Fig 6
%
%  continuation to NEGATIVE strat.  connection to convective rolls. BLOW-UP

\begin{figure}  
    \begin{center}   %, width
\adjincludegraphics[angle=0,width=12.0cm,trim={{.05\width} {.022\height} {.05\width}  {.04\height}},clip]{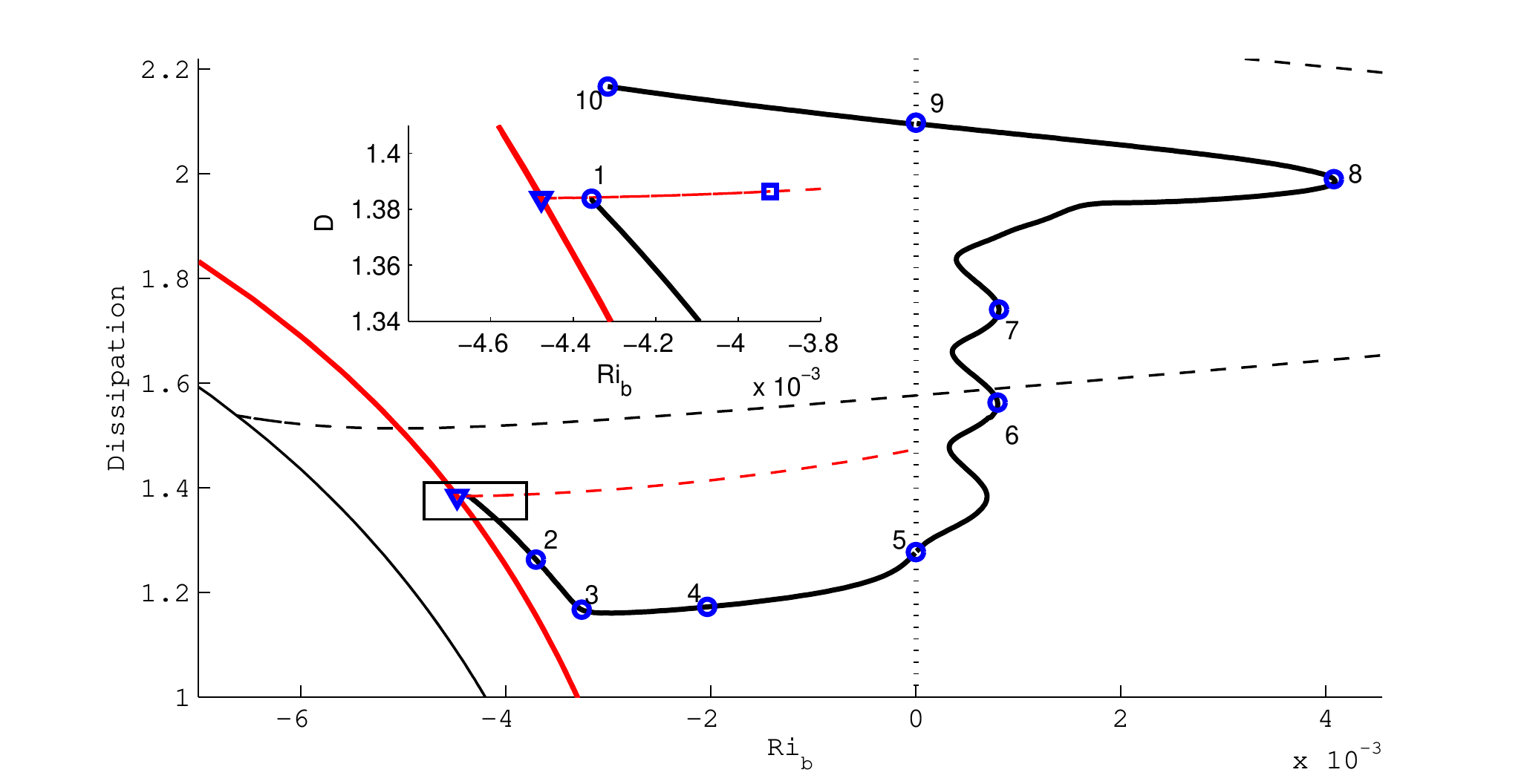}%
 \caption[]{
Continuation in $Ri_b$ of the equilibrium snake solution at fixed $Re=180$ extended to unstable stratification in the $4 \pi \times 2 \times 16 \pi$ box. 
The solution at each circle marked from 1 to 10 is visualised in figure \ref{Snaking_points}. The inset shows the connection
of the snake solution to the Nagata solution with spanwise wavenumber $\beta=1.5$ (the black dashed line is the Nagata solution with spanwise wavenumber $\beta=1$ shown in figure \ref{SNAKES_1_2}). the triangle marks the connection to the 2D convective rolls of Rayleigh-Benard convection.  }
      \label{SNAKE2_full_continuation}
    \end{center}
  \end{figure}

Once the snake solutions, EQ and TW, had been found, they could be traced around in $(Re,Ri_b)$ parameter space ($Pr=1$ to keep things manageable) using the Newton-GMRES algorithm. Fixing $Ri_b=0$ and varying $Re$ in the $16\pi$ wide box reproduced  the `snakes and ladders' plot of \cite{Schneider10b} (their figure 2) confirming the identity of the solutions: see figure \ref{SNAKES_1_2}(left). Interestingly, fixing $Re=180$ and varying $Ri_b$ also shows snaking in EQ. Further continuing this solution to negative $Ri_b$ (unstable stratification) - see figure \ref{SNAKE2_full_continuation} - reveals that EQ connects to a  Nagata's solution (of different spanwise wavenumber $\beta=1.5$  to that which the TW connects in figure 2 of \cite{Schneider10b} where $\beta=1$\,) just before this bifurcates off a 2D convective roll solution familiar from the Rayleigh-Benard problem (the extra shear from the boundaries does not affect this solution except to determine its orientation).  Salewski et al. \cite{Salewski} have also recently  found this same bifurcation sequence in rotating plane Couette flow which is known to be closely related to the Rayleigh-Benard problem.

\begin{figure}
    \begin{center}
\begin{tabular}{c}
\textsuperscript{1}\adjincludegraphics[angle=0,height=2.cm,width=7cm    ,trim={{.075\width} {.025\height} {.075\width}  {.06\height}},clip]{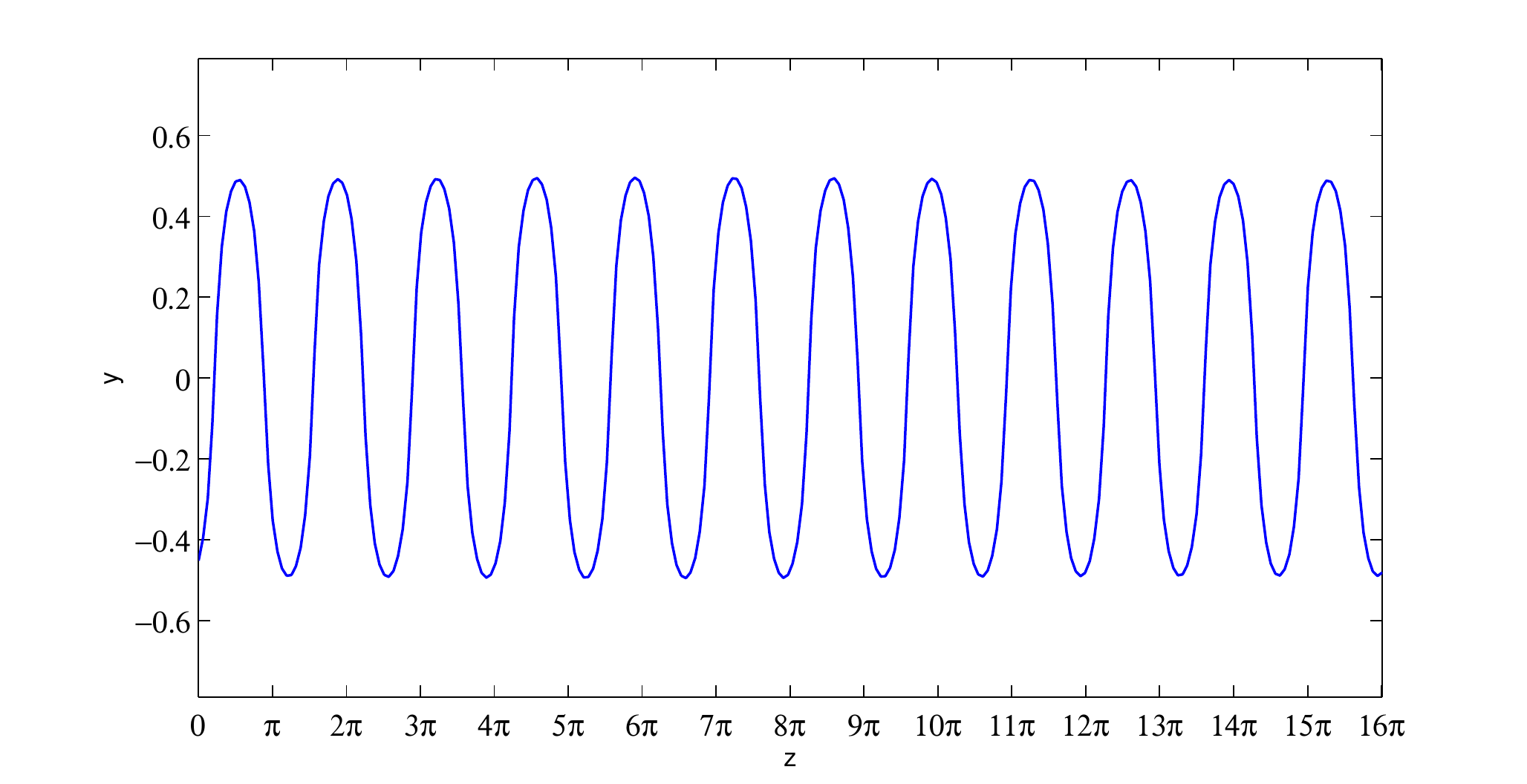}%
\textsuperscript{\; 2}\adjincludegraphics[angle=0,height=2cm,width=7cm    ,trim={{.075\width} {.025\height} {.075\width}  {.06\height}},clip]{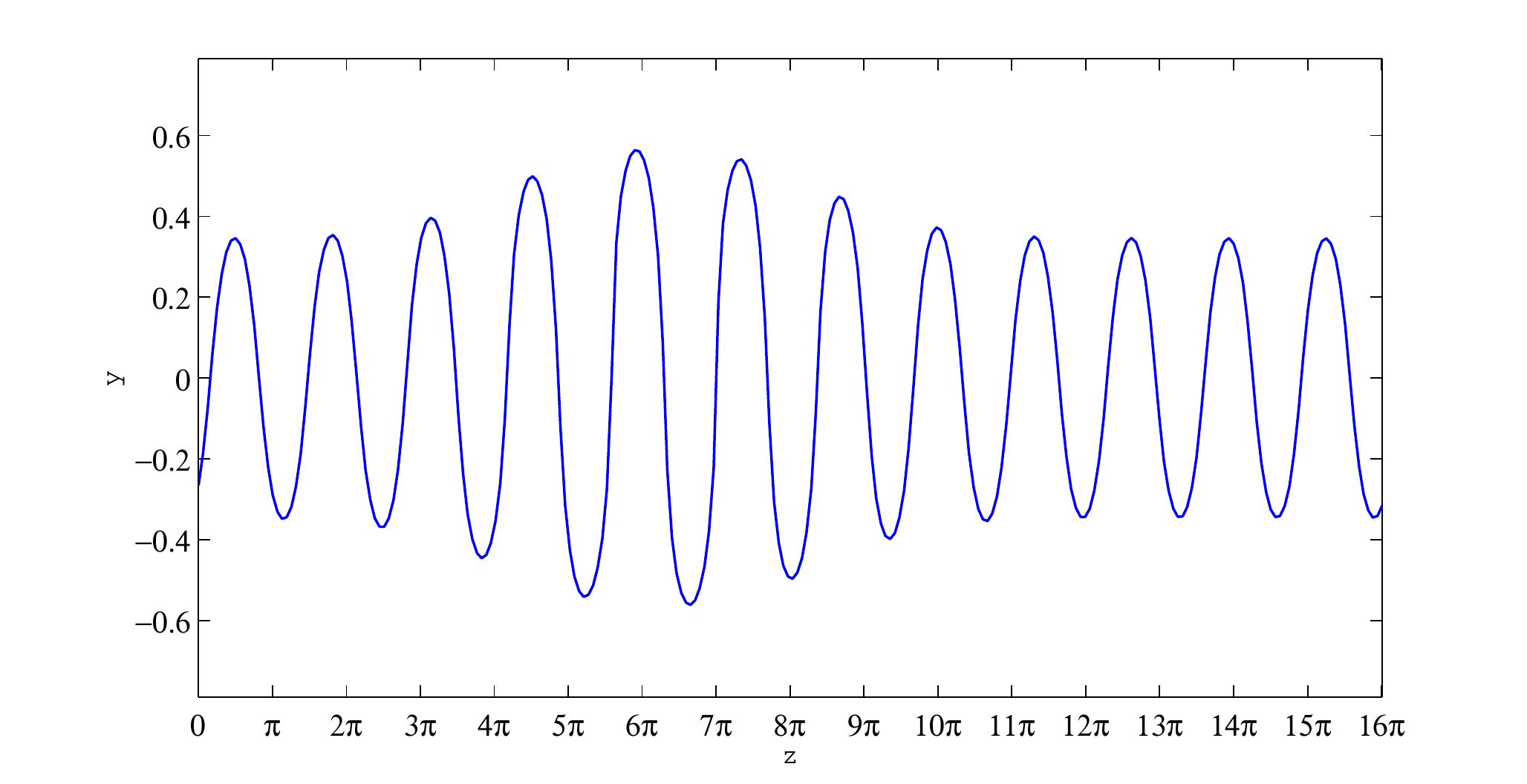}\\%
\end{tabular}
\begin{tabular}{c}
\textsuperscript{3}\adjincludegraphics[angle=0,height=2.cm,width=7cm    ,trim={{.075\width} {.025\height} {.075\width}  {.06\height}},clip]{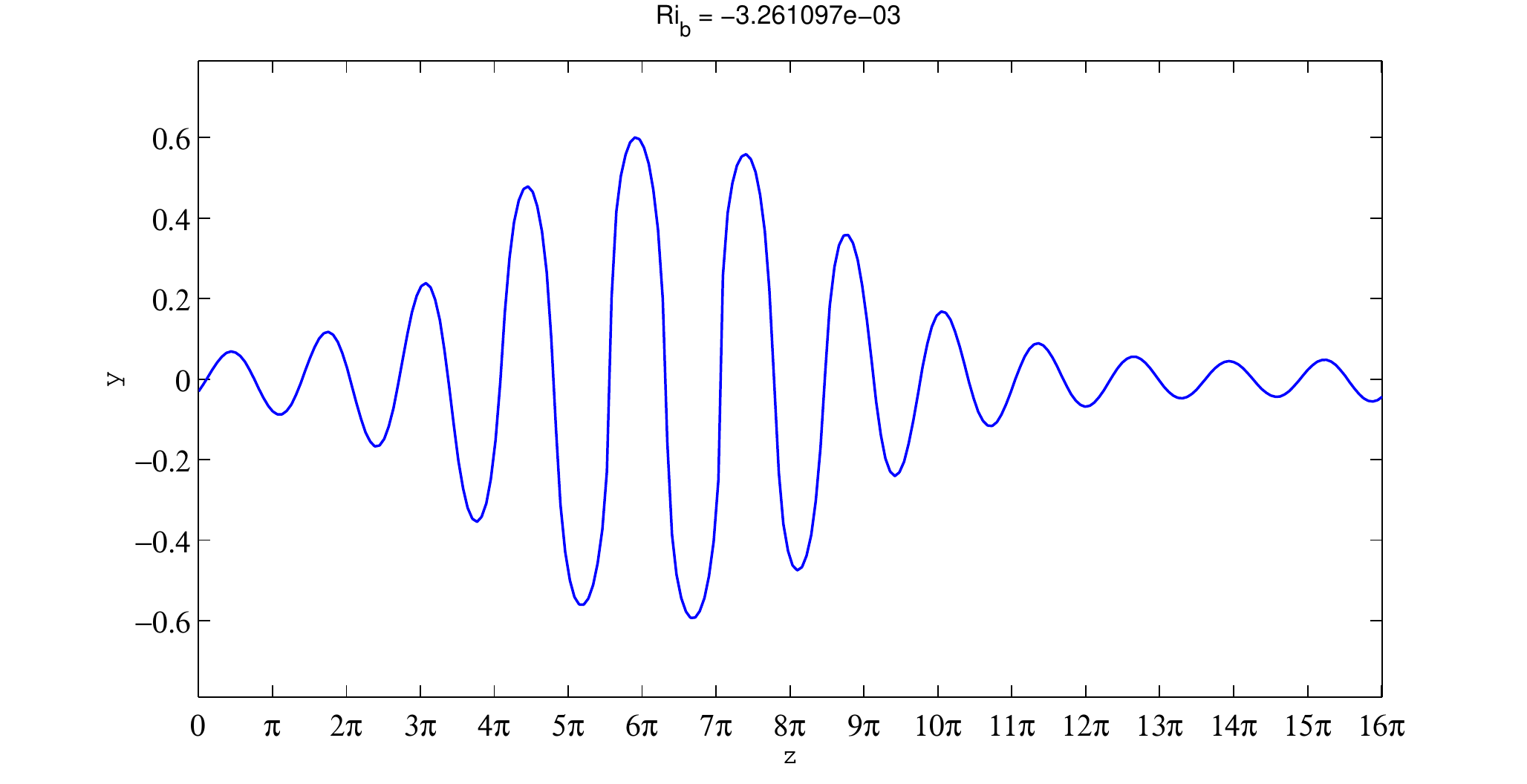}% 3
\textsuperscript{\; 4}\adjincludegraphics[angle=0,height=2.cm,width=7cm    ,trim={{.075\width} {.025\height} {.075\width}  {.06\height}},clip]{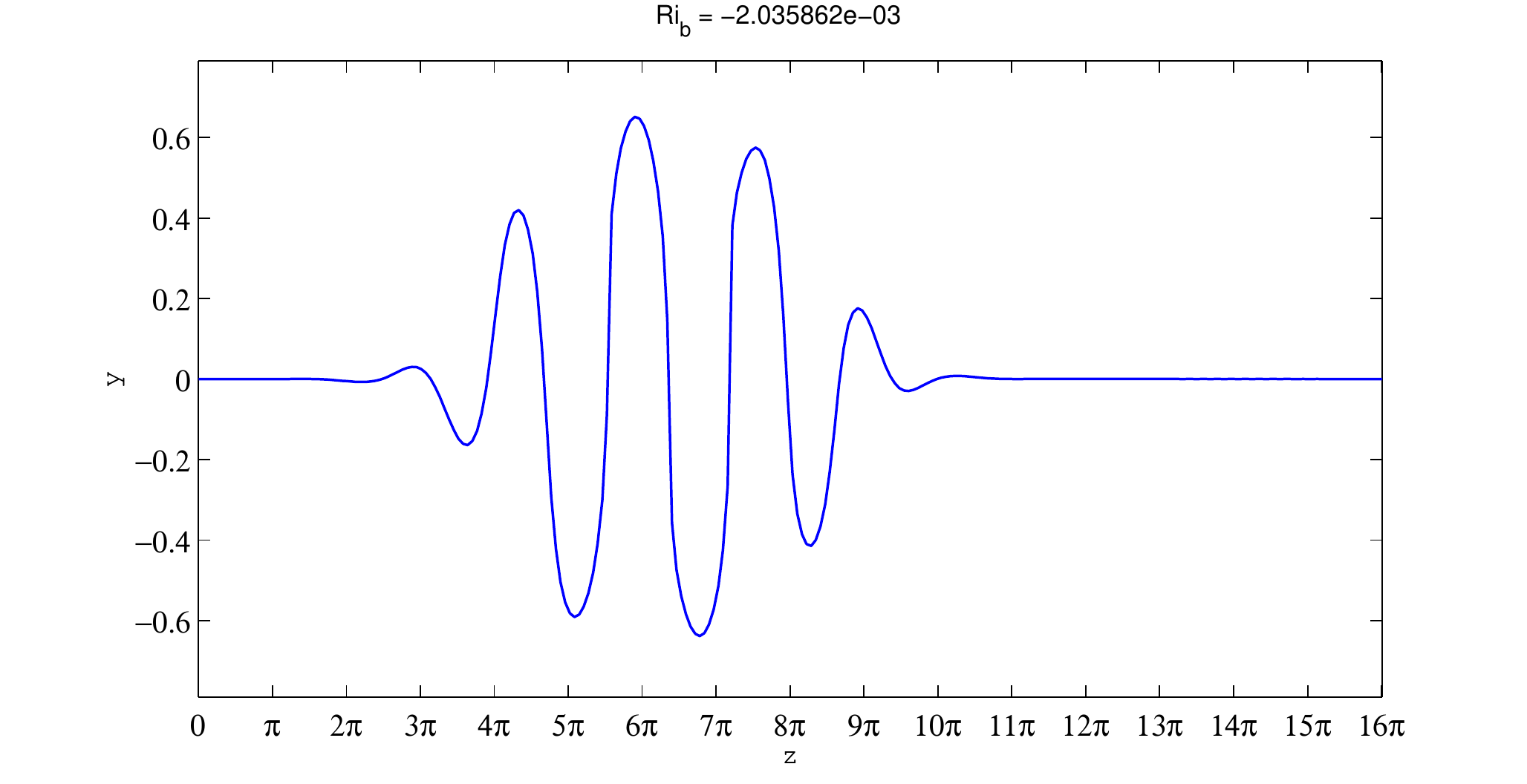} \\  %4
\end{tabular}
\begin{tabular}{c}
\textsuperscript{5}\adjincludegraphics[angle=0,height=2cm,width=7cm    ,trim={{.075\width} {.025\height} {.075\width}  {.06\height}},clip]{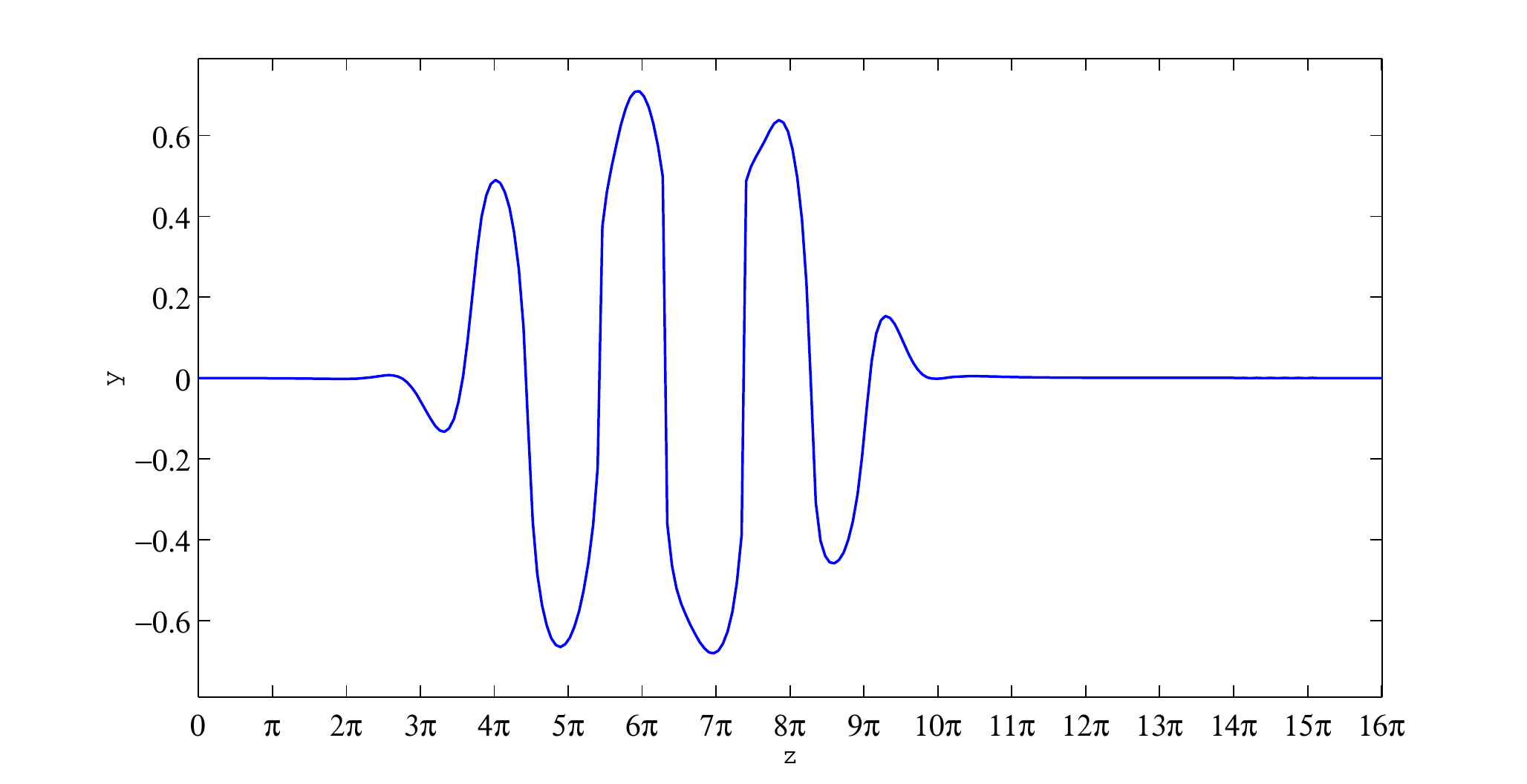}% 5
\textsuperscript{\; 6}\adjincludegraphics[angle=0,height=2cm,width=7cm    ,trim={{.075\width} {.025\height} {.075\width}  {.06\height}},clip]{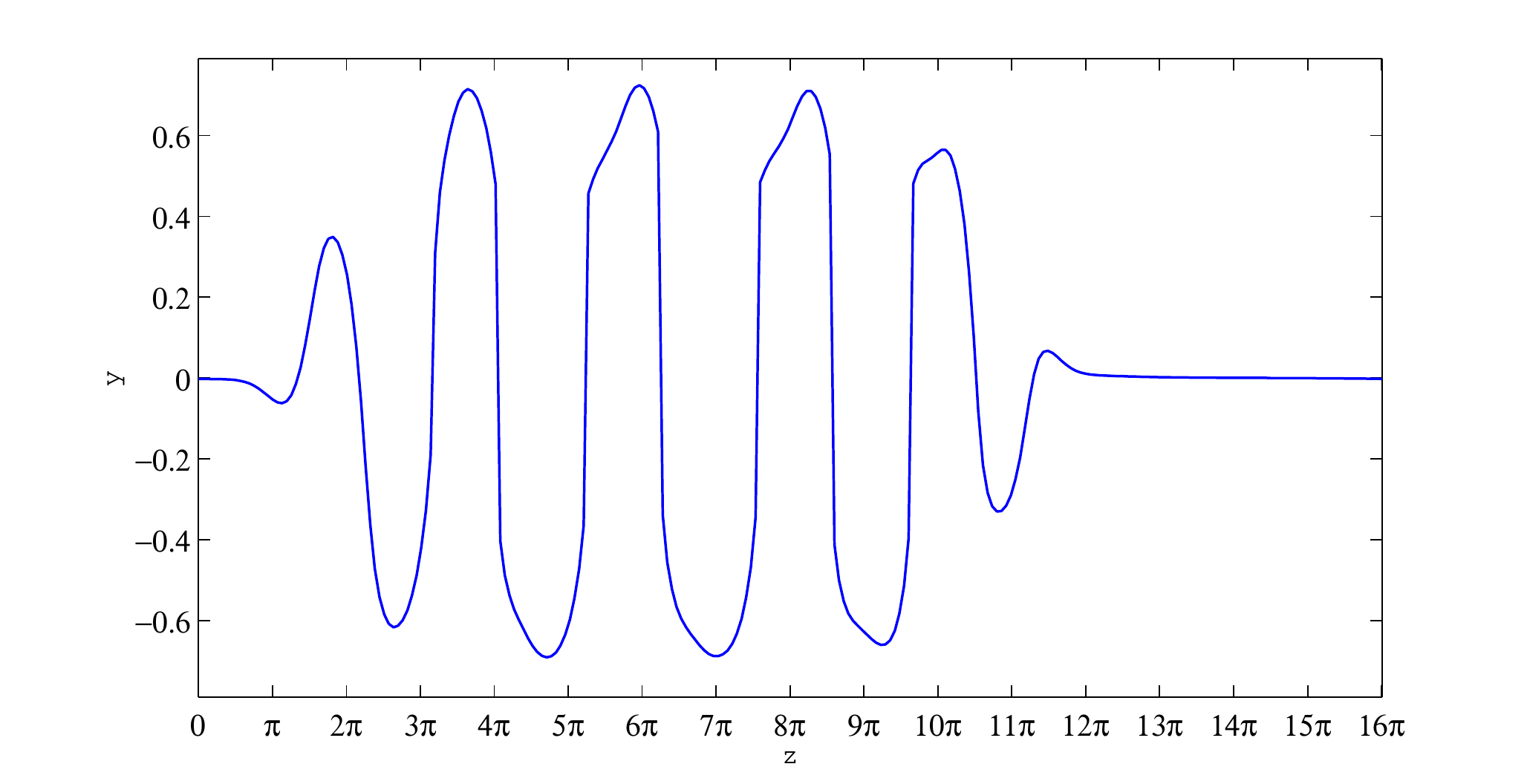}% 6
\end{tabular}
\begin{tabular}{c}
\textsuperscript{7}\adjincludegraphics[angle=0,height=2cm,width=7cm    ,trim={{.075\width} {.025\height} {.075\width}  {.06\height}},clip]{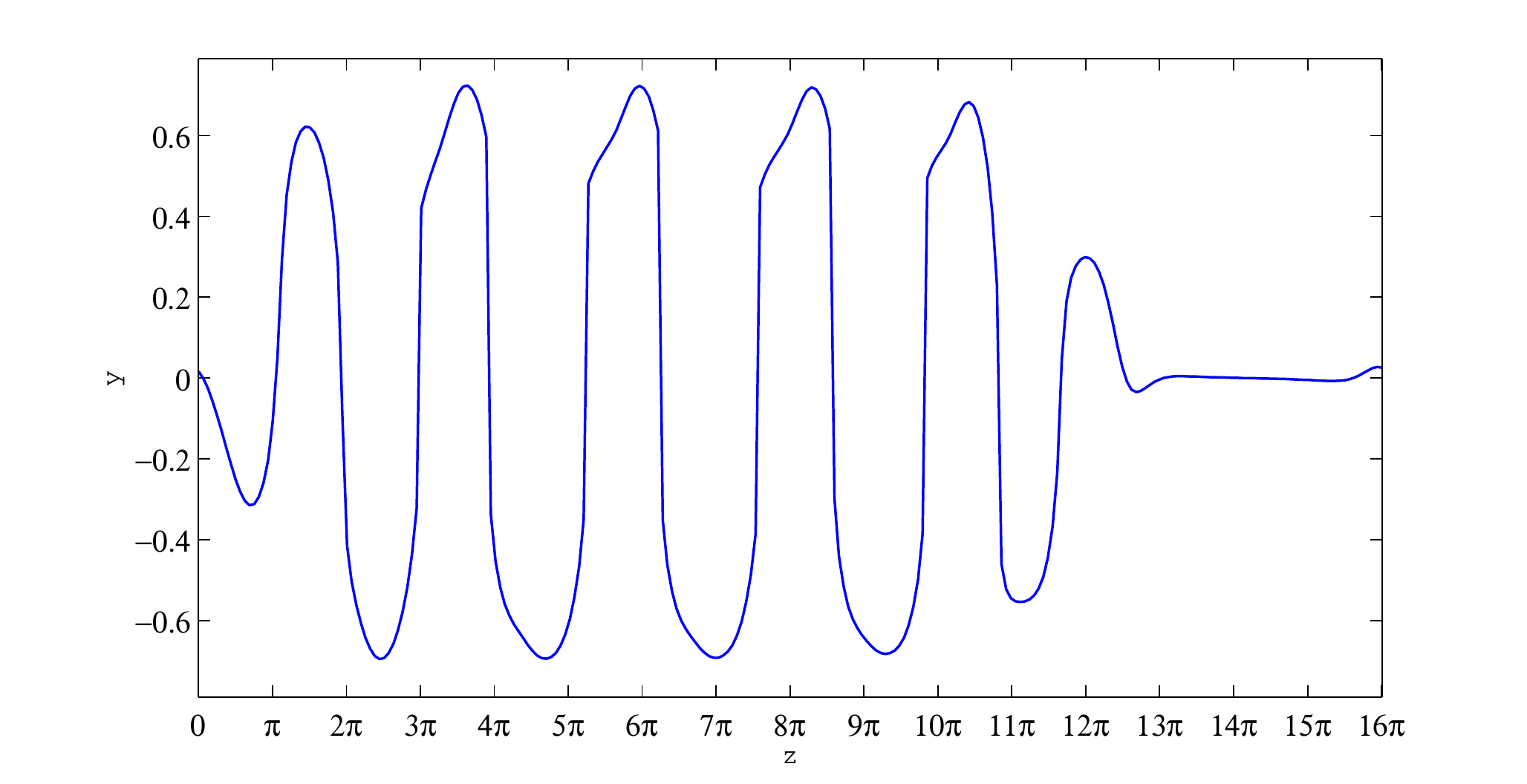}% 7
\textsuperscript{\;8}\adjincludegraphics[angle=0,height=2cm,width=7cm    ,trim={{.075\width} {.025\height} {.075\width}  {.06\height}},clip]{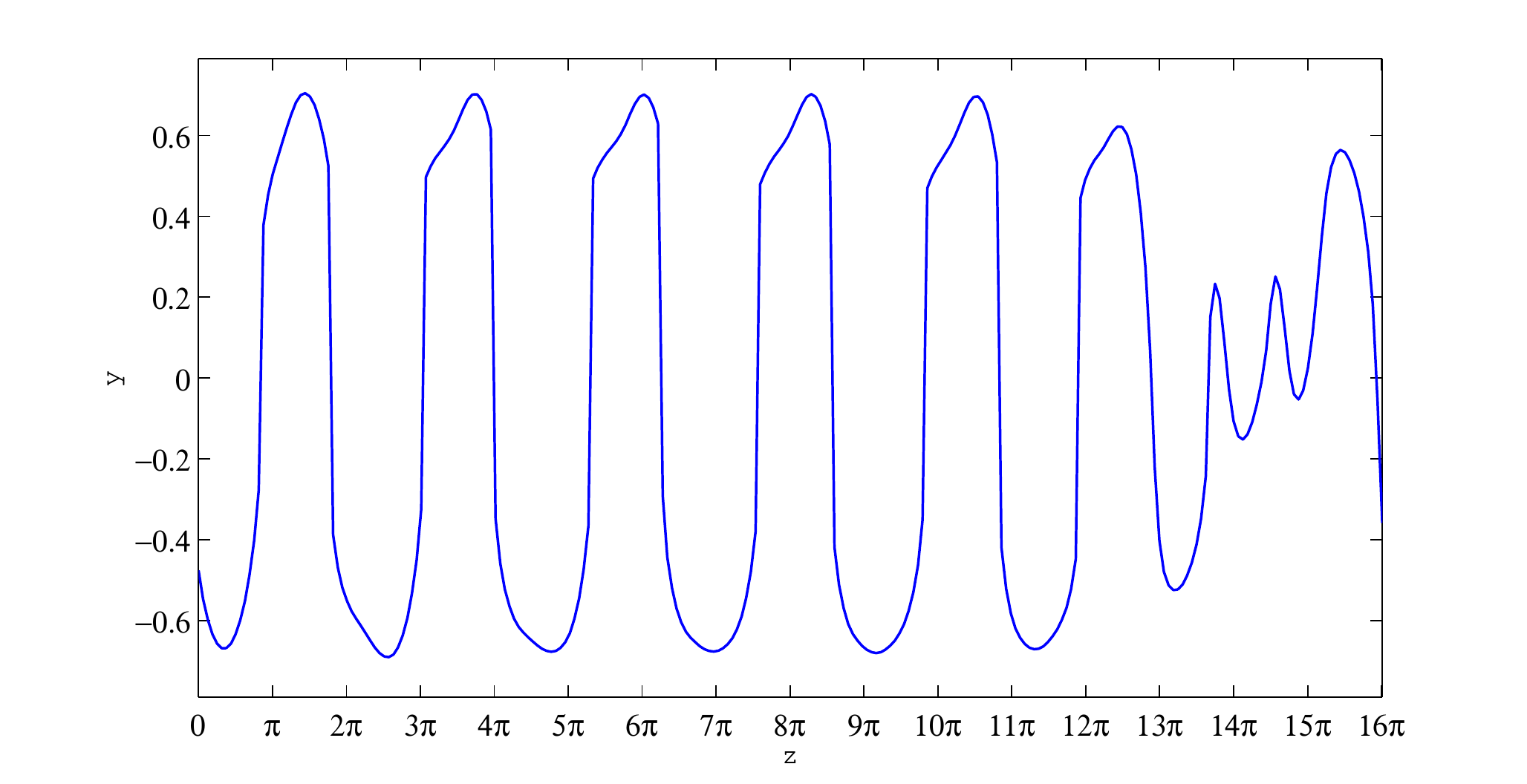} % 8
\end{tabular}
\begin{tabular}{c}
\textsuperscript{9}\adjincludegraphics[angle=0,height=2cm,width=7cm    ,trim={{.075\width} {.025\height} {.075\width}  {.06\height}},clip]{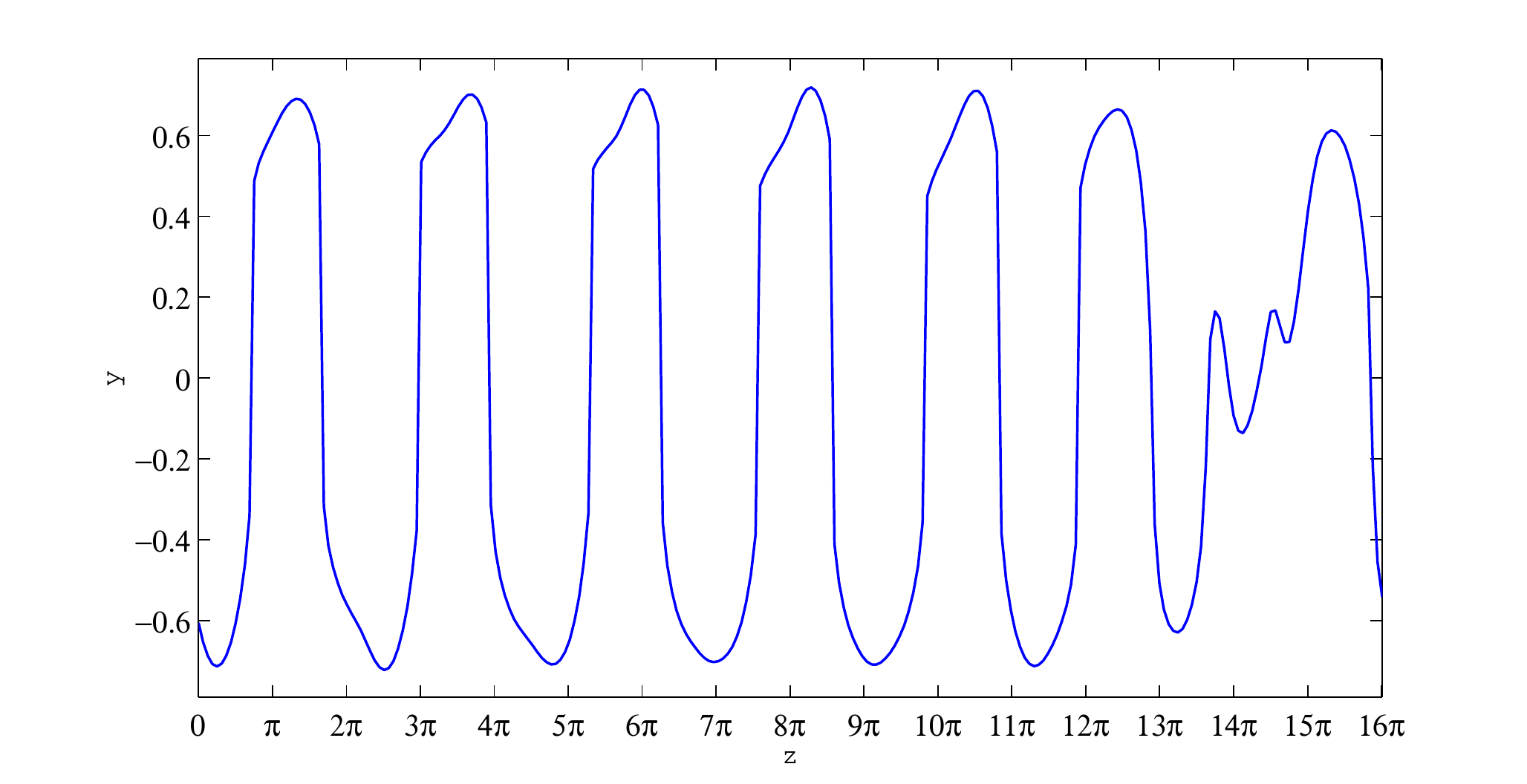}% 9 
\textsuperscript{\;10}\adjincludegraphics[angle=0,height=2cm,width=7cm    ,trim={{.075\width} {.025\height} {.075\width}  {.06\height}},clip]{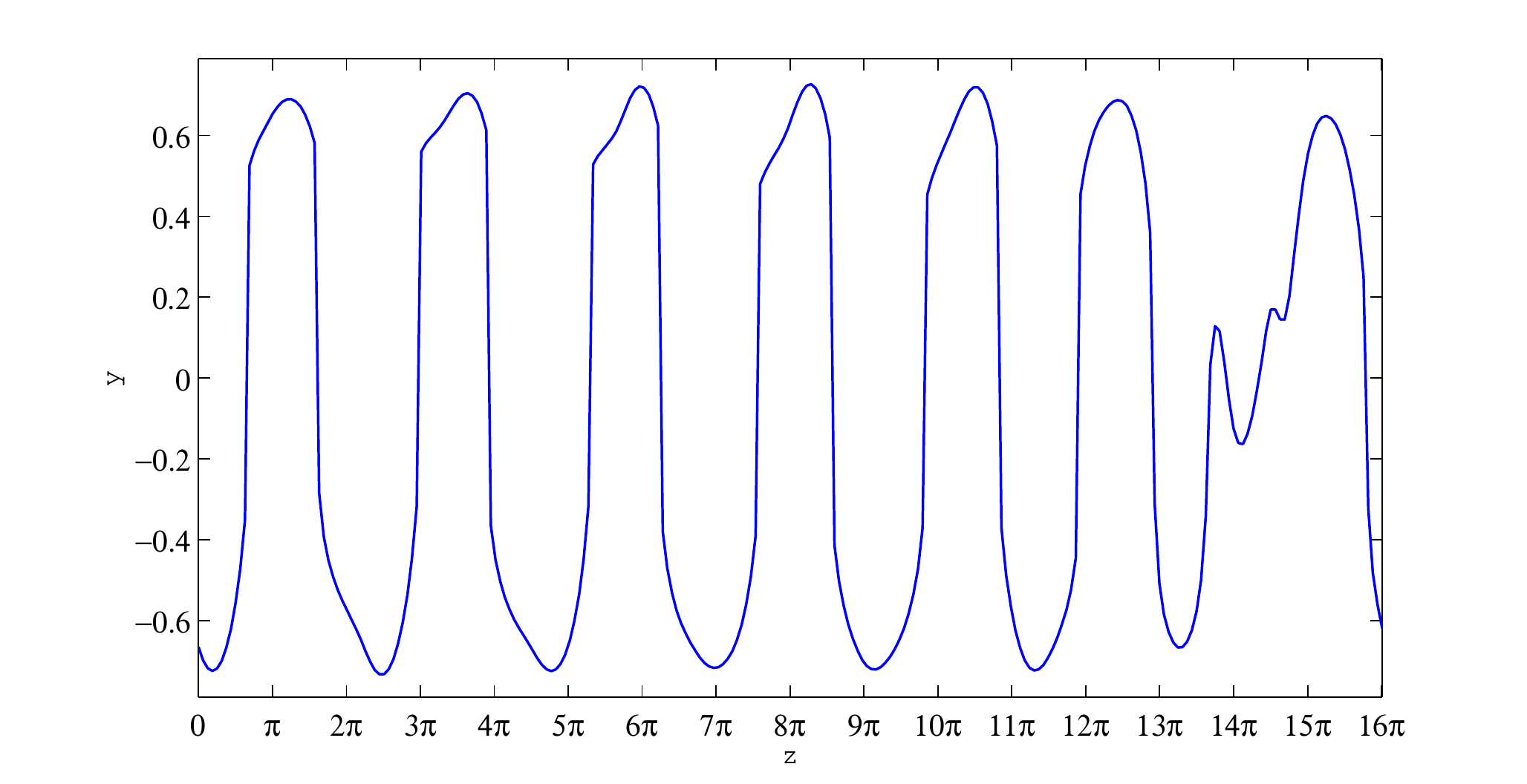} % 10
\end{tabular}
\caption{
Isolines for $u=0$ of the streamwise component of the total flow of the equilibrium  snake solution shown in figure \ref{SNAKE2_full_continuation}. The labels refer to states at the numbered circles on the solution curve.
}
\label{Snaking_points}
\end{center}
\end{figure}

\begin{figure} 
\begin{center}  
\adjincludegraphics[angle=0,width=8.5cm,trim={{.12\width} {.31\height} {.12\width}  {0.3\height}},clip]{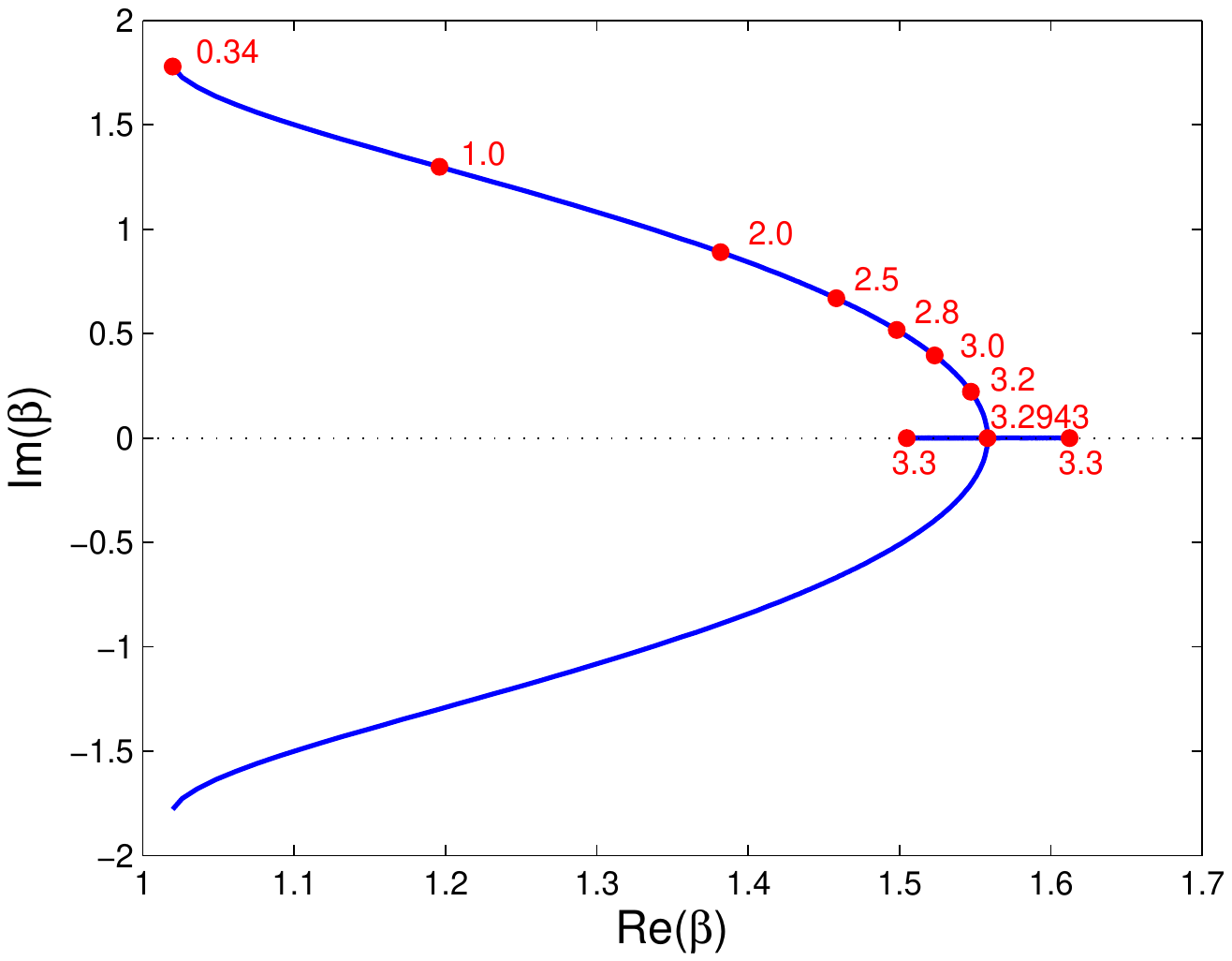}
\adjincludegraphics[angle=0,width=8.5cm,trim={{.12\width} {.31\height} {.12\width}  {0.3\height}},clip]{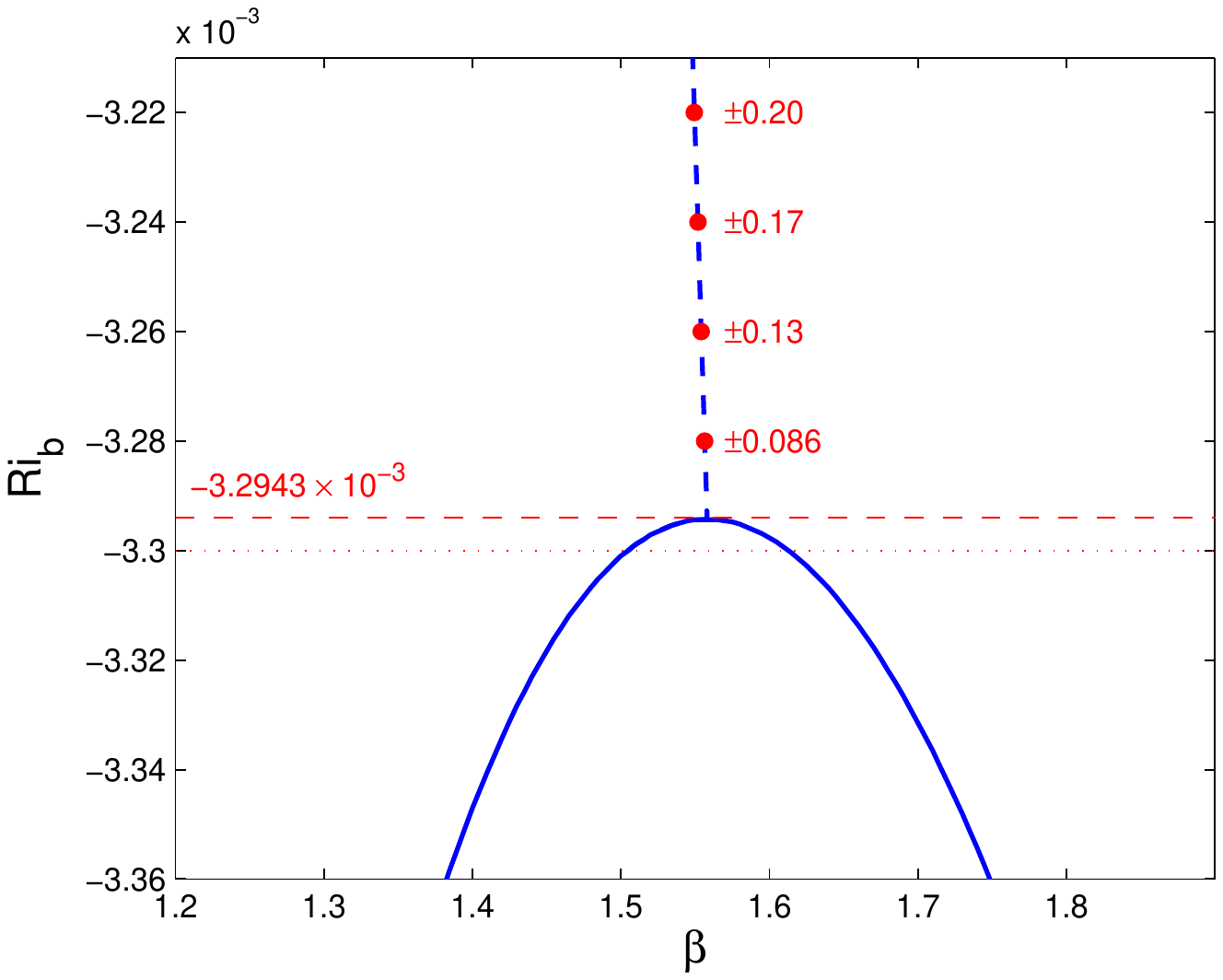}
 \caption[]{Left: the complex wavenumber for steady 2D streamwise-independent disturbances as a function of $Ri_b$ (labels are $-1000Ri_b$) at $Re=180$. $\Im m (\beta)$ indicates the spatial decay rate in the spanwise direction and $\Re e (\beta)$ the spatial frequency. Neutral spatially-periodic disturbances can exist for $Ri_b \leq -3.2943 \times 10^{-3}$ (the Rayleigh-Benard instability threshold) but have to become localised for larger (less negative) $Ri_b$ indicated by $\Im m(\beta) \neq 0$. Right: the neutral stability curve for Rayleigh-Benard convection ($Ra:=-Ri_b Re^2 Pr$ see the Appendix).  For $Ri_b \leq -3.2943 \times 10^{-3}$ (solid blue line), temporally neutral disturbances exist for real wavenumbers $\beta$ whereas for $Ri_b > -3.2943 \times 10^{-3}$ (dashed blue line), the wavenumber has to be complex for temporal neutrality. In this case, the non-vanishing imaginary part is indicated in red at selected points along the dashed blue line). The left and right plots show the neutral curve in $(\Re e(\beta), \Im m (\beta),Ri_b)$ space from two different perspectives. }
\label{LinProblem}
\end{center}
\end{figure}

\begin{figure} % --------------------->      A B            C  D                    E  F
    \begin{center}   %, width
\adjincludegraphics[angle=0,width=8cm,trim={{.05\width} {.01\height} {.05\width}  {.04\height}},clip]{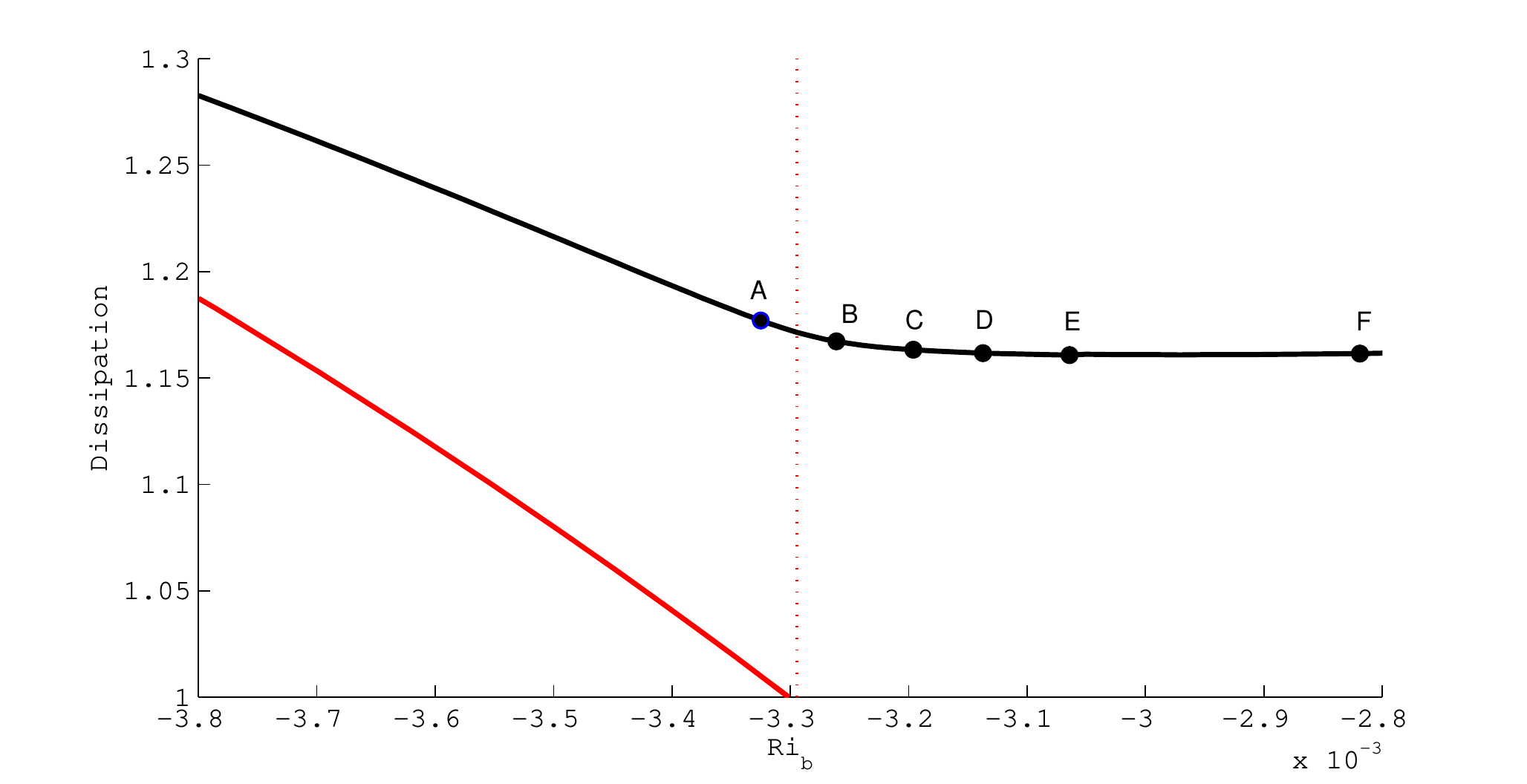}\\% A,B,C,D   F STRATsnake2_BLOWUP
\begin{tabular}{c}
\textsuperscript{A}\adjincludegraphics[angle=0,height=2.cm,width=7cm    ,trim={{.075\width} {.025\height} {.075\width}  {.06\height}},clip]{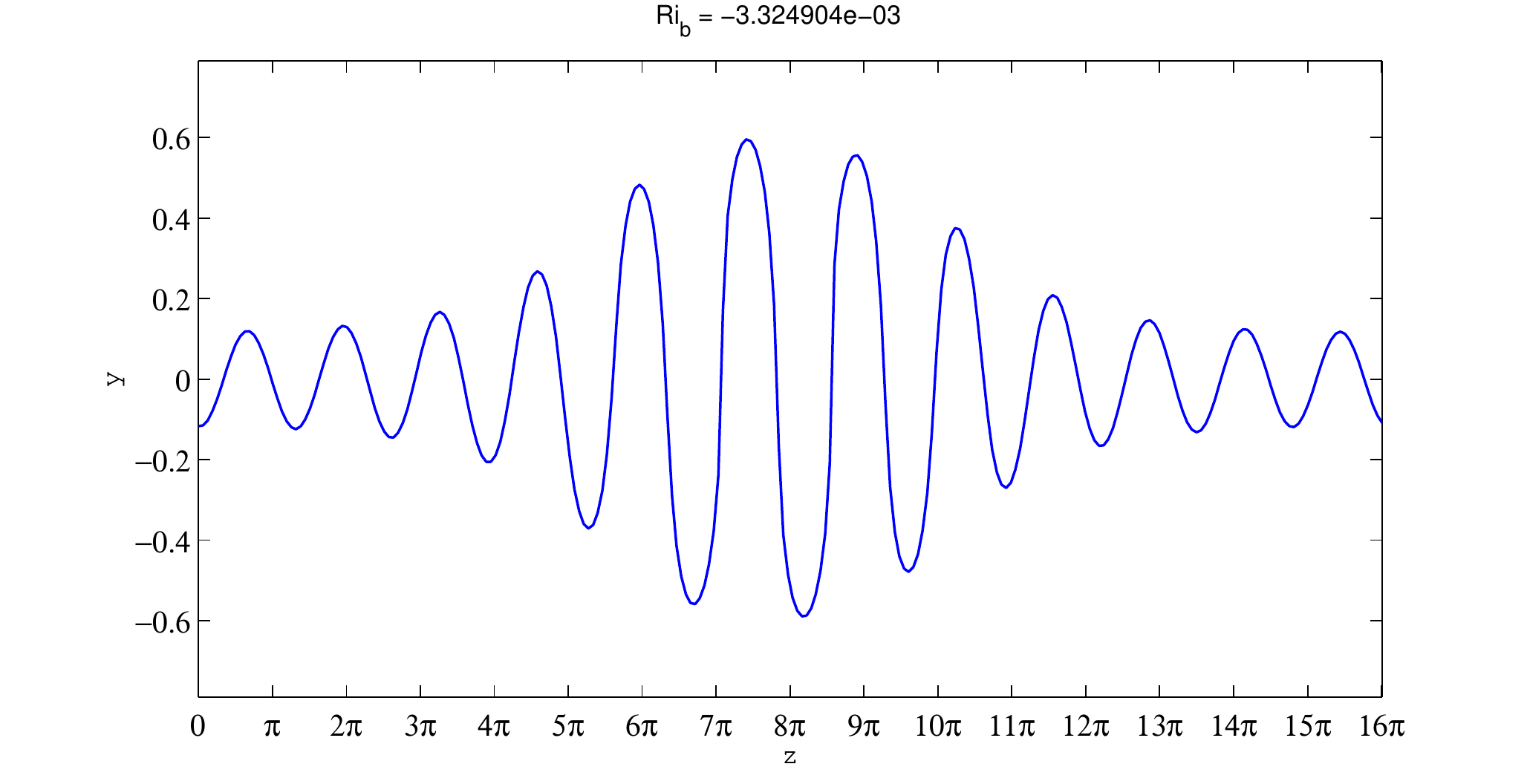}   % POINT_A_33249
\textsuperscript{\; B}\adjincludegraphics[angle=0,height=2.cm,width=7cm    ,trim={{.075\width} {.025\height} {.075\width}  {.06\height}},clip]{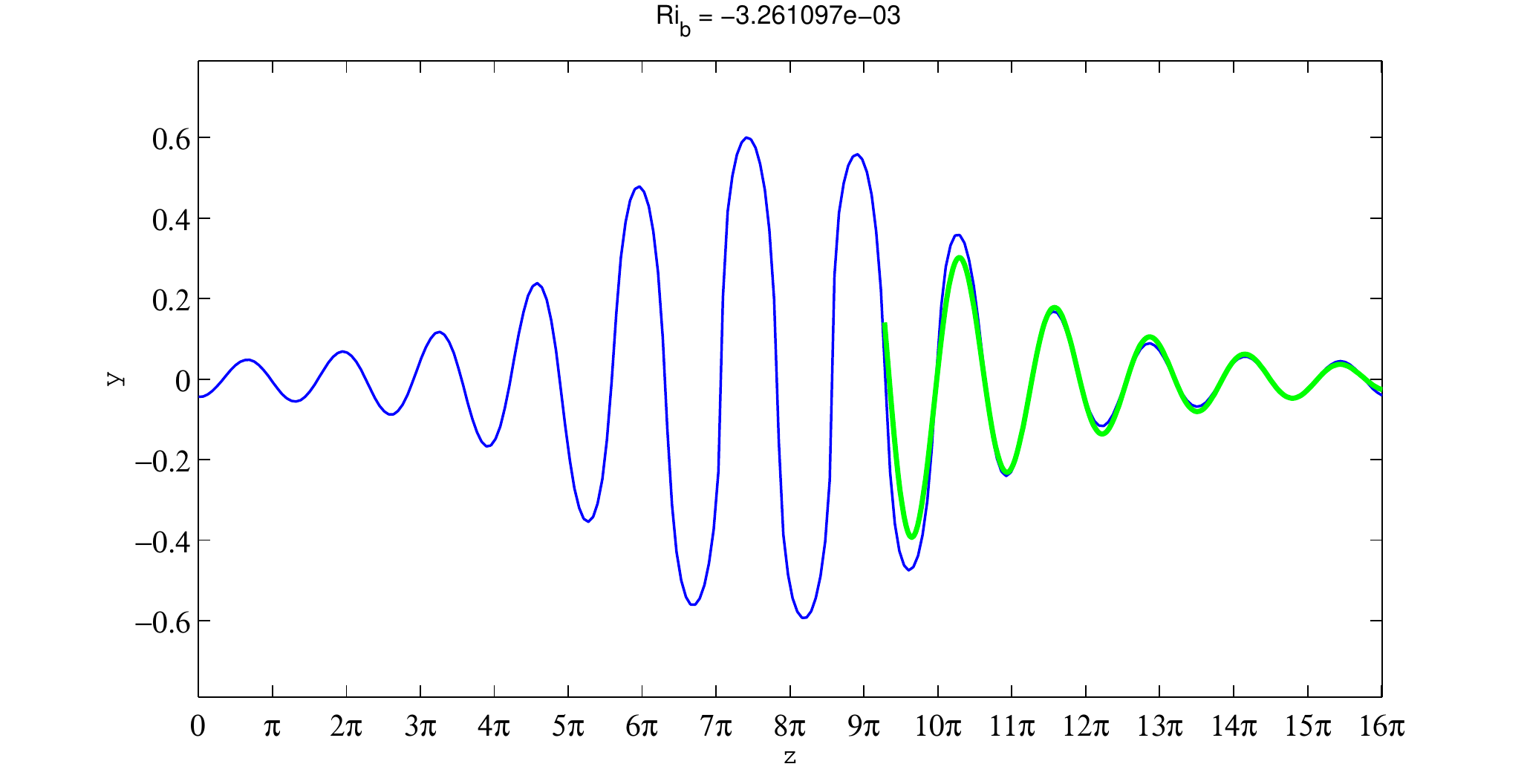}\\   %POINT_B_326
\textsuperscript{C}\adjincludegraphics[angle=0,height=2.cm,width=7cm    ,trim={{.075\width} {.025\height} {.075\width}  {.06\height}},clip]{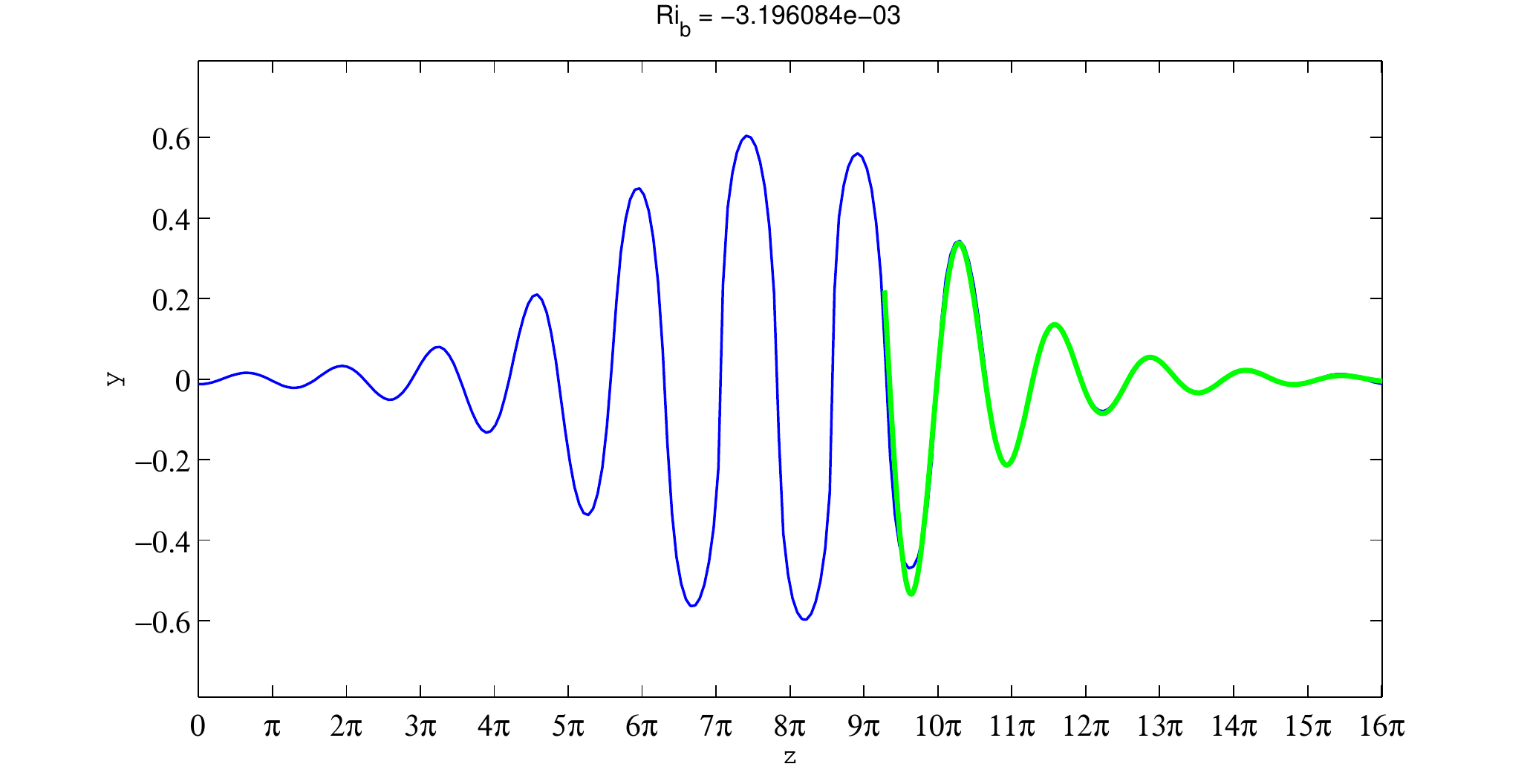}  
\textsuperscript{\; D}\adjincludegraphics[angle=0,height=2.cm,width=7cm    ,trim={{.075\width} {.025\height} {.075\width}  {.06\height}},clip]{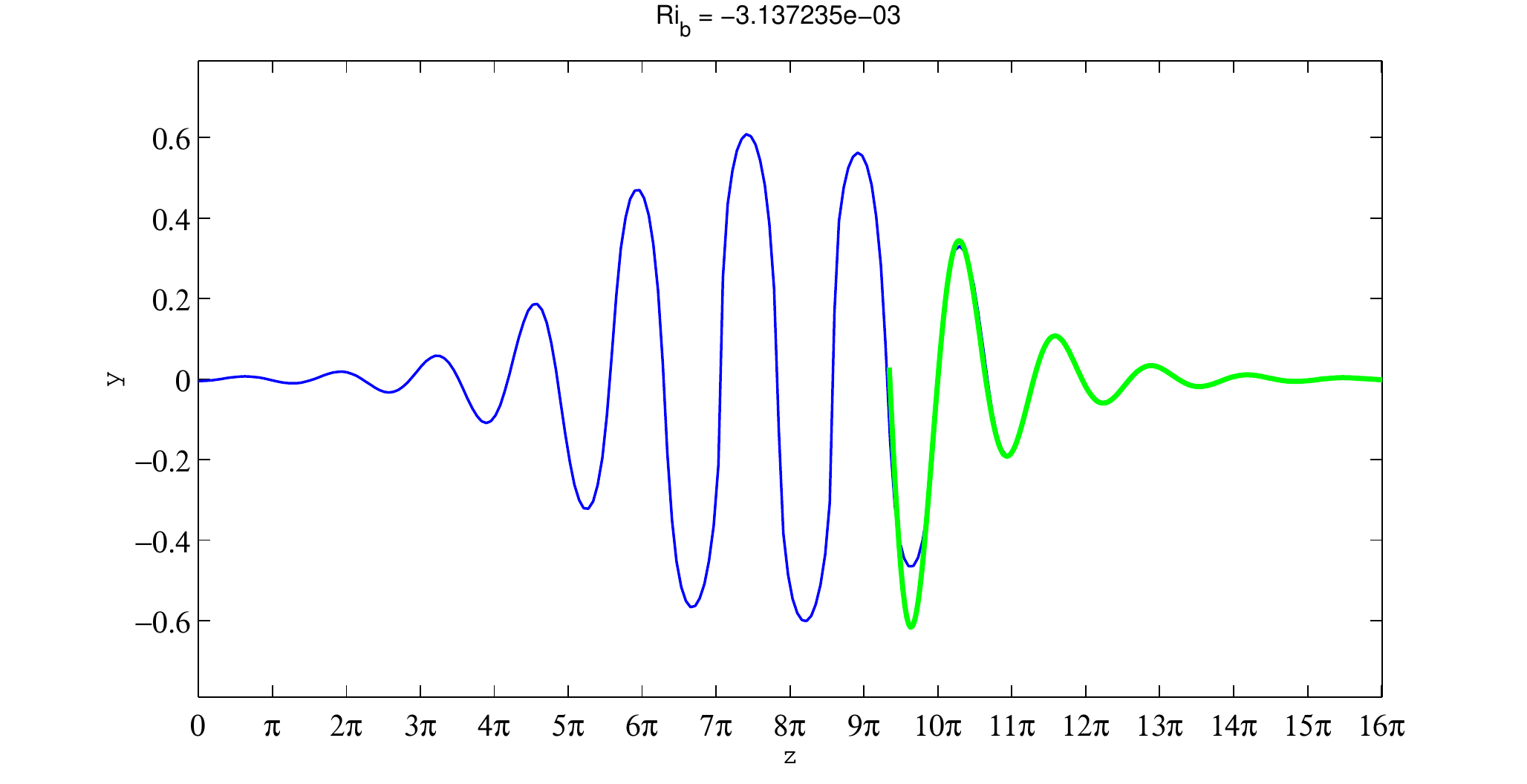}\\
\textsuperscript{\; E}\adjincludegraphics[angle=0,height=2.cm,width=7cm    ,trim={{.075\width} {.025\height} {.075\width}  {.06\height}},clip]{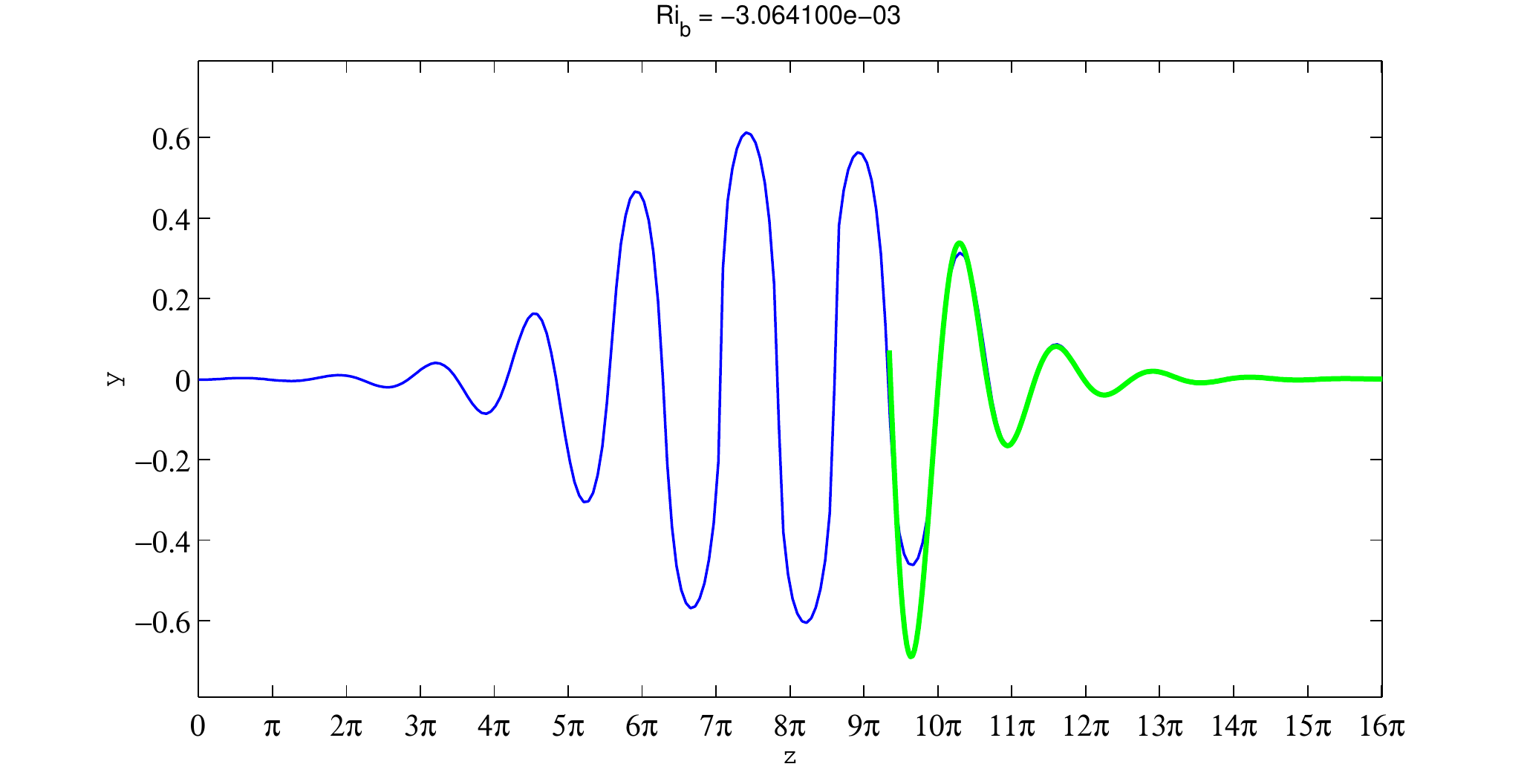} 
\textsuperscript{\; F}\adjincludegraphics[angle=0,height=2.cm,width=7cm    ,trim={{.075\width} {.025\height} {.075\width}  {.06\height}},clip]{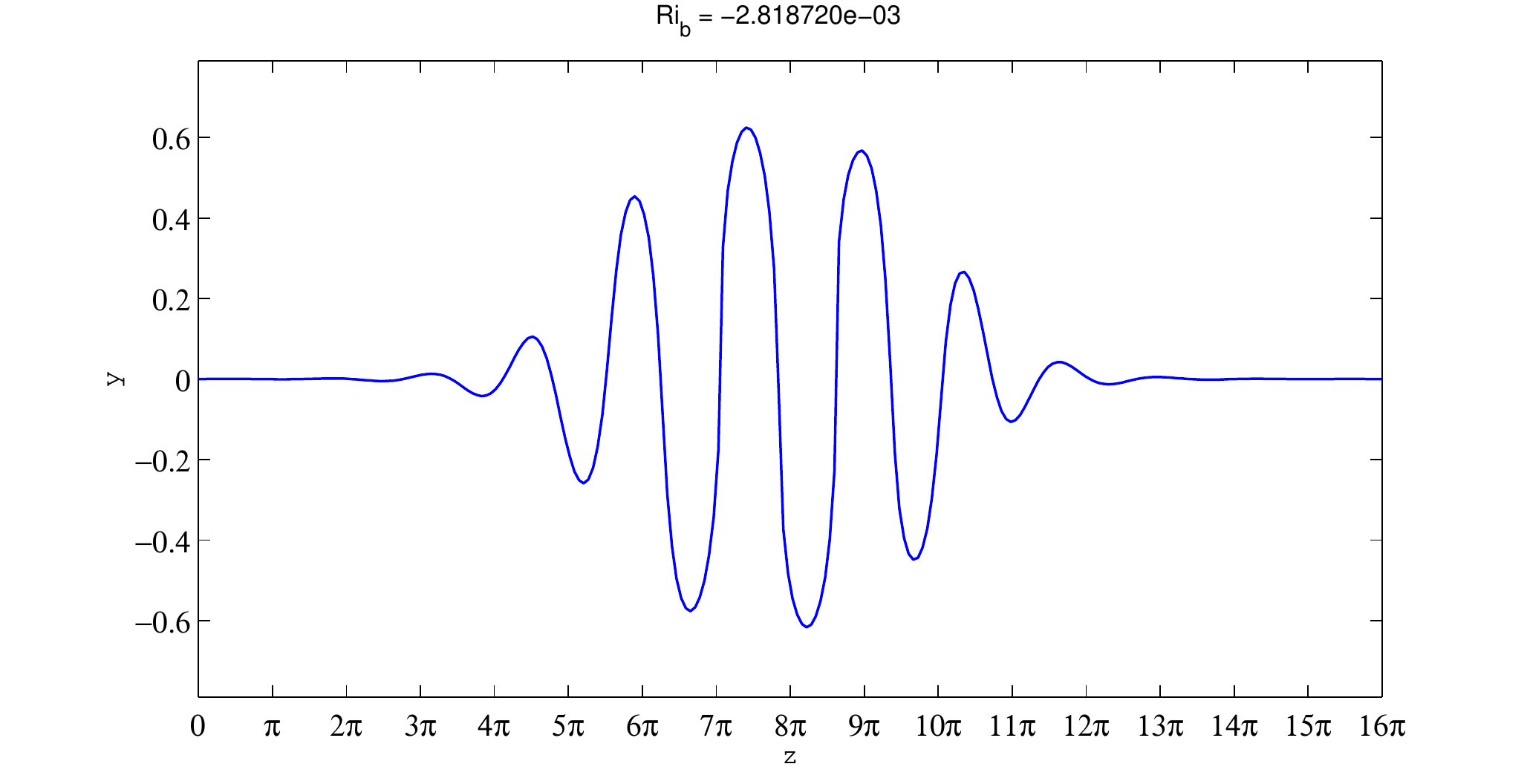} 
\end{tabular}
 \caption[]
{Top: blow-up of figure \ref{SNAKE2_full_continuation} centred on $Ri_b=-3.2943 \times 10^{-3}$ (red dotted line). Note there is a very small gap between where the roll solution (red line) bifurcates and the critical value of $Ri_b$ (the dotted red line) because the wavenumber which fits in a $16 \pi$ wide box is slightly non-optimal. Lower plots show isolines of $u=0$ at 
A) $Ri_b=-3.325\times10^{-3}$,  B) $ Ri_b=-3.261\times10^{-3}$, 
C) $Ri_b=-3.196\times10^{-3}$,  D) $ Ri_b=-3.137\times10^{-3}$, 
E) $Ri_b=-3.064\times10^{-3}$,  F) $Ri_b=-2.819\times10^{-3}$. 
}
      \label{Linear_decay}
    \end{center}
  \end{figure}

In connecting to the $\beta=1.5$ Nagata solution, the EQ snake solution has to delocalise and  figure \ref{Snaking_points} shows this is a gradual process as  $Ri_b$ decreases from 0 as opposed to that found for $Ri_b$ increasing from 0 when snaking occurs. The key observation for $Ri_b<0$ is that the (spatial) spanwise decay of the snake disappears  once the threshold for convective instability at $Ri_b=-3.2943 \times 10^{-3}$  is crossed  (the critical Rayleigh number$\,:= -Ri_b Re^2  Pr=1708/16$; Salewski et al. \cite{Salewski} see the same phenomenon in rotating plane Couette flow - see their figure 4). This can be understood by examining the linear operator about the linearly sheared state for the least-(spatially)-damped, {\em temporally steady} eigenfunction since the deviation away from this linearly-sheared state becomes vanishingly small in the spanwise tails of the snake \cite{Gibson14}. The snake becomes streamwise-independent in its tail regions suggesting analysis of the linear eigenvalue problem for 2D disturbances independent of the streamwise direction i.e. 
\beq
(\bu,\rh,\tilde{p})= (\,\bu(y),\rh(y), \tilde{p}(y) \, ) e^{i \beta z+\sigma t}
\eeq
so that
\begin{align}
\sigma \tilde{u} &= \frac{1}{Re}( \tilde{u}^{''}- \beta^2 \tilde{u})-\tilde{v},          \\
\sigma \tilde{v} &= \frac{1}{Re}( \tilde{v}^{''}- \beta^2 \tilde{v})-\tilde{p}^{'}-Ri_b \rh, \\
\sigma \tilde{w} &= \frac{1}{Re}( \tilde{w}^{''}-\beta^2 \tilde{w})-i \beta \tilde{p},\\
0 &=\tilde{v}^{'}+i \beta \tilde{w},\\
\sigma \rh &= \frac{1}{Re Pr} ( \rh^{''}-\beta^2 \rh)+ \tilde{v}.
\end{align}
Normally, $\beta$ is assumed real and the eigenvalue problem is scrutinised for complex $\sigma$ with $\Re e (\sigma)=0$ to find spatially-periodic, neutral eigenfunctions. Here, instead, the interest is in (real) $\sigma=0$ and complex $\beta$ to find spatially-decaying steady eigenfunctions (since EQ is steady). Of primary  interest is the eigenfunction with the smallest amplitude of $\Im m(\beta)$ where $\sigma=0$ which indicates the probable rate of spatial evanescence in the spanwise direction of a steady ECS when the amplitude gets small (see \S4.1 of \cite{Gibson14}). Figure \ref{LinProblem} shows this neutral curve in the complex $\beta$ plane on the left and the usual neutral curve as viewed in the wavenumber$-$control parameter plane is shown on the right. For $Ri_b \leq -3.2943 \times 10^{-3}$, neutral spatially-periodic eigenfunctions can exist but otherwise $\Im m (\beta) \neq 0$. 
Figure \ref{Linear_decay} shows that in the tail regions there is indeed good correspondence between the expected spatial decay and the  numerically observed decay  close to the linear instability threshold at $Ri_b=-3.2943 \times 10^{-3}$.

So, in summary, by using  optimal energy growth, we have managed to rediscover the snake solutions of \cite{Schneider10a, Schneider10b}. Having shown that this approach  works, we now turn our attention to a region of parameter space in pCf where no solutions are currently known beyond the linearly-sheared base state.

%
% Fig 7
%

%
% Fig 8
%

%
% Fig 9
% 

%________________________________________________________________________
%
%
\subsection{Very low $Re$ in pCf \label{low}}
%
%________________________________________________________________________

\begin{figure}[htp]   %
           %       2 rolls          /M-11/T20/qlop-1e4 
          \begin{center}  
\includegraphics[angle=0,width=5cm, trim=1.8cm 0.28cm 1.8cm 0.3cm]{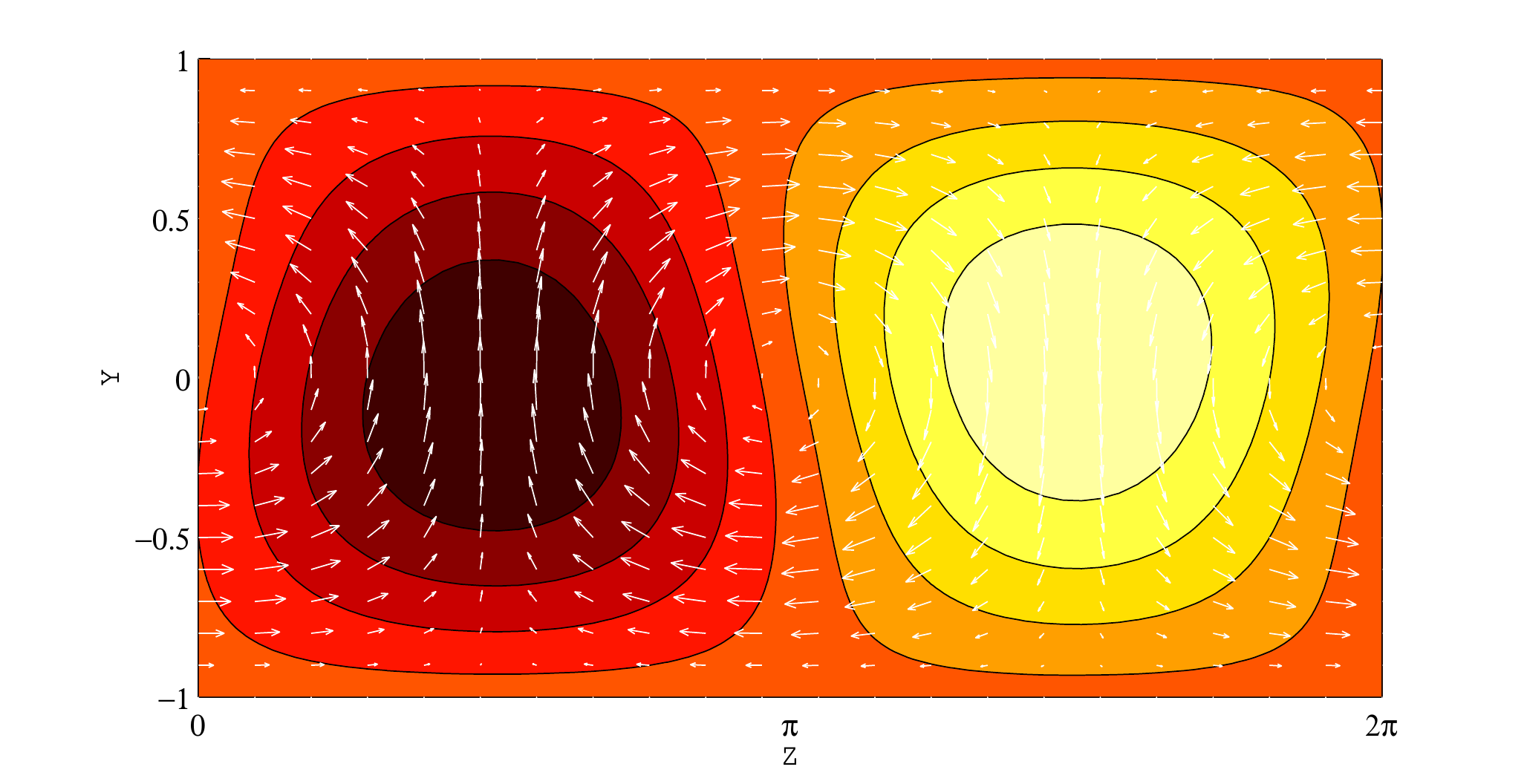} % QLOP_T20_1e4_4rolls_t0
\includegraphics[angle=0,height=2.5cm ,clip, trim=0.9cm 0.58cm 0.0cm 0.6cm]{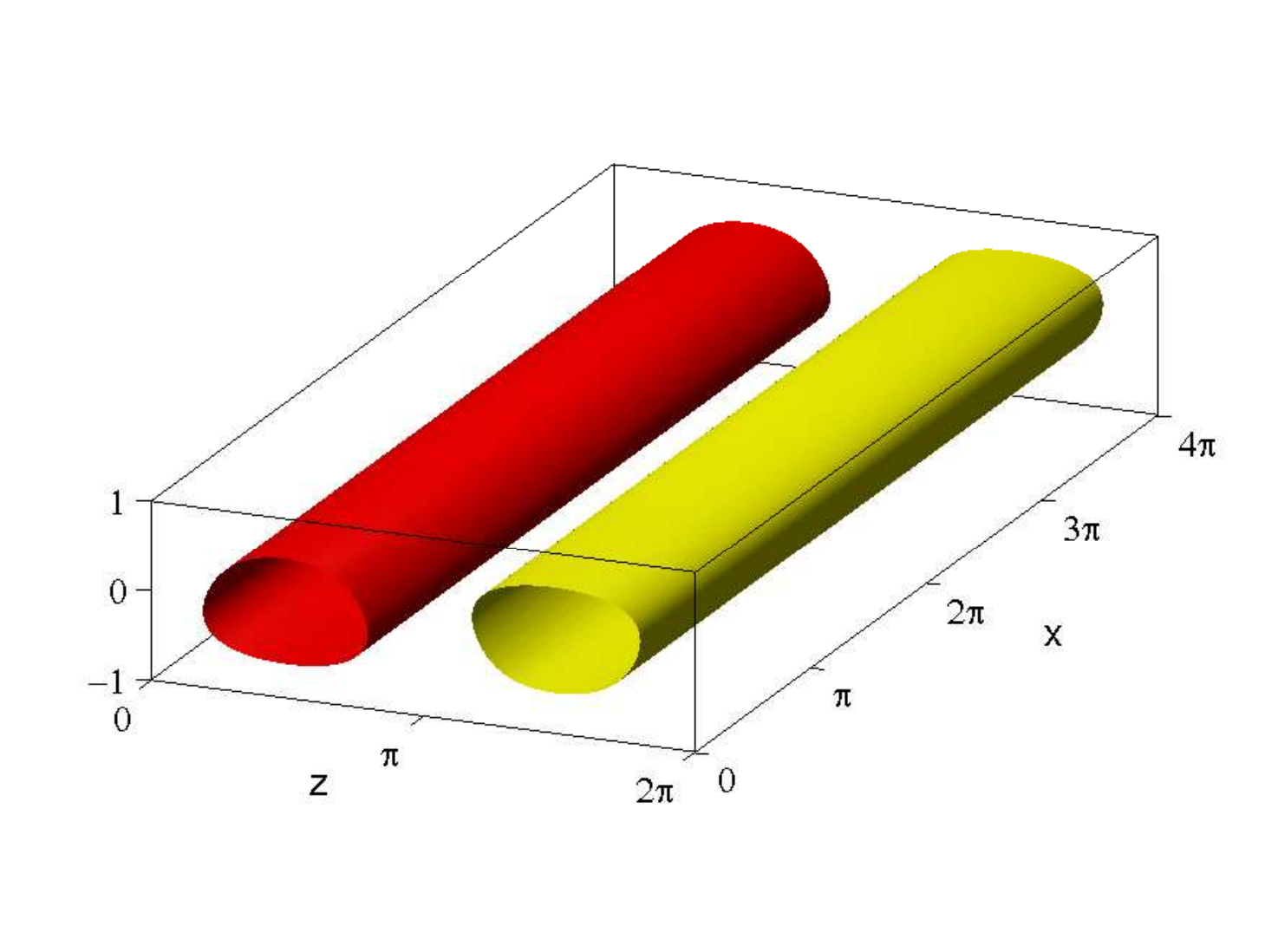} % QLOP_T20_1e4_4rolls_t0_ISOS.eps
\includegraphics[angle=0,width=5cm, trim=1.8cm 0.28cm 1.8cm 0.3cm]{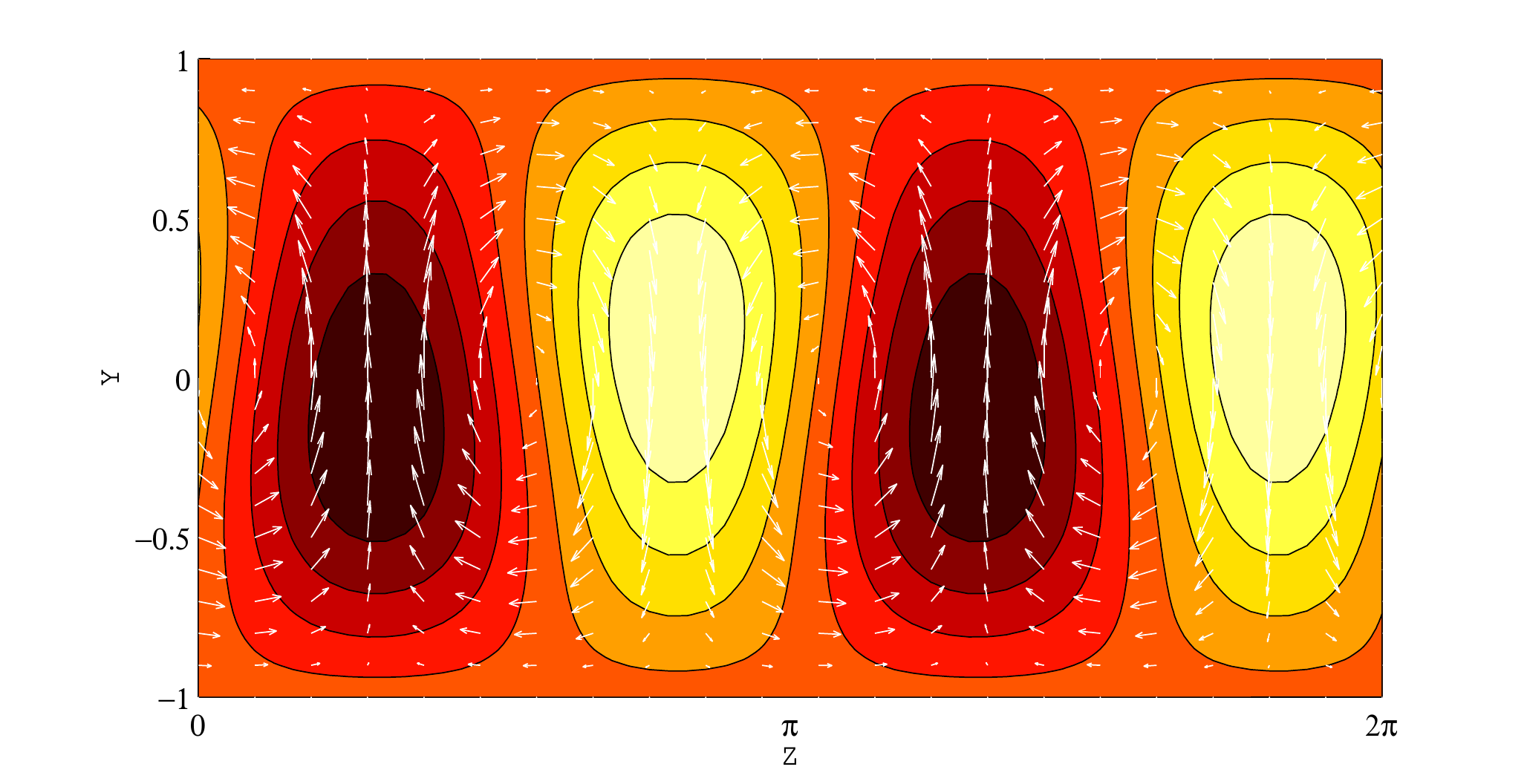} % QLOP_T20_1e4_2rolls_t0
\includegraphics[angle=0,height=2.5cm, clip, trim=0.9cm 0.58cm 0.0cm 0.6cm]{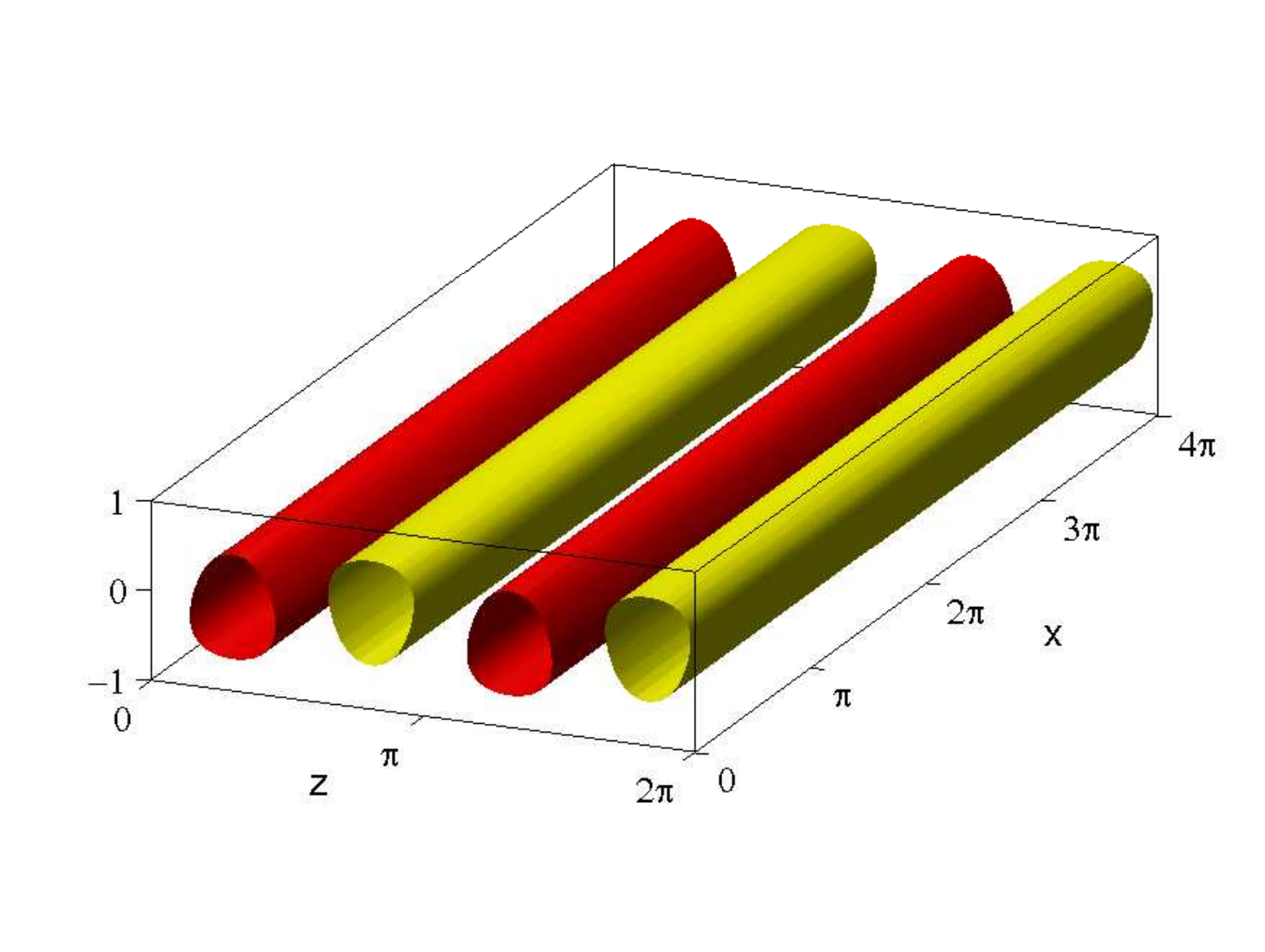} % QLOP_T20_1e4_2rolls_t0_ISOS
 \caption{$LOP_1$ (left)  and $LOP_2$ (right)  for   $T=20$ with $E_0=1.0 \times 10^{-4}$. $LOP_1$ is the 2D optimal shown in  figure 4 of \cite{Butler92}. The same contour levels are used for both plots (8 levels between -0.004 and 0.004) and isocontours are $\pm 60\%$ of maximum streamwise perturbation velocity.}
      \label{LOP}
    \end{center}
  \end{figure}

In this subsection, we turn stratification off ($Ri_b=0$) and set $Re=100$ which is above the energy stability threshold of $20.7$ \cite{Joseph66} up to which the basic sheared state is provably unique and below $127.7$ which is the current best estimate of when other solutions start to exist \cite{Nagata90,Waleffe03}. A geometry of $4 \pi \times 2 \times 2 \pi$ and a shortish target time of $T=20$  were chosen and $E_0$ gradually increased until the (nonlinear versions of the) linear optimal perturbations ($LOP_1$ and $LOP_2$) shown in figure \ref{LOP} were no longer found.
%
% fig 10:   LOP1 and LOP2 for T=20  ( which one resembles the LOP from Butler & Farrell 1992 ? )
%

%
%    thesis fig 5.4   to show NLOP and QLOP + no plateau
%
%\begin{figure}[htp]  
% \begin{center}   
%\adjincludegraphics[angle=0,height=5cm ,trim={{.05\width} {.02\height} {.08\width} {.058\height}},clip]{./low_Re/variational_output_T20_1e3.eps}%
% \caption[]{Time evolution 
%of optimal perturbation obtained in last iteration. Target time $T=20$ for $E_0=1.0 \times 10^{-3}$.  Thin line, QLOP, 
%red line, NLOP--1.
%For QLOP,  the variational method converges at optimal after 41 iterations. 
%NLOP reaches a higher level of energy, but decays faster to the laminar state.% Red circle, state shown in Figure \ref{YZ_T20_1e3}c at $t=11$, Blue square, state shown in Figure \ref{YZ_T20_1e3}d at $t=12$.  
%}
%\label{QLOP_NLOP_1e3}
 %   \end{center}
 % \end{figure}

%
% Fig 11  thesis fig 5.2    NLOP-1 and what it evolves into
%
\begin{figure}[htp]    %  --
    \begin{center}    
\adjincludegraphics[angle=0,width=5cm ,trim={{.075\width} {.029\height} {.075\width}  {.06\height}},clip]{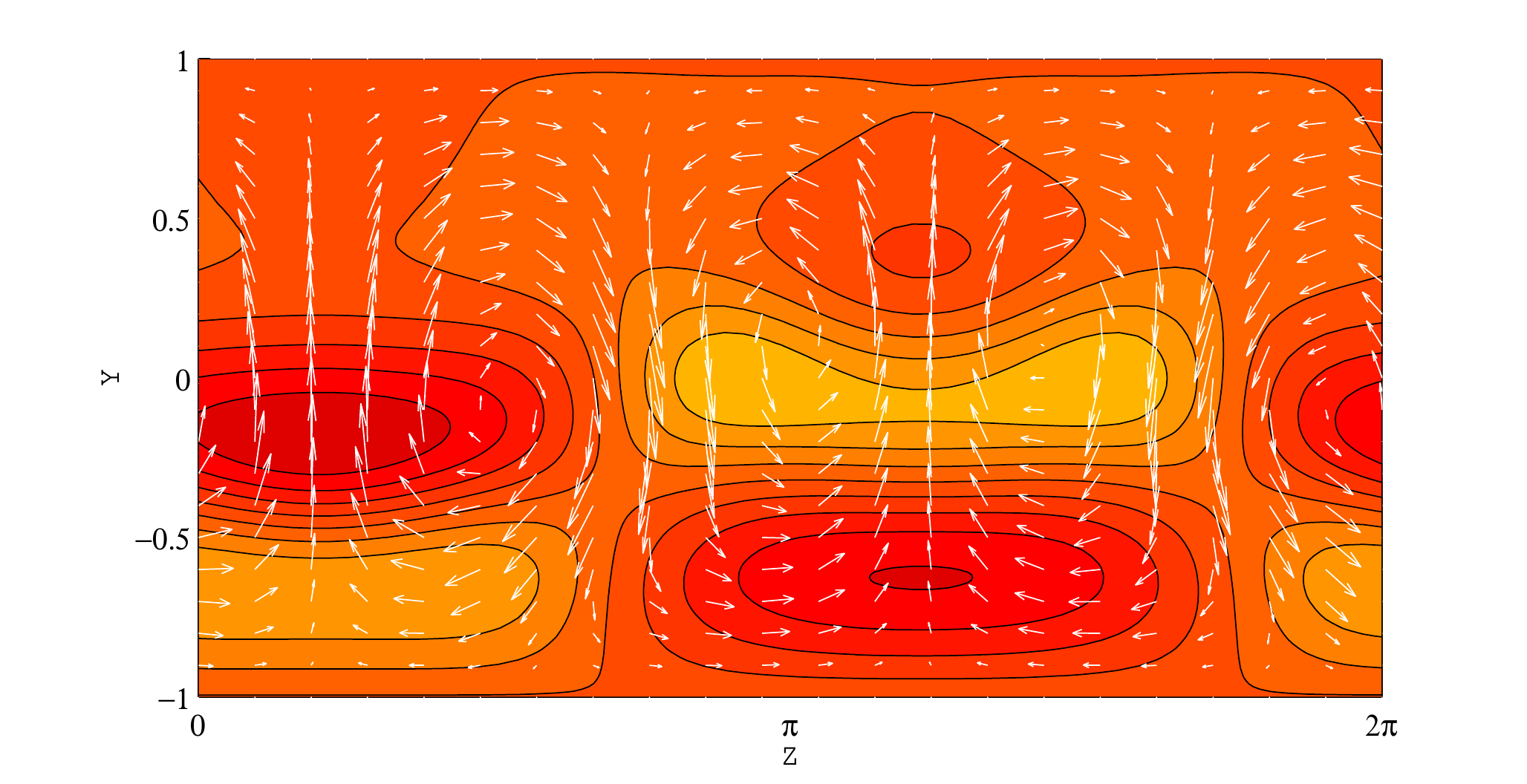}   % Chaotic_T20_1e3_4rolls_t0
\adjincludegraphics[angle=0,height=2.5cm ,trim={{.075\width} {.029\height} {.035\width}  {.06\height}},clip]{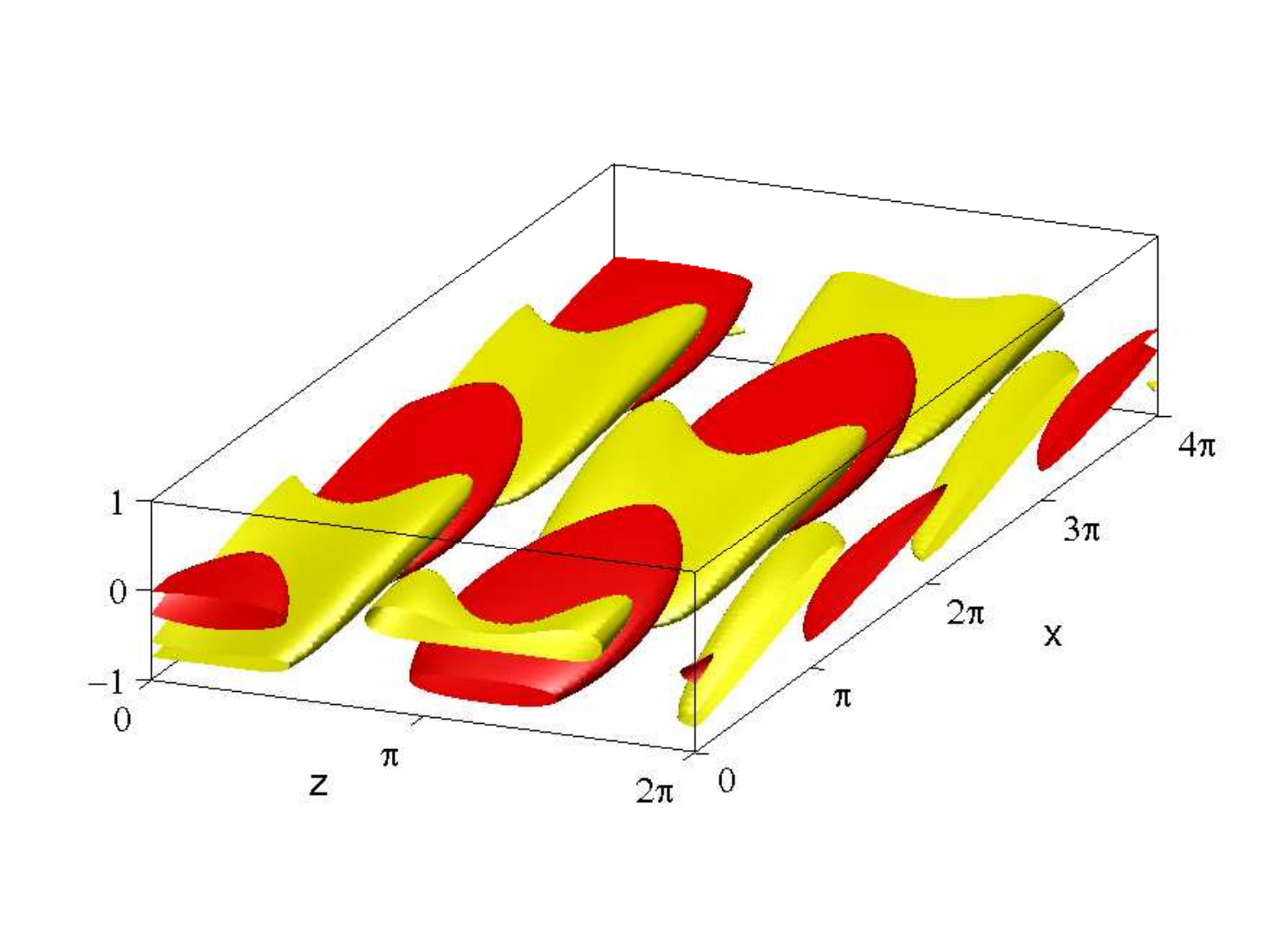}  % Chaotic_T20_1e3_4rolls_t0_ISOS
\adjincludegraphics[angle=0,width=5cm, trim={{.075\width} {.029\height} {.075\width}  {.06\height}},clip]{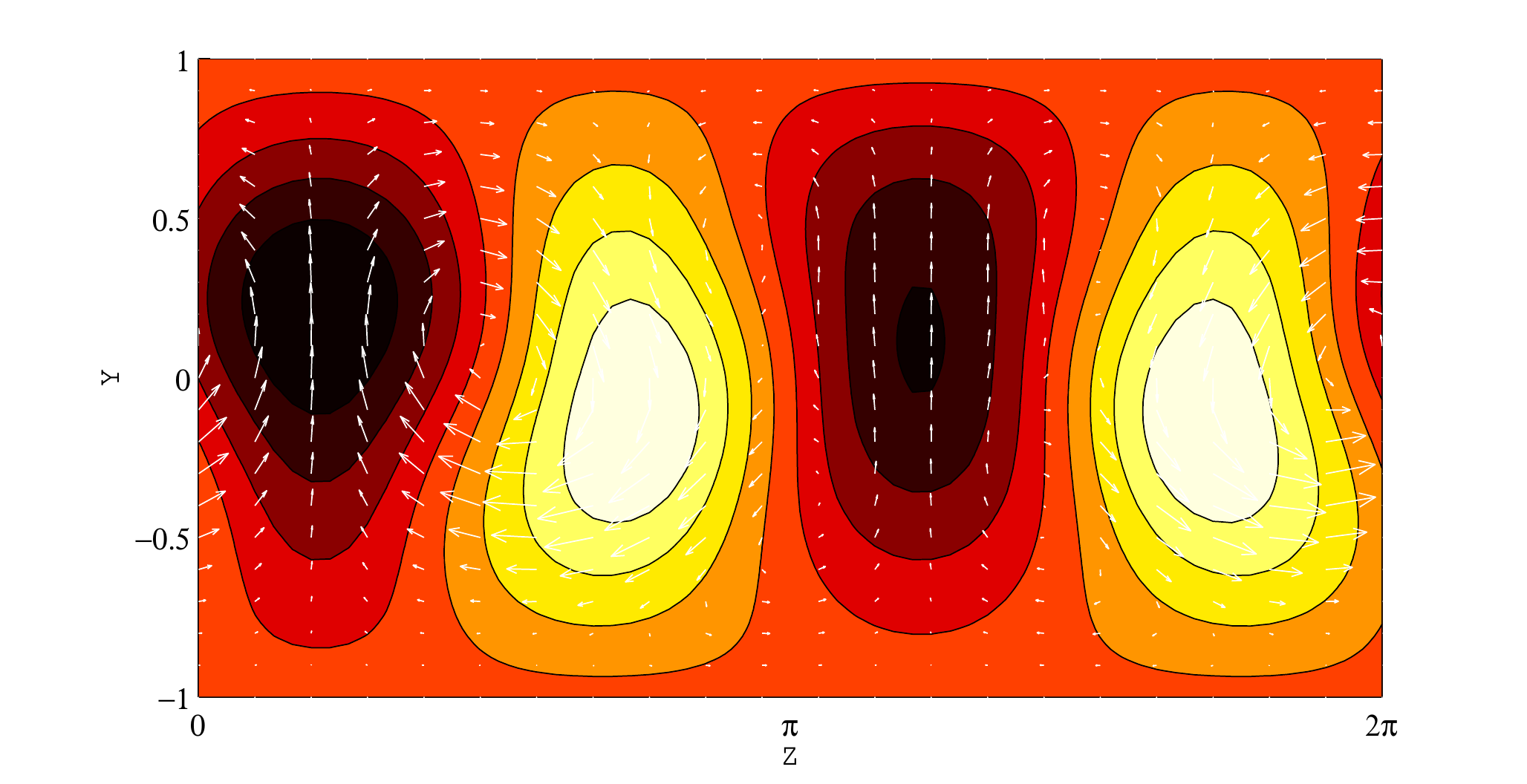}  % Chaotic_T20_1e3_4rolls_t11
\adjincludegraphics[angle=0,height=2.5cm ,trim={{.075\width} {.029\height} {.035\width}  {.06\height}},clip]{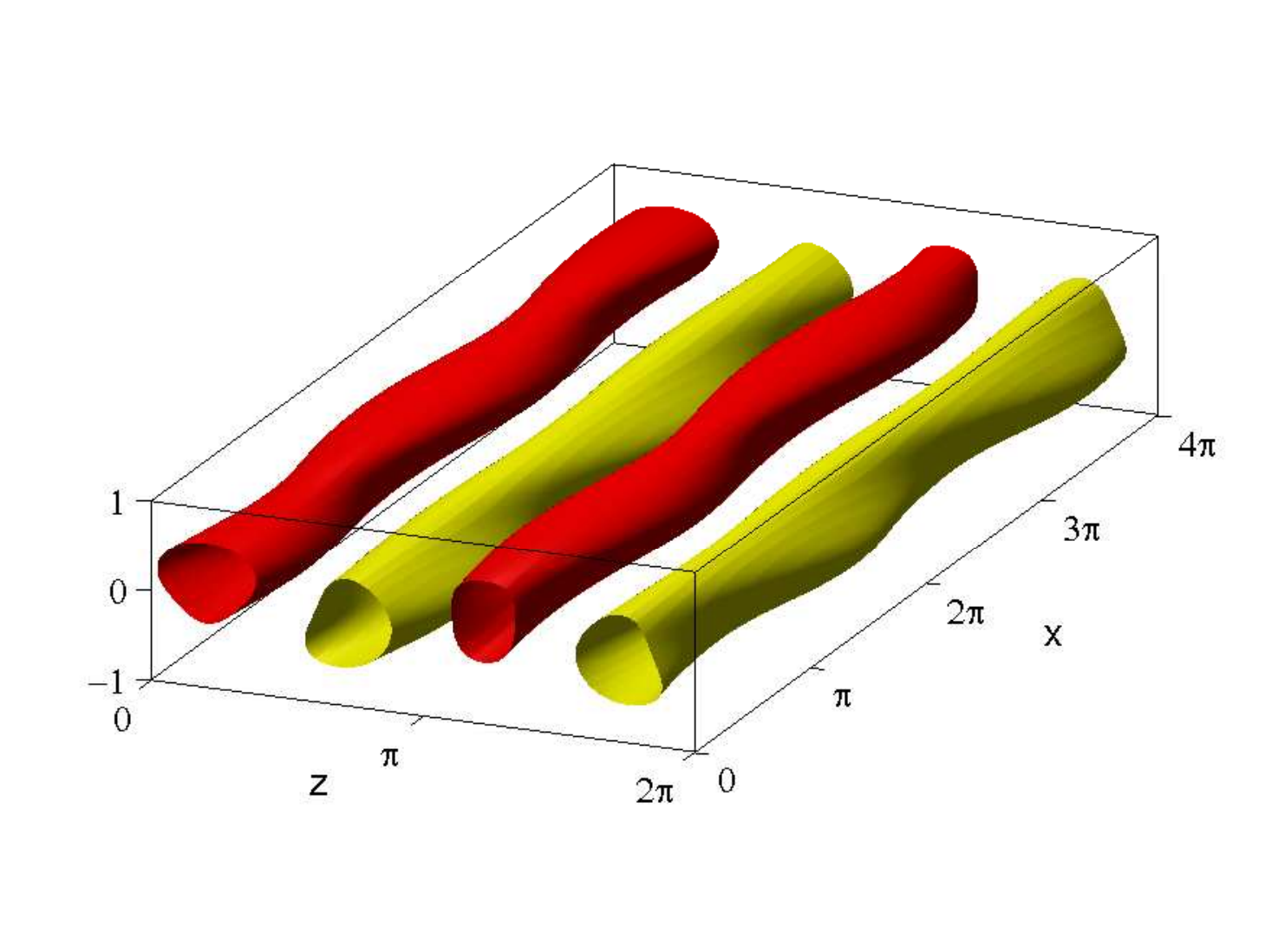} \\  % Chaotic_T20_1e3_4rolls_t11_ISOS
 \caption{ Contours of streamwise velocity for $NLOP_1$ (left) using  $T=20$ at $E_0=1.0 \times 10^{-3}$ and what it evolves into at $t=11$ (close to peak growth) on the right.  The same contour levels are used for both plots (8 levels between -0.315 and 0.30) and the isocontours are  $\pm 60\%$ of maximum streamwise perturbation velocity.   }
 \label{YZ_T20_E1e3}
    \end{center}
  \end{figure}

%
% Fig 12 thesis fig 5.5    NLOP_2 and what it evolves into.
%
\begin{figure}[htp]    %  --
    \begin{center}    
\adjincludegraphics[angle=0,width=5cm ,trim={{.075\width} {.029\height} {.075\width}  {.06\height}},clip]{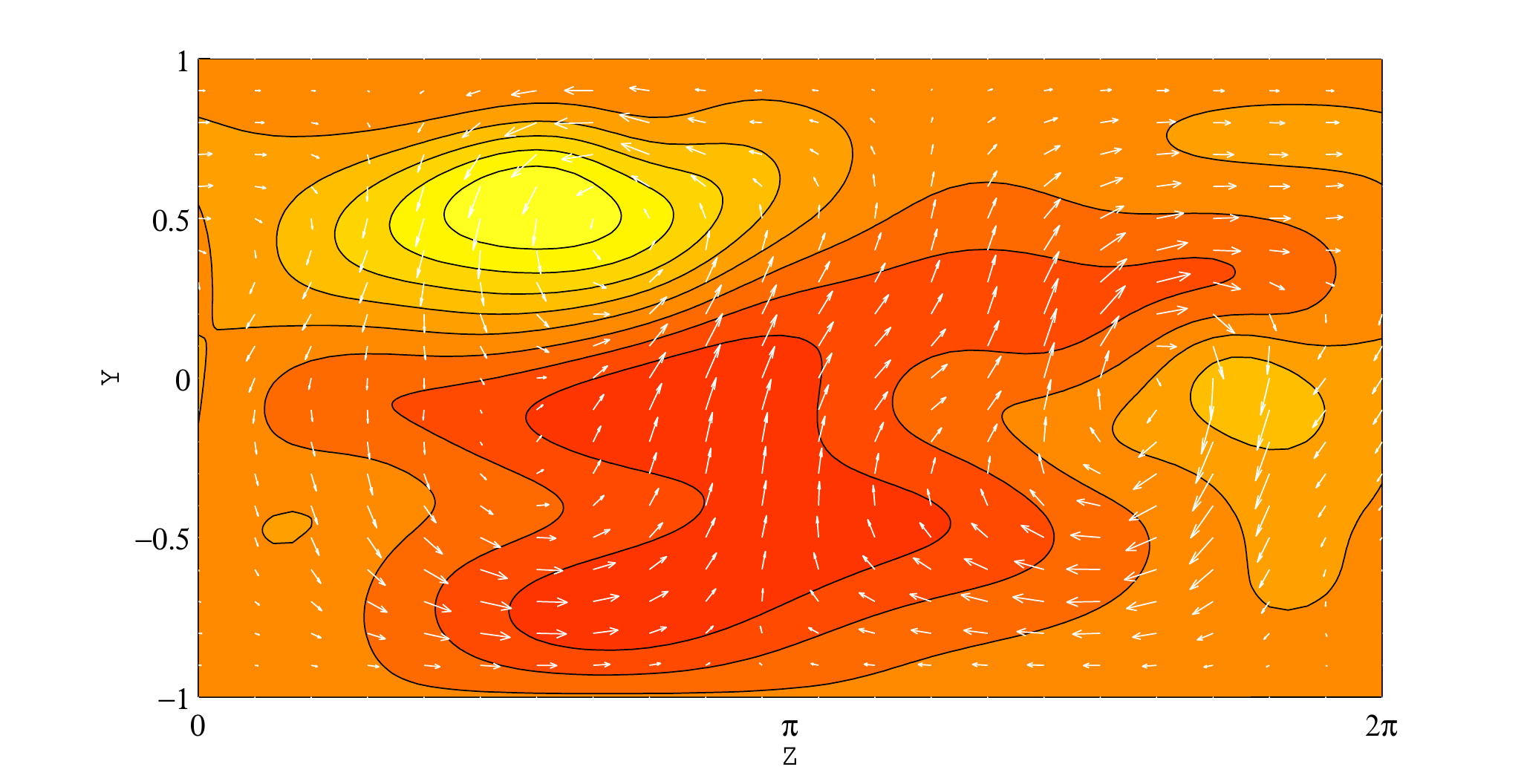}   % NLOP_T40_8e3_t0
\adjincludegraphics[angle=0,height=2.5cm ,trim={{.075\width} {.029\height} {.035\width}  {.06\height}},clip]{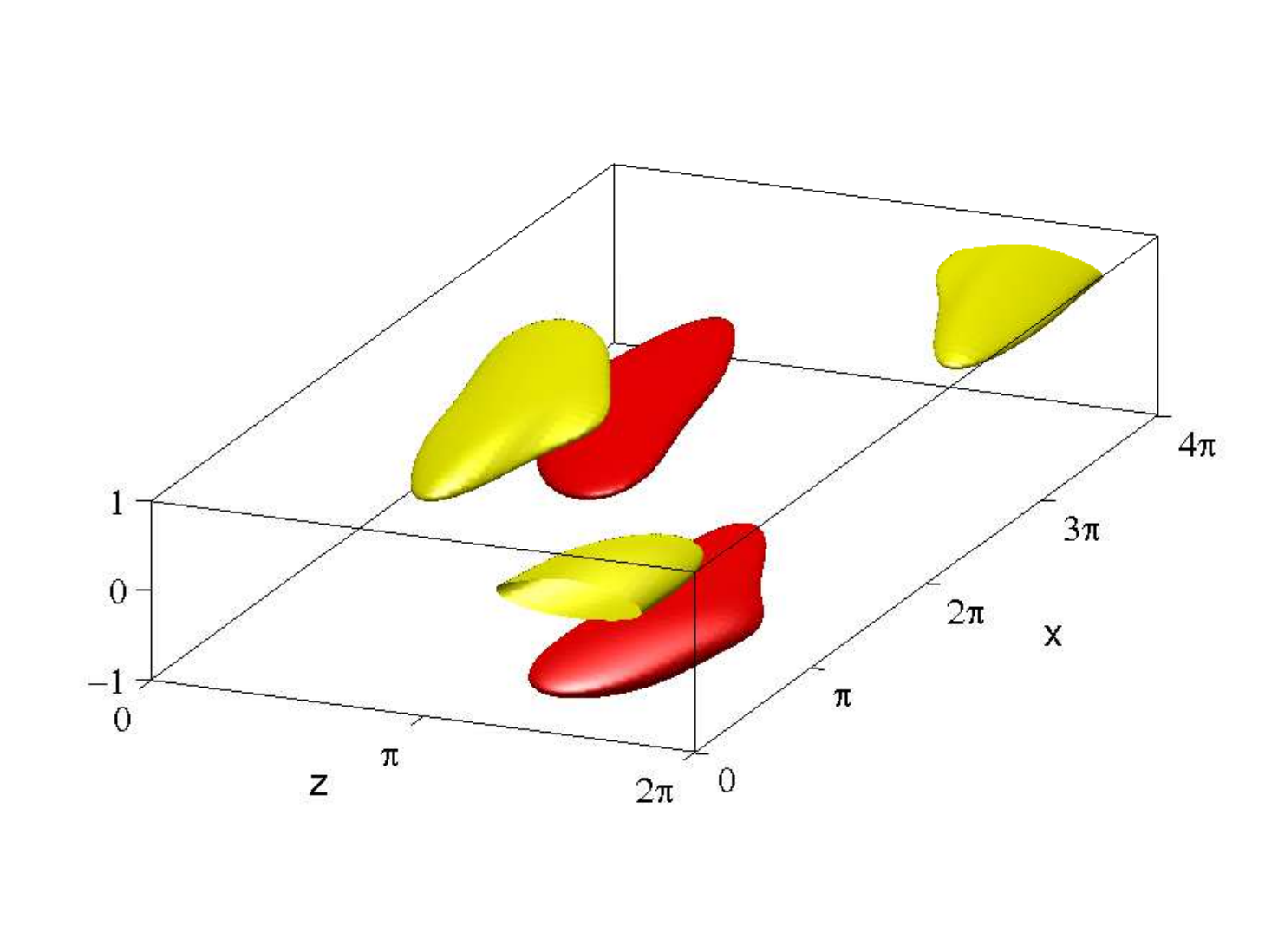} % NLOP_T40_8e3_t0_ISOS
\adjincludegraphics[angle=0,width=5cm, trim={{.075\width} {.029\height} {.075\width}  {.06\height}},clip]{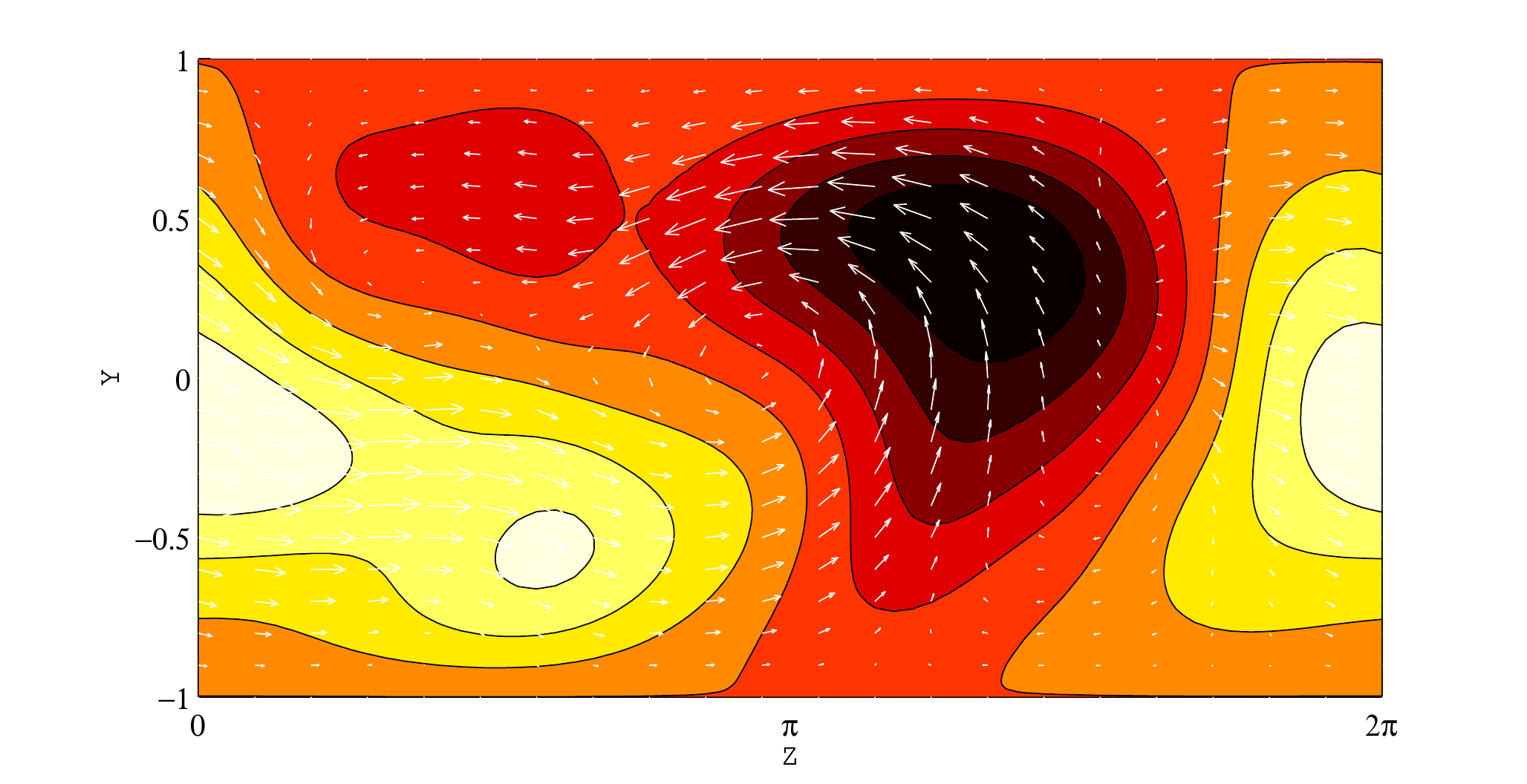}  % NLOP_T40_8e3_t16
\adjincludegraphics[angle=0,height=2.5cm ,trim={{.075\width} {.029\height} {.035\width}  {.06\height}},clip]{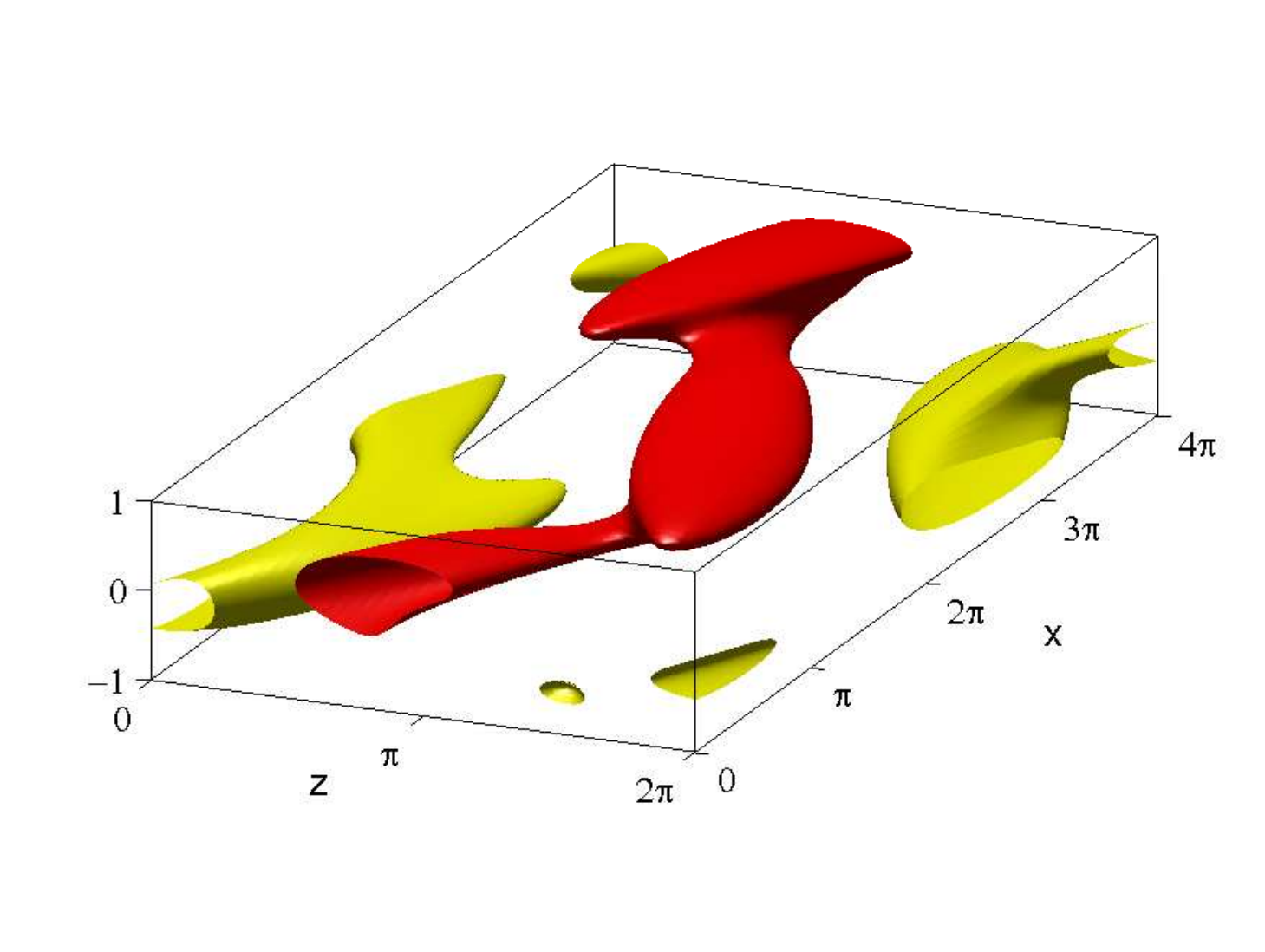} \\  % NLOP_T40_8e3_t16_ISOS
 \caption{ Contours of streamwise velocity for $NLOP_2$ (left)  using $T=40$ at $E_0=8.0 \times 10^{-3}$ and what it evolves into at  $t=16$ (maximum kinetic energy) on the right.  The same contour levels are used for both plots (8 levels between -0.72 and 0.58)  and the isocontours are  $\pm 60\%$ of maximum streamwise perturbation velocity.  }
 \label{T40_8e3}
    \end{center}
  \end{figure}

%
% Fig 13 thesis fig 5.7 (remove Ri_b=-0.005 line)  Evolution with unsucessful and successful convergence
%
\begin{figure}[htp]                
    \begin{center}%     
\adjincludegraphics[angle=0,height=4.5cm ,trim={{.05\width} {.02\height} {.08\width} {.055\height}},clip]{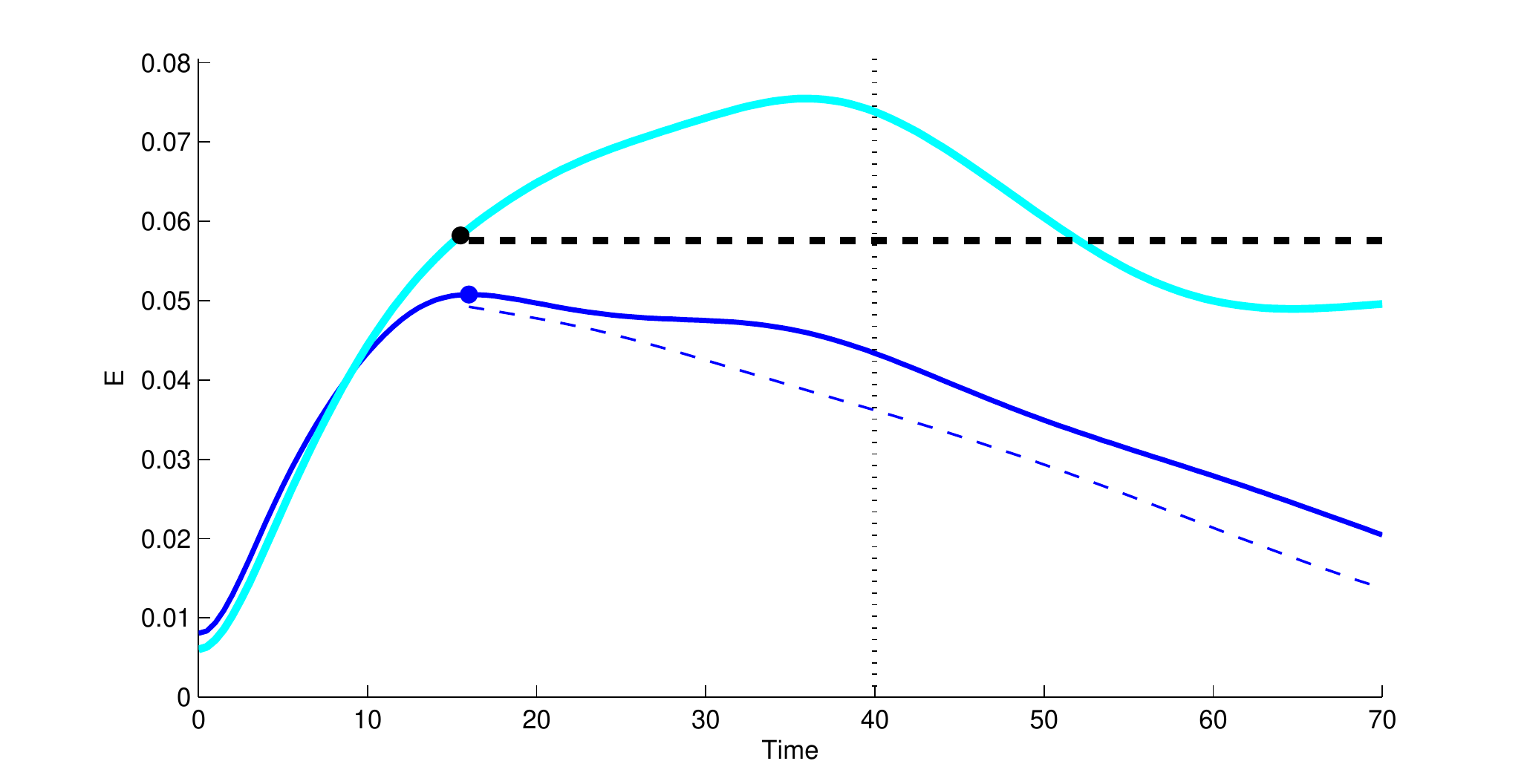}\\   % 
 \caption
{ 
Time evolution of optimal perturbations for $T=40$.
Blue line, evolution of $NLOP_2$ at $Ri_b=0$ for $Re=100$ at $E_0=8.0 \times 10^{-3}$; cyan line, evolution of $NLOP_2$ at $Ri_b=0$ for $Re=130$ at $E_0=6.0 \times 10^{-3}$.
The circles indicate states used as initial guess for the GMRES algorithm, the dashed lines show the evolution of the
GMRES output. When $Re=100$ at $E_0=8.0 \times 10^{-3}$ the GMRES iterations decays back to the laminar state.
At $Re=130$, however, and  $E_0=6.0 \times 10^{-3} $ GMRES converges to Nagata's solution. 
}
      \label{NLOP2_RE100_RE130}
    \end{center}
  \end{figure}

%
% Fig 14 thesis fig 5.8   starting guess and Nagata converged.
%
%\begin{figure}[htp]  %  T=40 
 %   \begin{center}  
%\adjincludegraphics[angle=0,width=7.5cm ,trim={{.075\width} {.029\height} {.08\width}  {.06\height}},clip]{./low_Re/GMRESguess_NLOP2_6e3.eps}
%\adjincludegraphics[angle=0,height=5cm ,trim={{.075\width} {.029\height} {.035\width}  {.06\height}},clip]{./low_Re/GMRESguess_NLOP2_6e3_ISOS.eps}\\
%\adjincludegraphics[angle=0,width=7.5cm ,trim={{.075\width} {.029\height} {.08\width}  {.06\height}},clip]{./low_Re/GMRESconveged_NLOP2_6e3.eps}
%\adjincludegraphics[angle=0,height=5cm ,trim={{.075\width} {.029\height} {.035\width}  {.06\height}},clip]{./low_Re/GMRESconveged_NLOP2_6e3_ISOS.eps}\\
% \caption[]{ Contours of streamwise velocity for $T=40$ and $E_0=6.0 \times 10^{-3}$ 
%taken in the middle of channel ($x=\pi$). 
%GMRES input and output using NLOP-2 at $Re=130$, (see figure \ref{NLOP2_RE100_RE130}).   Left,
%state at $t=16.0$ used as a initial guess for the GMRES algorithm. Left, output of GMRES, converged ECS (Nagata solution)   }
 %     \label{GMRES_Re130Converged}
 %   \end{center}
 % \end{figure}

%
% Fig 14   BIG  PICTURE   Target vs E_0
%
\begin{figure}
\begin{center}   
\adjincludegraphics[angle=0,width=8.5cm ,trim={{.08\width} {.02\height} {.06\width} {.055\height}},clip]{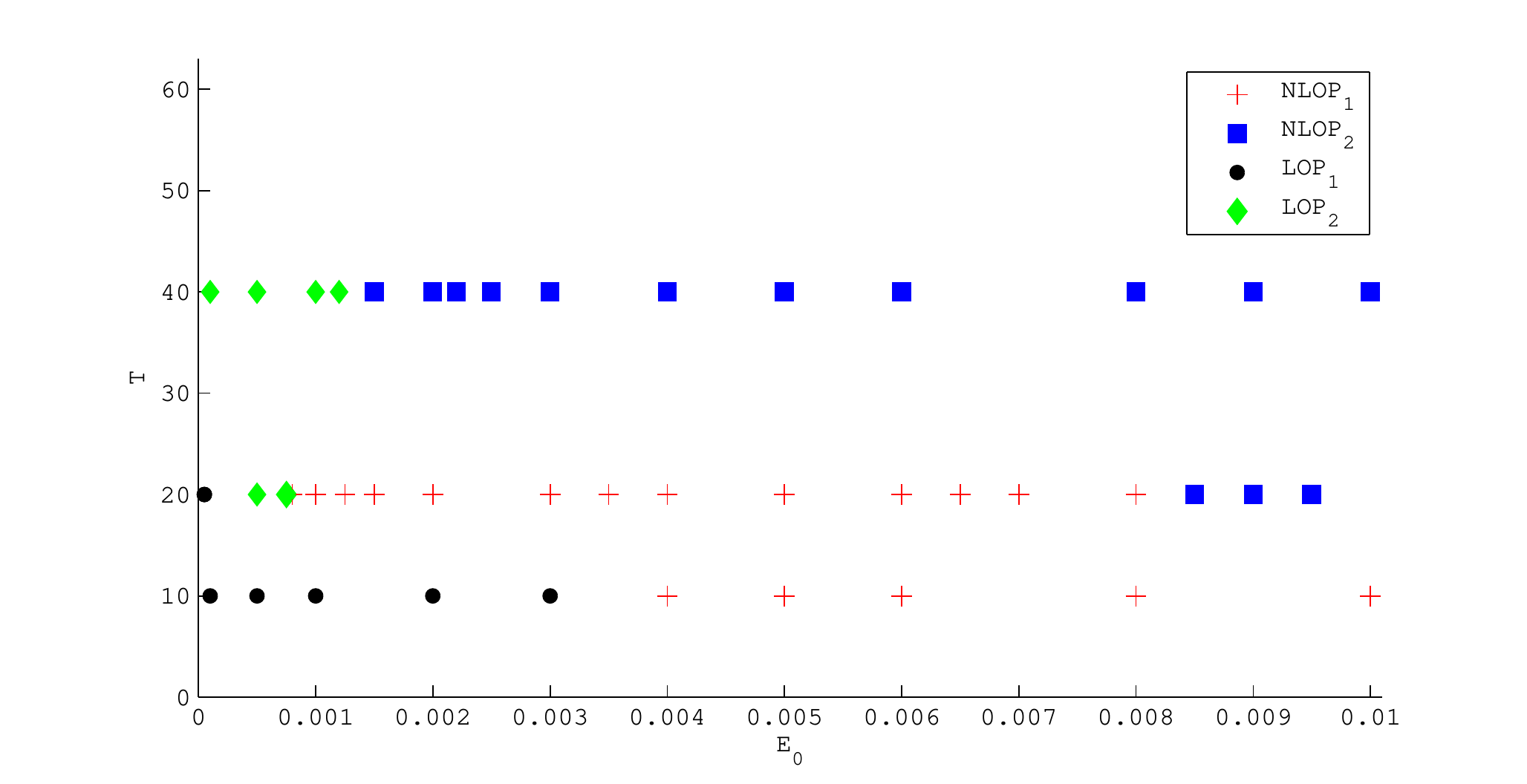}
\adjincludegraphics[angle=0,width=8.5cm ,trim={{.05\width} {.02\height} {.07\width} {.025\height}},clip]{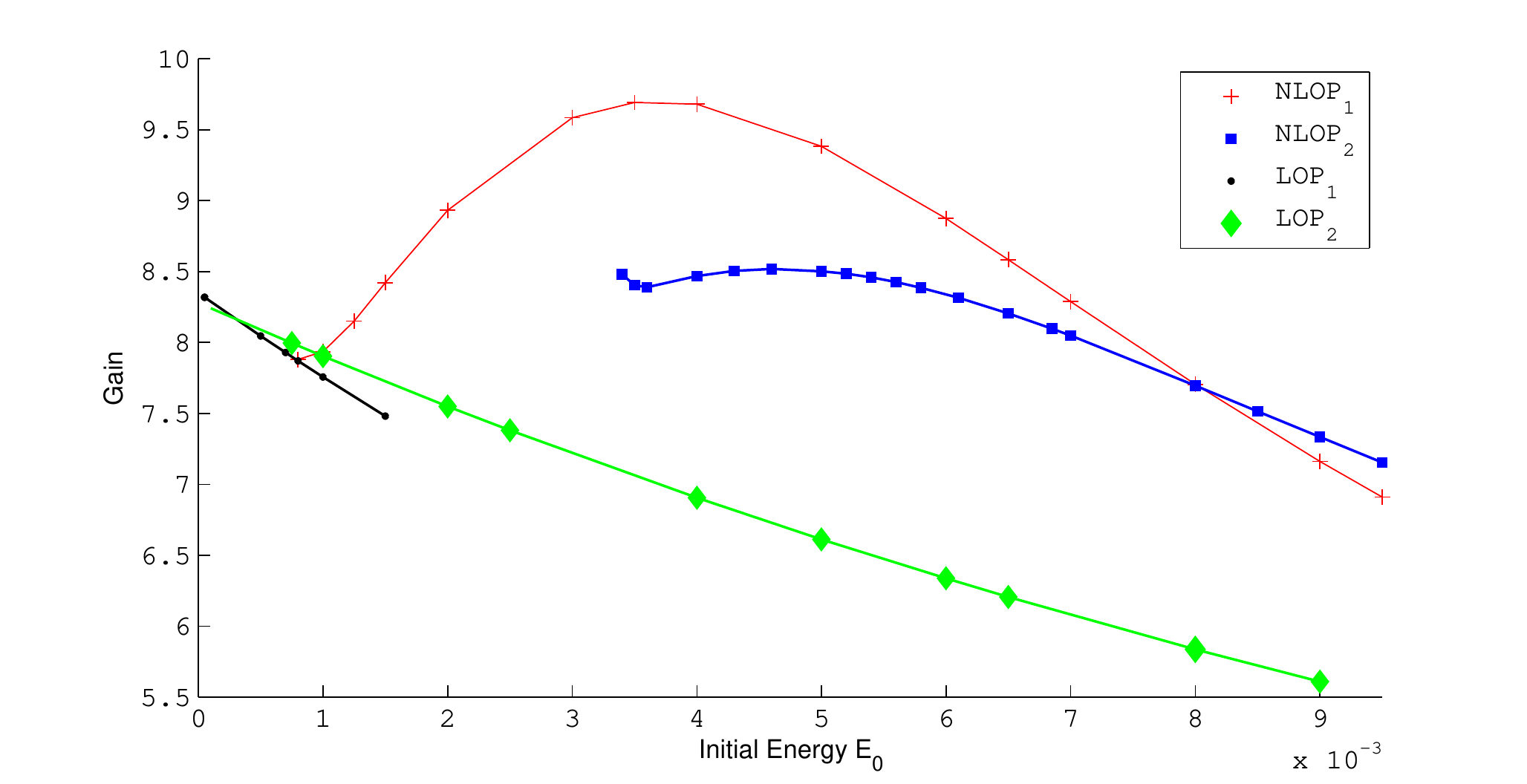}
 \caption[ ]
{
Left: $T$ against $E_0$ of optimal perturbations for $Re=100$ in  a $4 \pi \times 2 \times 2\pi$ box indicating which type of optimal is the global optimal. 
Right: a slice across the left plot at $T=20$ showing optimal and suboptimal gains against initial energy $E_0$. Note that the preferred optimal quickly changes from $LOP_1$ (black line) to $LOP_2$ (green line) as $E_0$ increases from 0 and that $NLOP_1$ and then $NLOP_2$ eventually win out as global optimals. 
 }
\label{BP_optimals}
\end{center}
\end{figure}

At $E_0=10^{-3}$,  a new nonlinear optimal perturbation ($NLOP_1$) emerges which actually experiences larger growth than both the $LOP_1$ or $LOP_2$ at earlier times.
 This initial condition is 3D, localised towards one wall and evolves into 2 pairs of wavy fast-slow streaks: 
see figure \ref{YZ_T20_E1e3}. % {\color{red} (thesis 5.2)}. 
However, there is no discernable energy plateau in its evolution so  $T$ was increased to 40 whereupon  a different nonlinear optimal ($NLOP_2$) emerges  
at  $E_0 \approx 1.5 \times 10^{-3}$: see figure   \ref{T40_8e3} for its structure at $E_0=8 \times 10^{-3}$. 
This optimal gives rise to an energy plateau in its subsequent evolution as the 
initial energy is increased:  see figure  \ref{NLOP2_RE100_RE130} 
for the situation at $E_0=8 \times 10^{-3}$. In this, a good candidate to initiate a convergence attempt  is the flow state at $t=16$, however,  this simply converges to the linearly sheared base state.  Repeating the calculation at $Re=130$ (with $E_0=6 \times 10^{-3}$), again taking the flow state at $t=16$,  does converge but  to 
Nagata's solution. %: see figure \ref{GMRES_Re130Converged}
A number of other  searches were done for $Re<128$ with all guesses converging to the base state and for  $Re>128$ where all attempts converged smoothly to Nagata's solution (the geometry is slightly sub-optimal in that the saddle node value for Nagata's solution is $\approx 128$ rather than $127.7$).  No evidence emerged of any other state beyond Nagata's solution existing during these computations adding further weight to the view that  the linearly-sheared base state is a global attractor up to $Re=127.7$. 
 
 Despite this apparent simplicity, the results of a systematic optimal energy growth analysis over the $(E_0,T)$ plane are still quite rich. Figure \ref{BP_optimals} indicates the various global optimals found at $Re=100$ and $T=10$, $20$ and $40$ over the interval $E_0 \in [0,0.01]$. At $T=20$, for example,  4 different optimals emerge with all  being the global optimal at some $E_0$: see Figure \ref{BP_optimals}(right).    That the nonlinear energy growth problem is nontrivial even in the absence of any exact coherent structures, is presumably because phase space is already starting to structure itself to incorporate such states at slightly higher $Re$.

%
% Fig 15                Time evolution of perturbation at Ri=1e-6  ( Thesis figure 3.5 )
%
   \begin{figure}[htp]  % 
    \begin{center}   
\adjincludegraphics[angle=0,height=4cm ,trim={{.06\width} {.02\height} {.08\width} {.058\height}},clip]{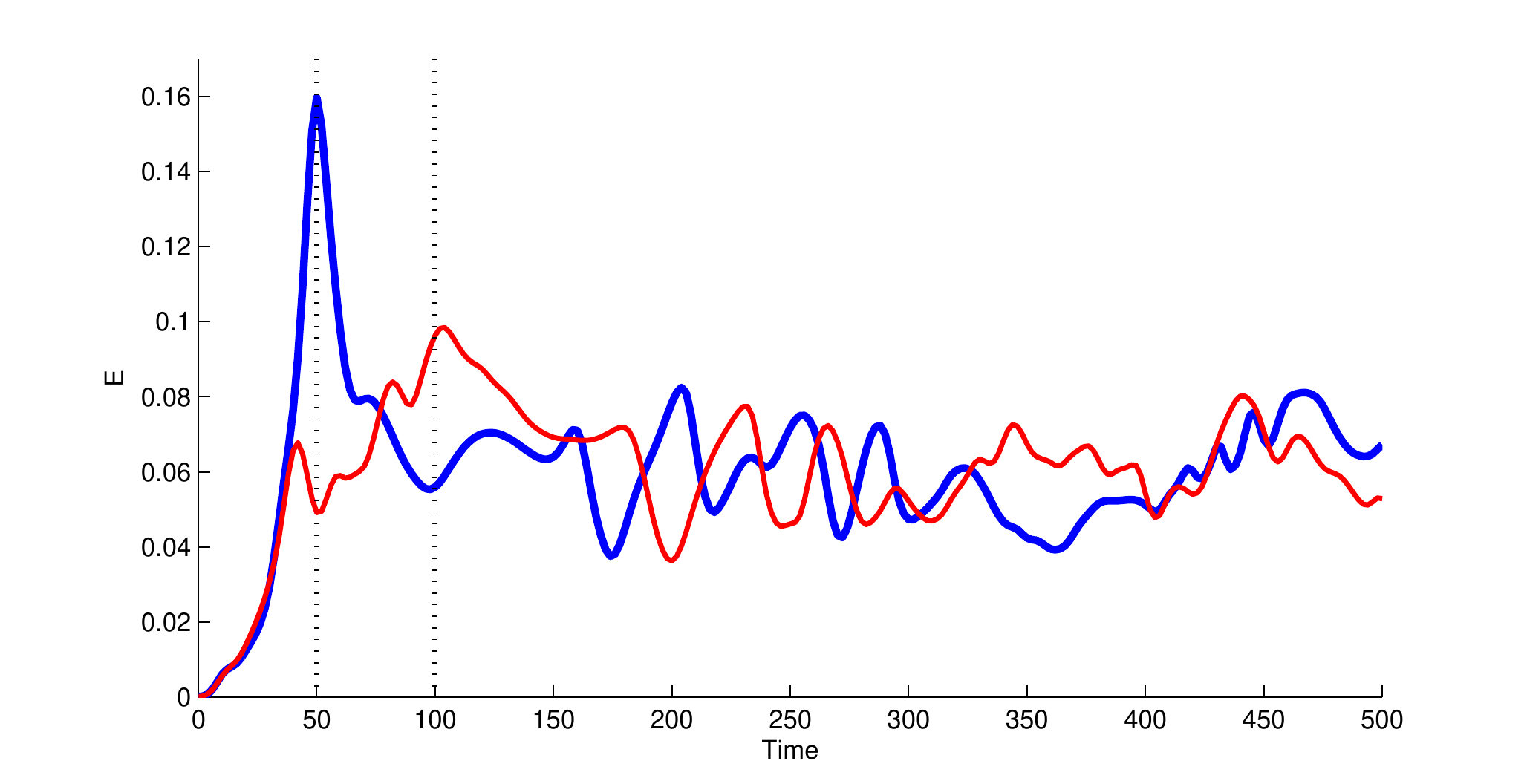}
      \caption[Bursts]{
Time evolution of initial conditions that lead to turbulence in very weakly stratified plane Couette flow at $Re=400$ and $Ri_b=1.0 \times 10^{-6}$ in box size $2\pi \times 2 \times \pi$. The initial energy is  $E_0=1.5\times10^{4}$ with the 
red line for target time $T=100$ and the thicker blue line for $T=50$. 
). 
}
      \label{Turbulence_T100_T50}
    \end{center}
  \end{figure}

%
% Fig 16  ( thesis figure 3.6) 
%
\begin{figure}[htp] 
     \begin{center}   
\adjincludegraphics[angle=0,height=4cm ,trim={{.06\width} {.01\height} {.08\width} {.058\height}},clip]{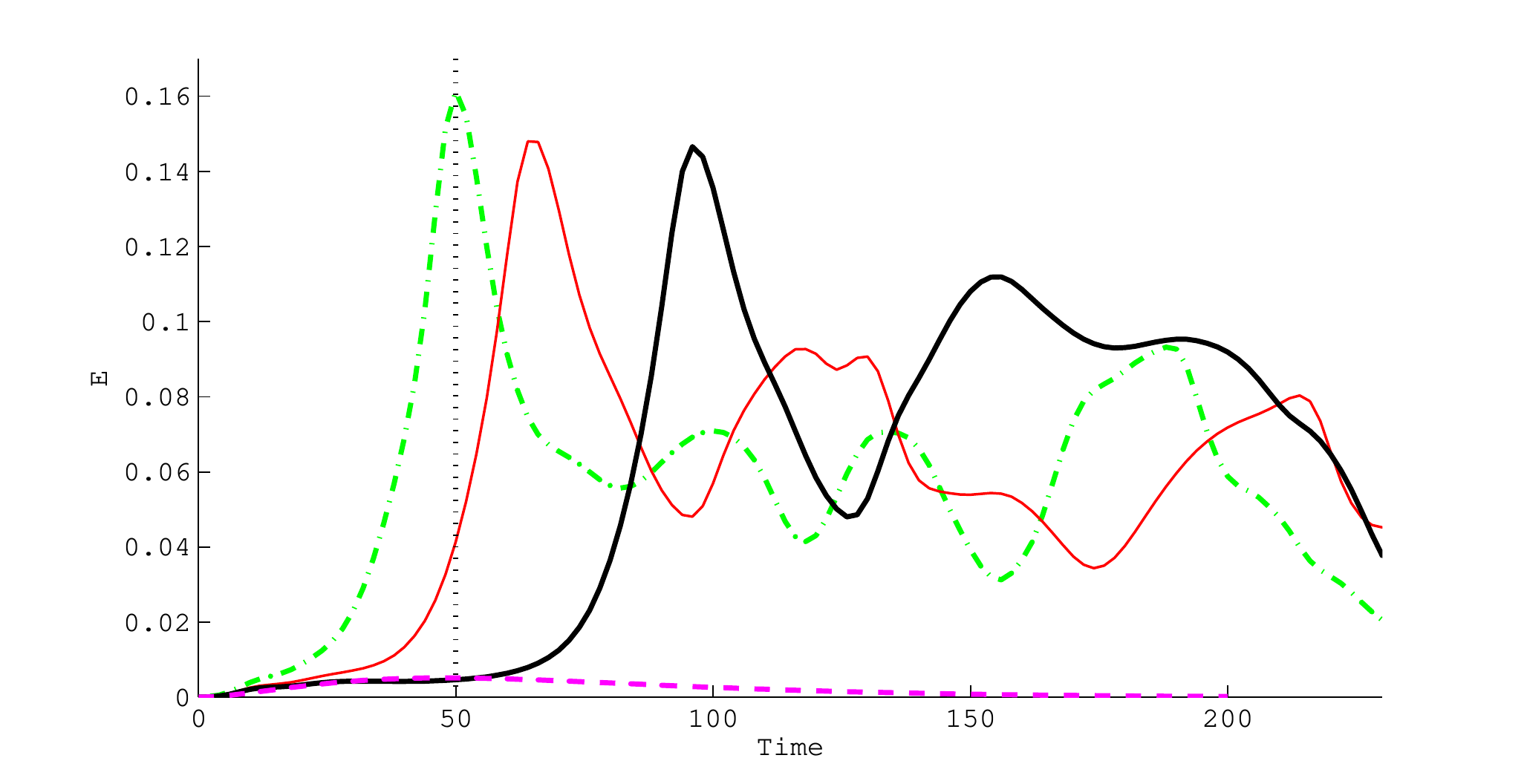} % E_CRITICAL_Ri1e6_T50
\adjincludegraphics[angle=0,height=4cm, trim={{.02\width} {.01\height} {.08\width} {.058\height}},clip]{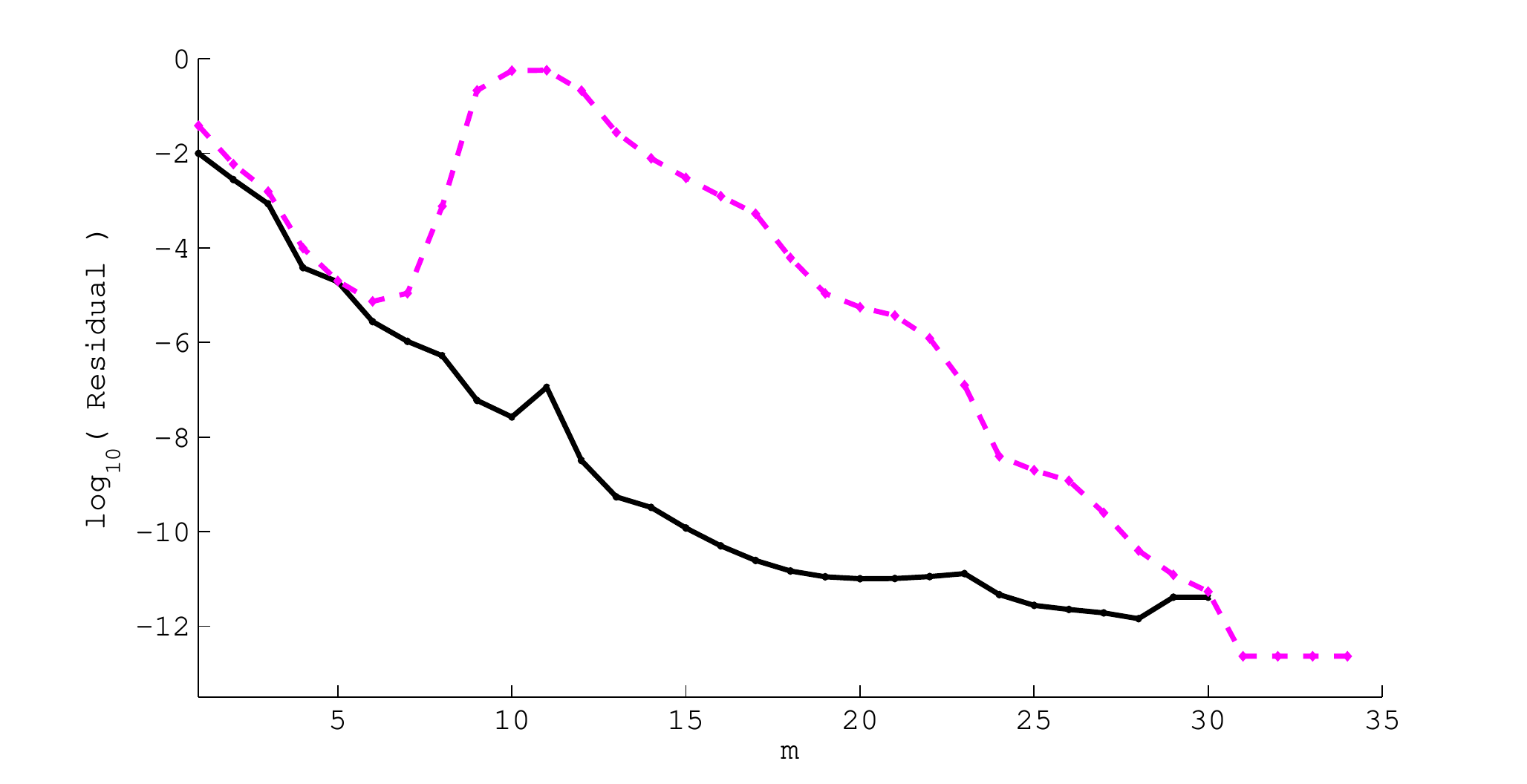} \\ % Ri1e6_Residuals_Variational
% \adjincludegraphics[angle=0,height=3.cm,width=5.5cm, trim={{.075\width} {.029\height} {.075\width}  {.06\height}},clip]{./high_RI/Ri1e6_T50_E_5e5_time40_NLOP.eps}    %  t=40
%\adjincludegraphics[angle=0,height=3.cm,width=5.5cm, trim={{.075\width} {.029\height} {.075\width}  {.06\height}},clip]{./high_RI/Ri1e6_T50_E_3e5_QLOP.eps} \\
 \adjincludegraphics[angle=0,height=2.5cm,width=5.5cm, trim={{.075\width} {.029\height} {.075\width}  {.06\height}},clip]{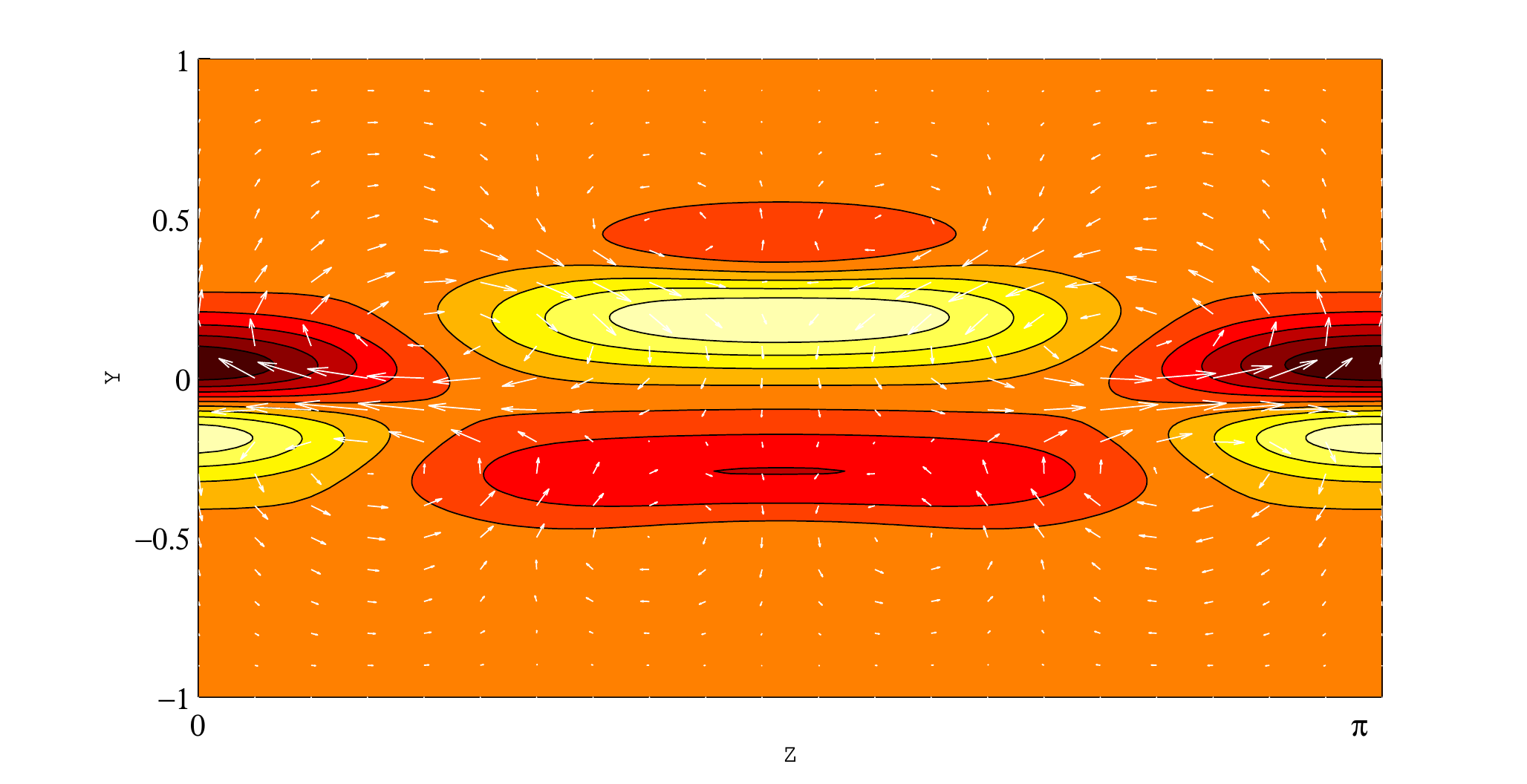} %  I.C. t=0  Ri1e6_T50_E_5e5_time0_NLOP
\adjincludegraphics[angle=0,height=2.5cm,width=5.5cm, trim={{.075\width} {.029\height} {.075\width}  {.06\height}},clip]{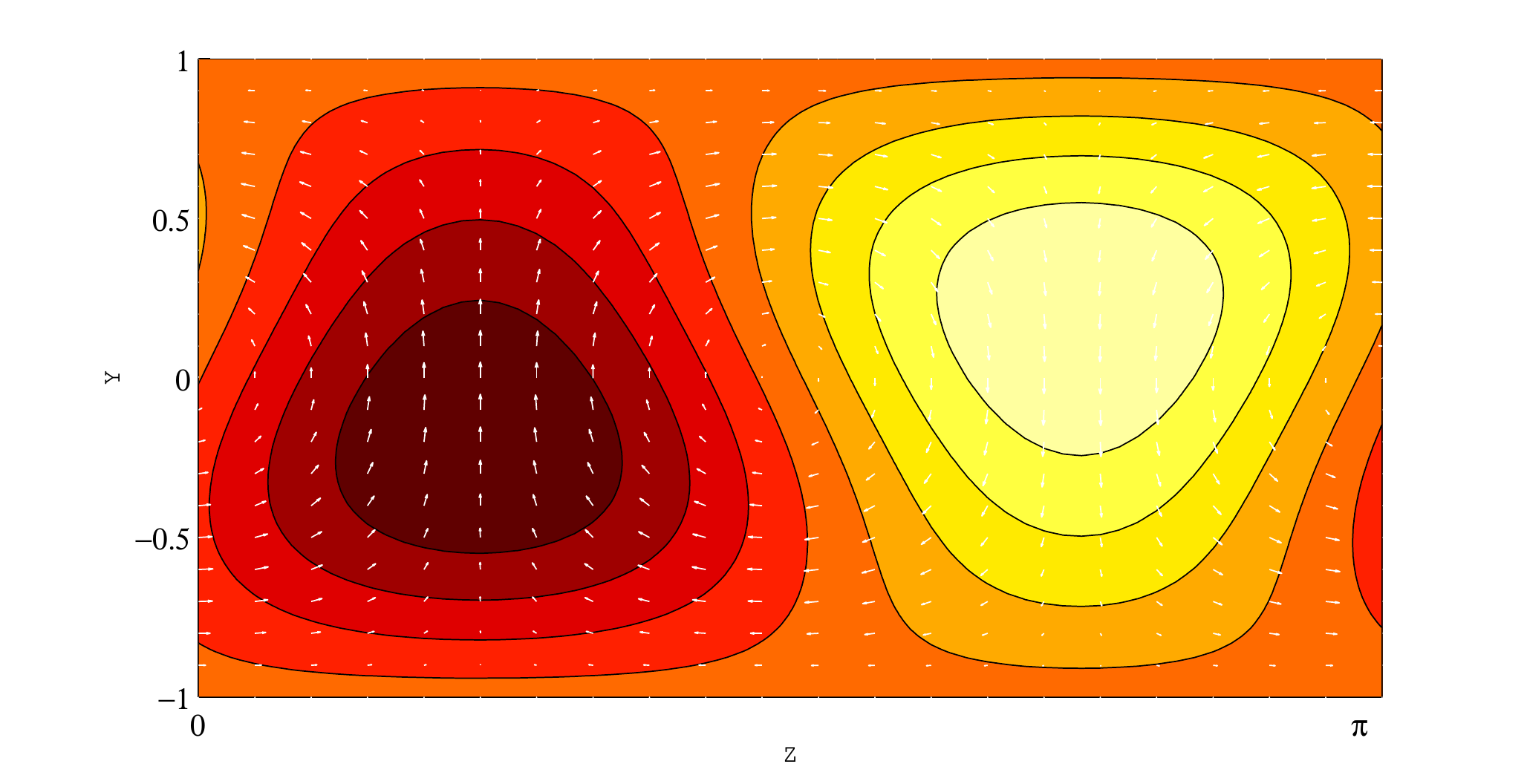} %  I.C.  t=0  Ri1e6_T50_E_3e5_time0_QLOP
%%%%%%%%%%%
      \caption[Bursts]{Top left: time evolution of  initial conditions at $Ri_b=1.0 \times 10^{-6}$, $Re=400$ and $T=50$
for different initial energies:
    $E_0 =9.0 \times 10^{-5}$ (green dash-dot line), 
    $E_0 =6.0 \times 10^{-5}$ (black line),
    $E_0 =5.0 \times 10^{-5}$ (red line),
    $E_0 =3.0 \times 10^{-5}$(magenta dashed line). 
The critical energy $E_c$  of the minimal seed lies between  $E_0 =3.0 \times 10^{-5}$ and $E_0 =5.0 \times 10^{-5}$. 
Top right: Residual as a function of iterate $m$  for the $E_0=5.0\times10^{-5}$ (solid black line) and $E_0=3.0\times10^{-5}$ (dashed magenta line). 
Bottom, optimal perturbations (initial conditions) for $E_0 =5.0 \times 10^{-5}$ (left),
 and $E_0 =3.0 \times 10^{-5}$ (right) to maximise growth at $T=50$ (the same 8 contour levels are used for both plots).  }
\label{Critical_E_Ri_1e6}
\end{center}
\end{figure}

%
%
%----------------------------------------------------------------------
\subsection{Stratified pCf for high $Ri_b$ \label{burst}}
%----------------------------------------------------------------------
%
%

As our third (and final) application of the optimal energy growth technique, we consider what happens to the edge, which in phase space separates states that become turbulent from those that relaminarise, when the addition of stable stratification kills the turbulent attractor.  Direct numerical simulations \cite{OK17} indicate that large energy growth can still occur and so we seek an explanation why by examining how the optimal energy growth perturbation changes as $Ri_b$ increases from 0 in a  geometry of $2 \pi \times 2 \times \pi$ where $Re=400$.

There are two important values of $Ri_b$: the value $Ri_b^t(Re)$ beyond which no turbulent attractor exists and the limiting value $Ri_b^m(Re)$ for the existence of the global equilibrium EQ7 \cite{Gibson09} which is the edge state in this geometry (\, $10^{-6} < Ri_b^t(400) \lesssim 10^{-2}$ and $Ri_b^m(400) \approx 0.057$). Previous work at $Re=1000$ by Rabin et al. \cite{Rabin12} and Eaves \& Caulfield \cite{Eaves15} has identified the minimal seed for transition at $Ri_b=0$ and $0< Ri_b< Ri_b^t$ respectively. Our focus here is $Ri_b^t \, < Ri_b $ but we start by looking at one value of $Ri_b=10^{-6} < Ri_b^t$ to make contact with this earlier work. 

Figure \ref{Turbulence_T100_T50}  % thesis figure 3.5
shows partially-converged optimal perturbations at $E_0=1.5 \times 10^{-4}$ using two target times $T=50$ and $100$ which both indicate that turbulence is  triggered. 
Taking  $T=50$ and reducing $E_0$ until turbulence is no longer triggered by any iterate or the final converged optimal suggests that $E_c$, the critical energy threshold for transition is between $E_0=3 \times 10^{-5}$ and $5 \times 10^{-5}$: see figure \ref{Critical_E_Ri_1e6}.   
The form of the  optimal at $3 \times 10^{-5}$ is still the nonlinearly adjusted $LOP$ whereas the optimal at $5 \times 10^{-5}$ resembles the minimal seed found by \cite{Rabin12} at  $Ri_b=0$ (see figure 5 of \cite{Rabin12}).
The optimal for $E_0=5 \times 10^{-5}$ leads to an evolution  from which a good enough starting guess can be extracted to converge EQ7 (not shown). This is another case where the approach has worked to identify an unstable state albeit only one which is  minimally unstable because, as  an edge state, it only has one unstable direction or eigenvalue.

\begin{figure}  
\begin{center}   
\adjincludegraphics[angle=0,height=5cm ,trim={{.06\width} {.02\height} {.08\width} {.058\height}},clip]{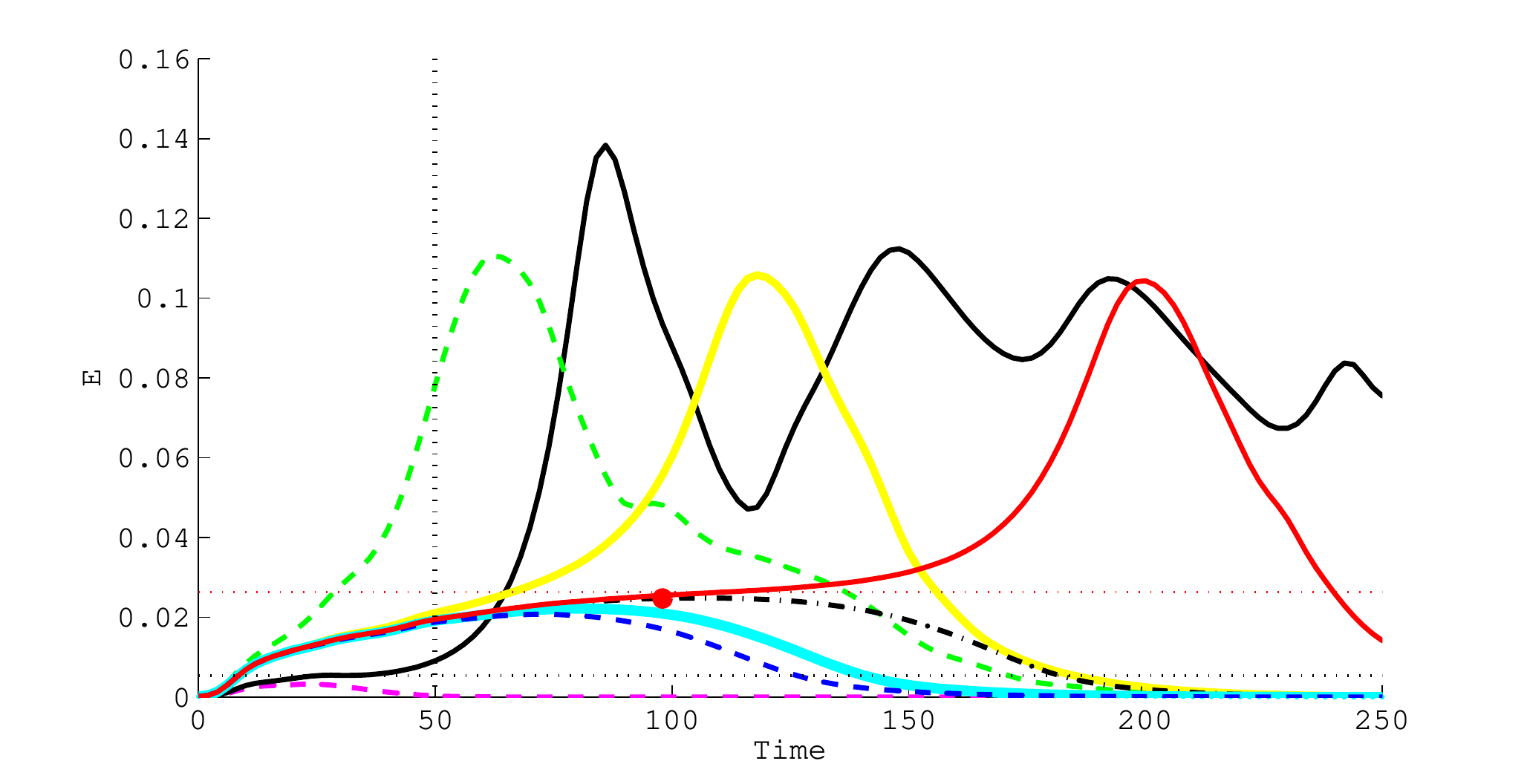}% BURST_RE400_RI01_04
\caption[Bursts]{The time evolution for optimal perturbations  at $Re=400$ and $T=50$ in a box $2\pi \times 2 \times \pi$ for  $Ri_b=0.01$ and $Ri_b=0.04$. For $Ri_b=0.01$, $E_c$ lies 
between $E_0 =6.0 \times 10^{-4}$ (dashed magenta line)  and $E_0 =7.5 \times 10^{-4}$(bold solid black line) with the lower horizontal black dotted line indicating  
the  energy of  EQ7 at $Ri_b=0.01$. A more detailed investigation at $Ri_b=0.04$ shows that $E_c$
lies between $E_0 =2.108 \times 10^{-4}$ (bold red line)
and $E_0 =2.102 \times 10^{-4}$ (dash-dotted black line). The state at $t=95$ and $E_0=2.108 \times 10^{-4}$ (indicated by a red dot) can be converged in 7 steps to EQ7, the energy of which is shown as the upper horizontal red dotted line.}
\label{BURST400}
\end{center}
\end{figure}

Moving to stronger stratification, figure \ref{BURST400} shows converged optimals for $E_0>E_c$ with  $Ri_b=0.01$ (which is just above $Ri_b^t$) and $Ri_b=0.04$. The former shows a transient turbulent episode while the latter only a  `burst' of energy growth which then decays away. Again if $E_0$ is near enough to $E_c$, the  evolution of the optimal transiently visits the neighbourhood of the edge state sufficiently closely to be able to select a flow state  which subsequently converges. For $Ri_b=0.04$, we find $2.102 \times 10^{-4} \,< \, E_c \, <\, 2.108 \times 10^{-4}$ and a flow state taken at $t=95$ from the $E_0=2.108 \times 10^{-4}$ optimal evolution converges in just 7 steps to EQ7 (not shown).

\begin{figure}
\begin{center}   
\adjincludegraphics[angle=0,height=4.8cm ,trim={{.05\width} {.02\height} {.08\width} {.058\height}},clip]{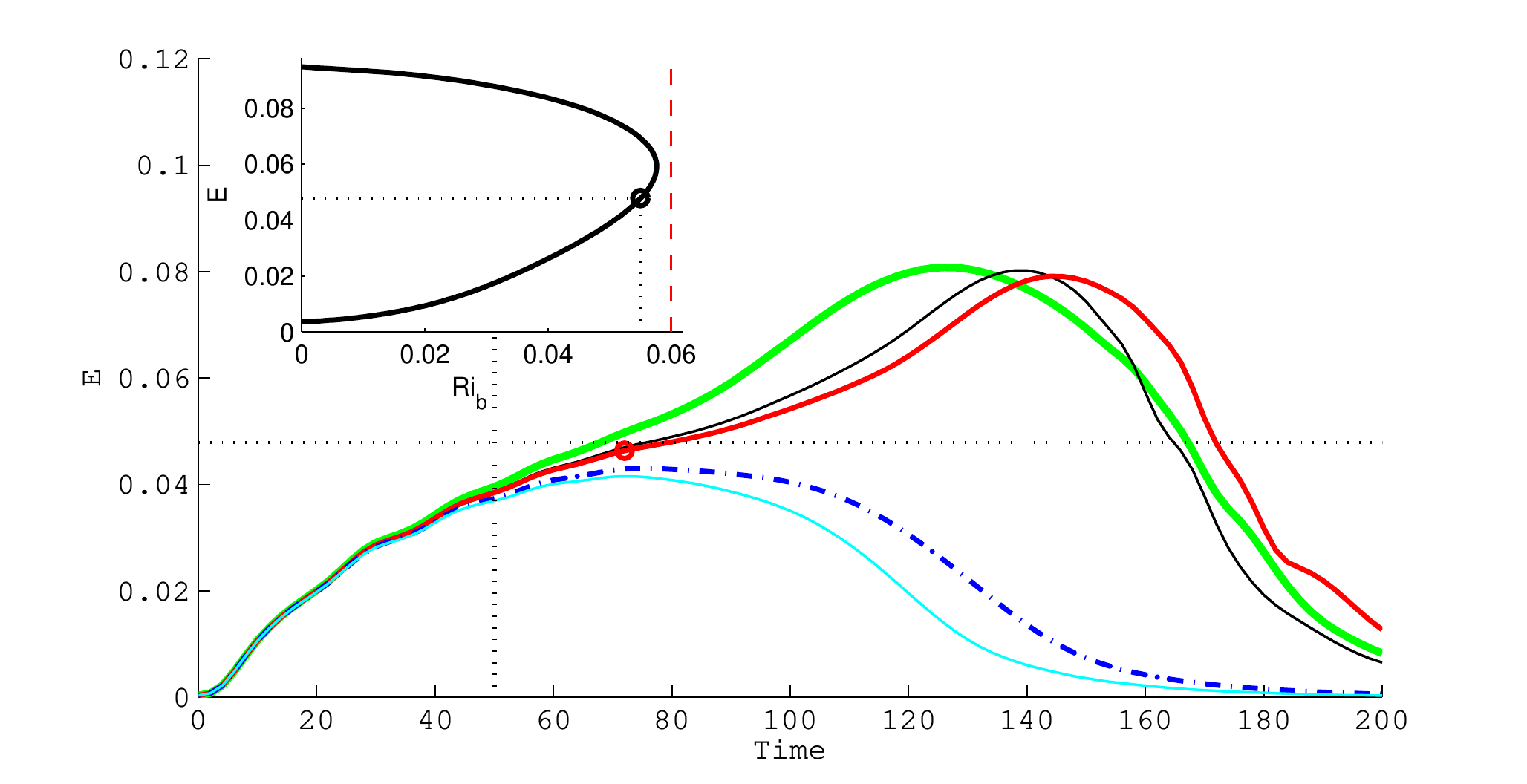}
\adjincludegraphics[angle=0,height=4.8cm ,trim={{.05\width} {.02\height} {.08\width} {.058\height}},clip]{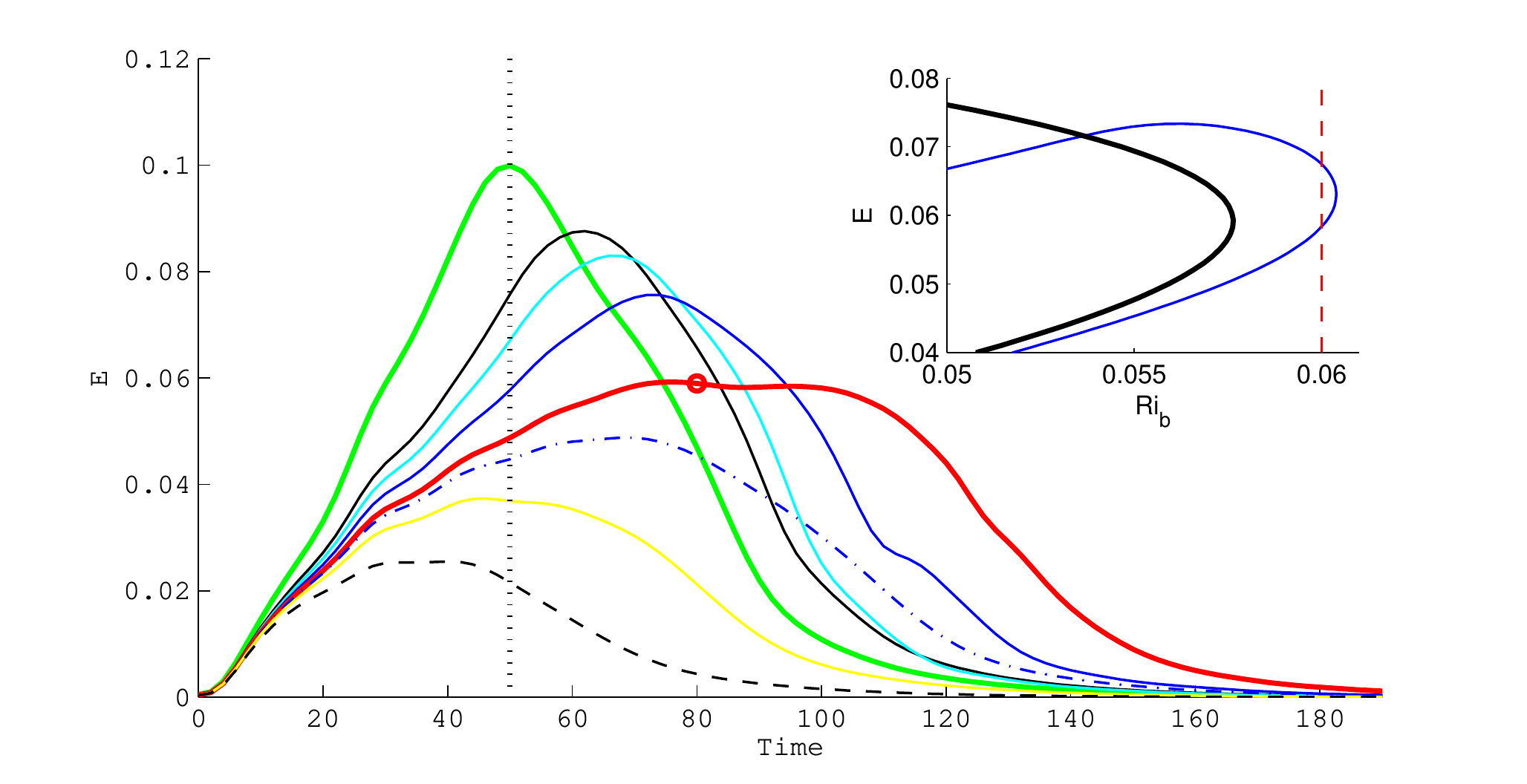}
\caption{
Left: The time evolution for optimal perturbations  at $Re=400$ and $T=50$ in a box $2\pi \times 2 \times \pi$ for  $Ri_b=0.055$ for various $E_0$. $E_c$ lies 
between $E_0 =3.53 \times 10^{-4}$ (bold red)  and $E_0 =3.525 \times 10^{-4}$(dash dot) with the lower horizontal black dotted line indicating  
the  energy of EQ7 at $Ri_b=0.055$.
Right.  Time evolution of optimal  perturbations for several $E_0$  in the same box for $Ri=0.06$.
No discernible critical $E_0$ can be detected as the magnitude of the burst decays monotonically. 
$E_0=5.0 \times 10^{-4}$, green line;
$E_0=4.5 \times 10^{-4}$, black line;
$E_0=4.4 \times 10^{-4}$, cyan line;
$E_0=4.3 \times 10^{-4}$, blue line;
$E_0=4.2 \times 10^{-4}$, red bold line;
$E_0=4.15 \times 10^{-4}$, blue dash-dot line;
$E_0=4.05 \times 10^{-4}$, yellow line;
$E_0=3.80 \times 10^{-4}$, black dashed line.
The flow state at $t=80$ (red circle) from  $E_0=4.2 \times 10^{-4}$ trajectory
was used as initial guess for the Newton-GMRES method but failed to converge.
Inset, continuation in $Ri_b$ at fixed $Re=400$ (black bold line) and $Re=500$ (blue line) of solution EQ7 making it clear that EQ7 does not exist at $Ri_b=0.06$ for $Re=400$ but does at $Re=500$.
}
\label{BURST}
\end{center}
\end{figure}

\begin{figure}[htp]  % 
    \begin{center}    %
\adjincludegraphics[angle=0,height=2.5cm,width=4.5cm ,trim={{.075\width} {.029\height} {.075\width}  {.06\height}},clip]{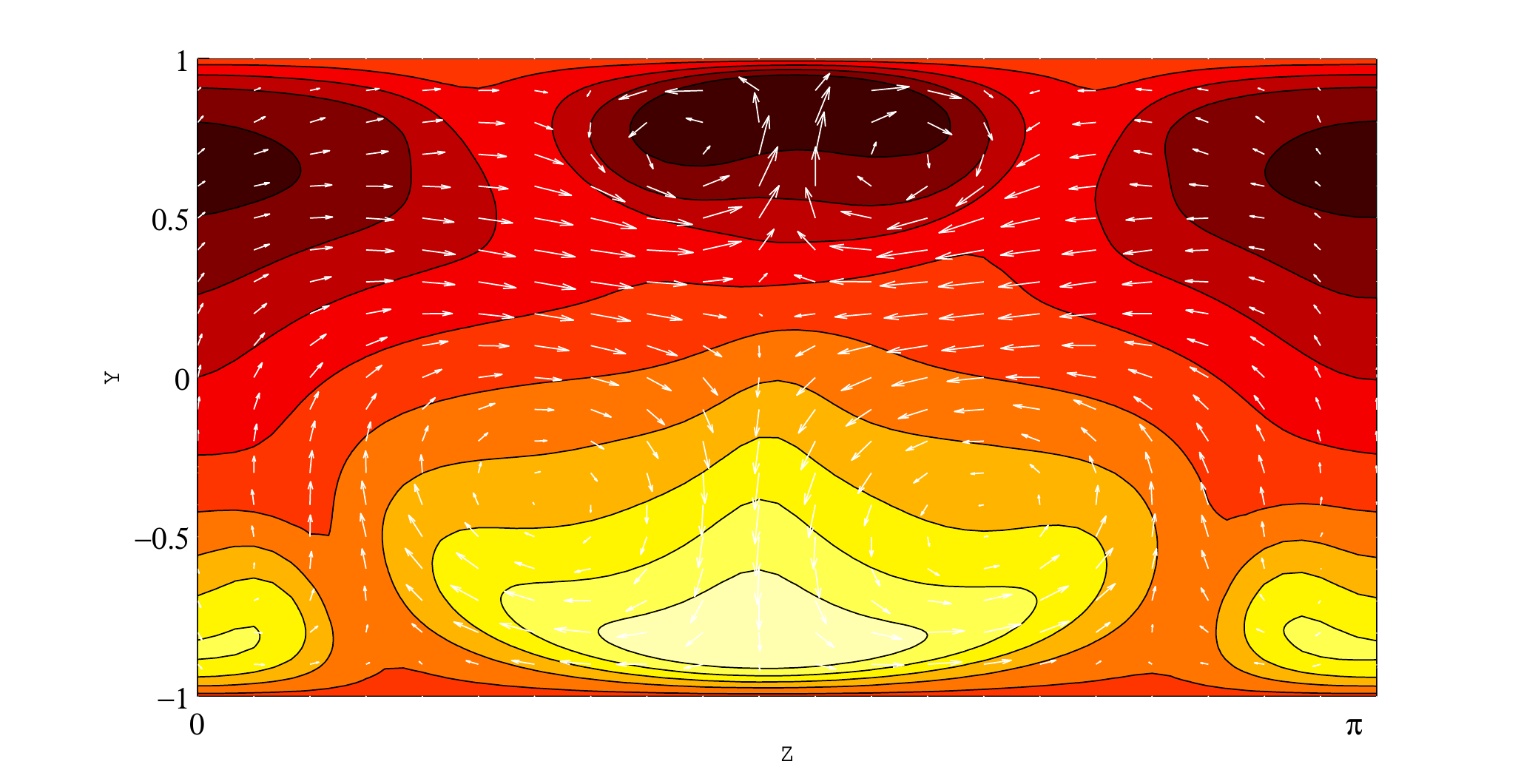}  % MAX_RI04
\adjincludegraphics[angle=0,height=3cm ,trim={{.075\width} {.029\height} {.085\width}  {.06\height}},clip]{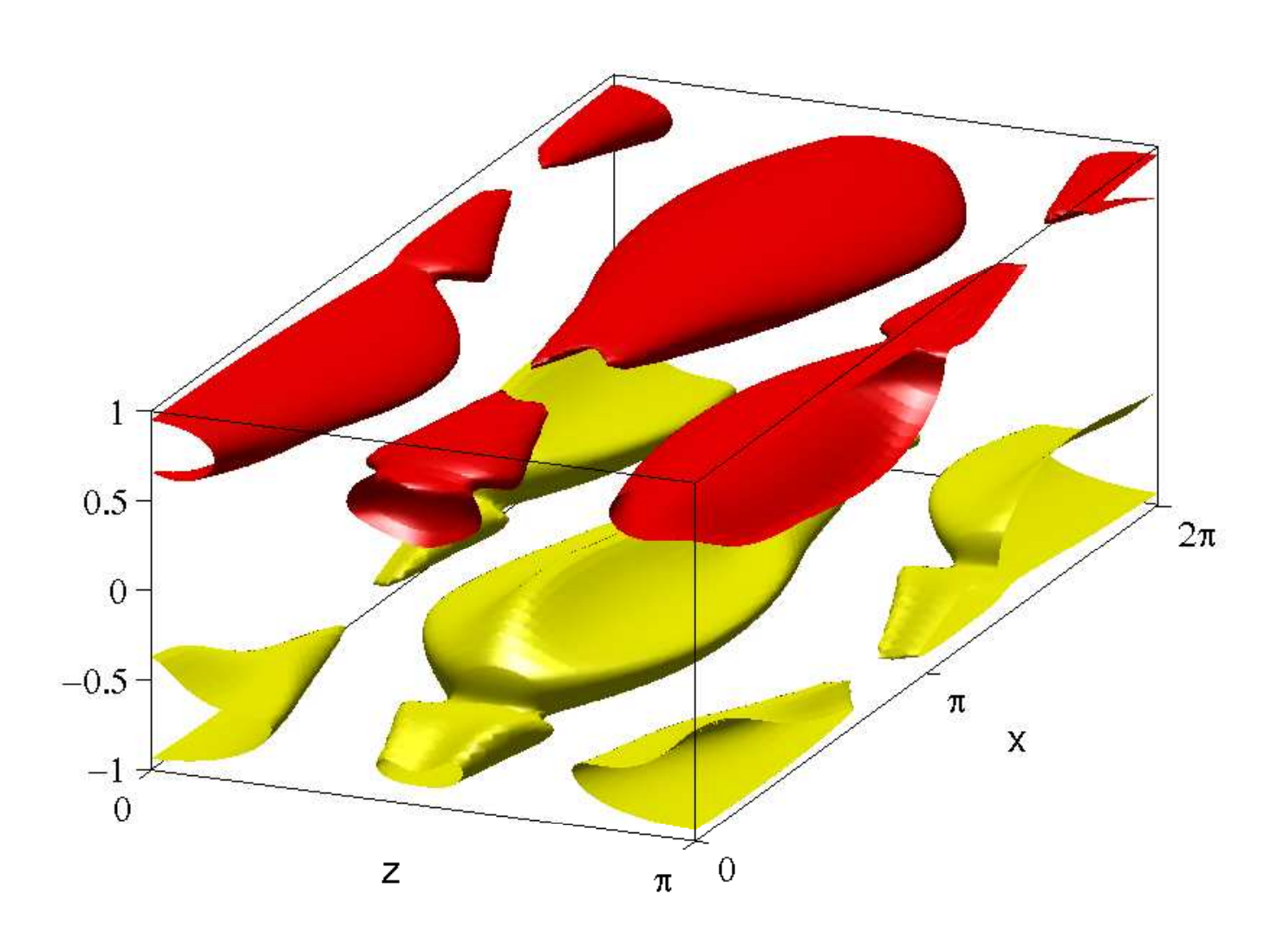}  % MAX_RI04_ISO
\adjincludegraphics[angle=0,height=2.5cm,width=4.5cm ,trim={{.075\width} {.029\height} {.075\width}  {.06\height}},clip]{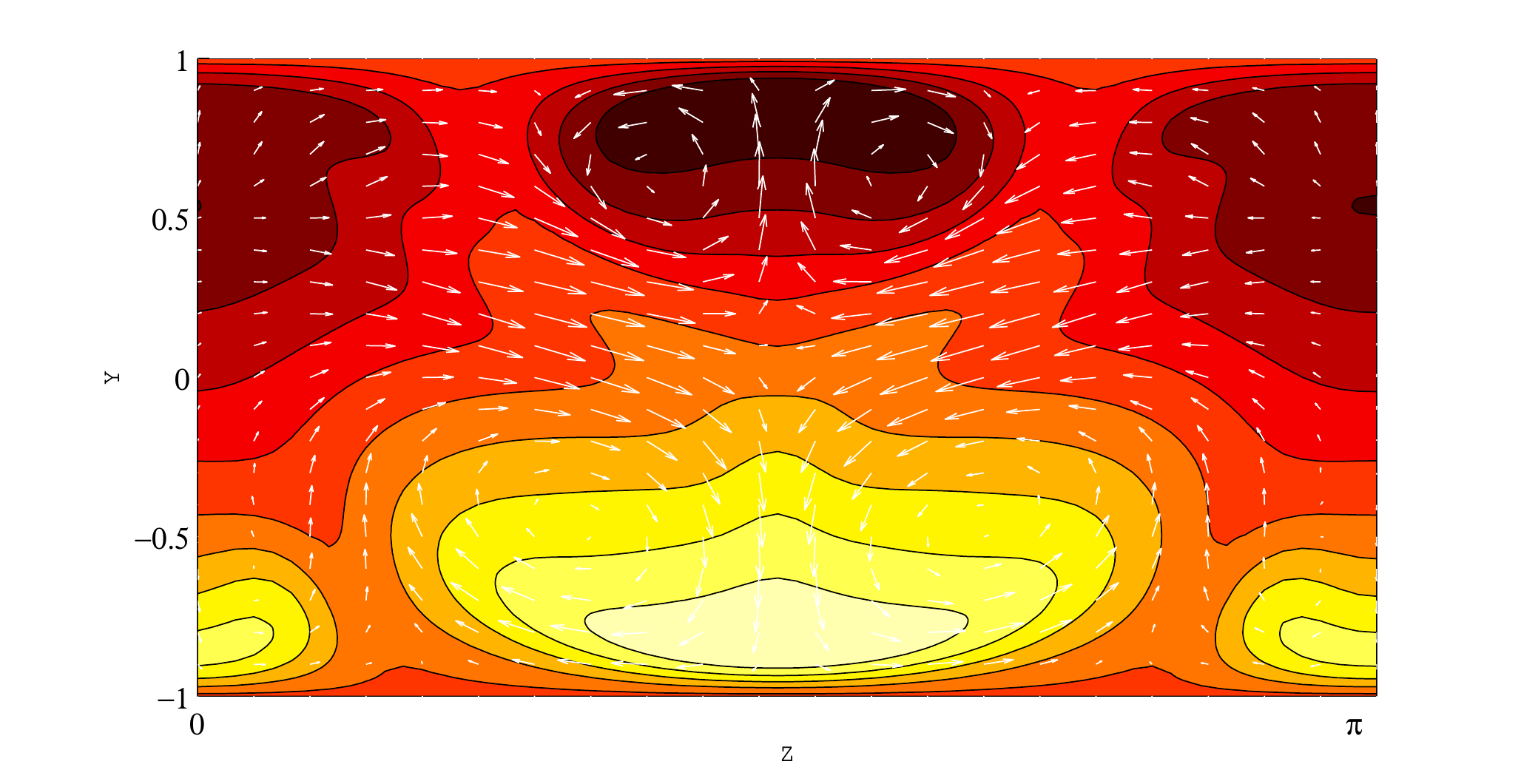}  % MAX_RI06_E5_t50
\adjincludegraphics[angle=0,height=3cm ,trim={{.075\width} {.029\height} {.085\width}  {.06\height}},clip]{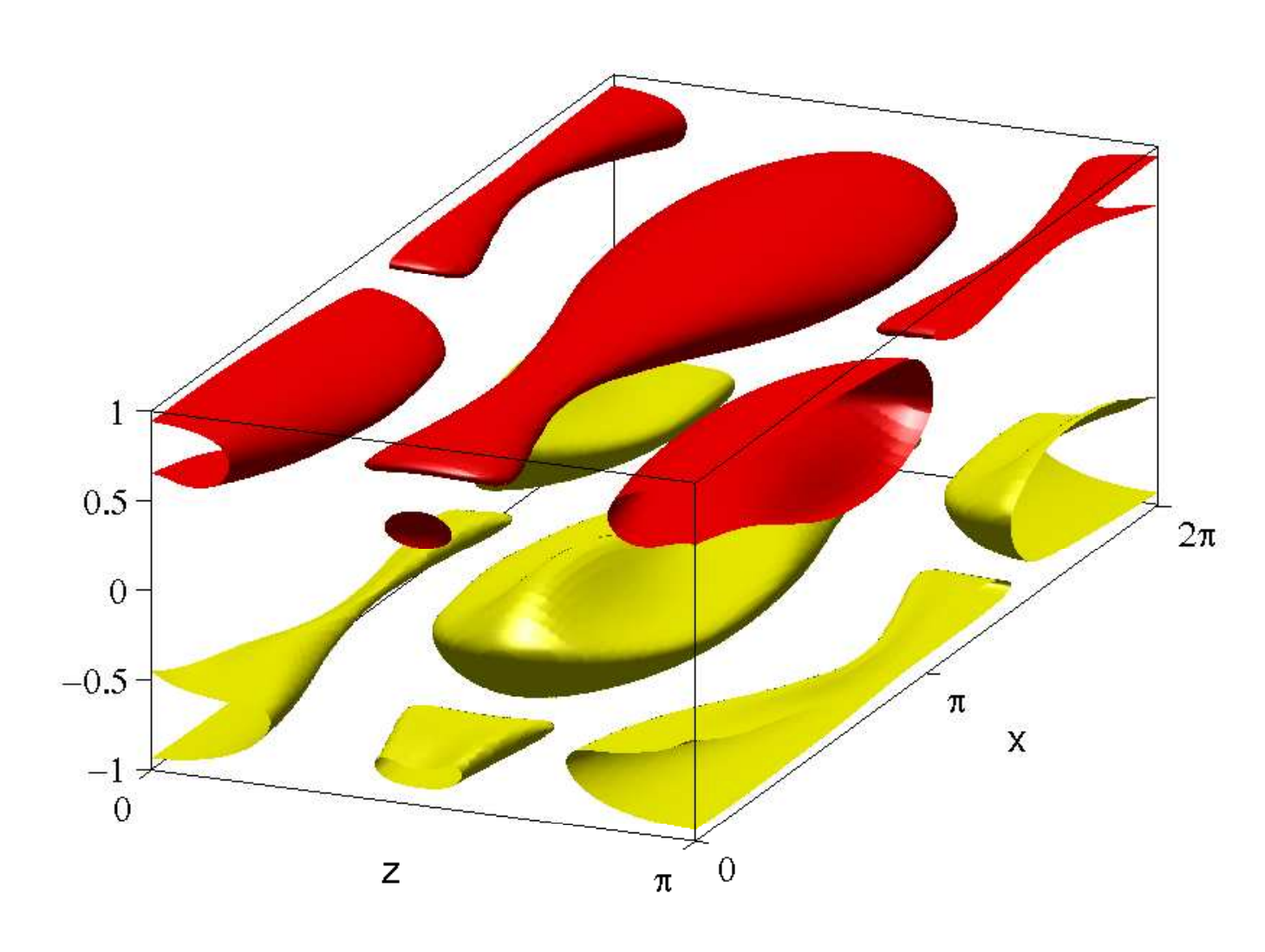} \\ % MAX_RI06_E5_t50_ISO
 \caption[Bursts]{  Peaks of energy.
Left plots: $Ri_b=0.04$ at $E_0=2.108 \times 10^{-4}$  $t=195$.
Right plots: $Ri_b=0.06$ at $E_0=5.0 \times 10^{-4}$, $t=50$.  The isocontours are  $\pm 60\%$ of maximum streamwise perturbation velocity and  9 contour levels are used between -0.75 and 0.925.}
 \label{B}
    \end{center}
  \end{figure}

Increasing the stratification further, figure \ref{BURST} compares the results of working at $Ri_b=0.055< Ri_b^m$ where a burst is still discernable (notice the upward curvature of the energy curve) and $Ri_b=0.06> Ri_b^m$ where it is not.
Clearly, the `bursting' is produced by the unstable manifold of the edge state (directed away in phase space from the linearly sheared state) and largely vanishes when the edge state disappears (i.e. the upward curvature disappears). In fact, phase space for $Ri_b \gtrsim Ri_b^m$ should still reflect the memory of the manifold but this ebbs away with increasing $Ri_b$. To confirm this simple explanation, we carried out a couple of checks. The first was to confirm that the state reached at  the energy peak  
at pre- and post-$Ri_b^m$ stratifications is roughly the same - see figure \ref{B}. And the second was attempting to converge a flow state on an optimal energy plateau for the $Ri_b=0.06$ (taken from the optimal evolution at $E_0=4.2 \times 10^{-4}$: see the red dot in figure \ref{BURST}(right)). This  failed to converge at $Re=400$ but did converge to EQ7 at $Re=500$ as $Ri_b^m(400)\,< \,Ri_b=0.06\, < \, Ri_b^m(500)$: see  the inset of figure \ref{BURST}(right).

\begin{figure}
    \begin{center}    %
\begin{tabular}{cc}
\adjincludegraphics[angle=0,height=2.5cm,width=4.5cm ,trim={{.075\width} {.029\height} {.075\width}  {.06\height}},clip]{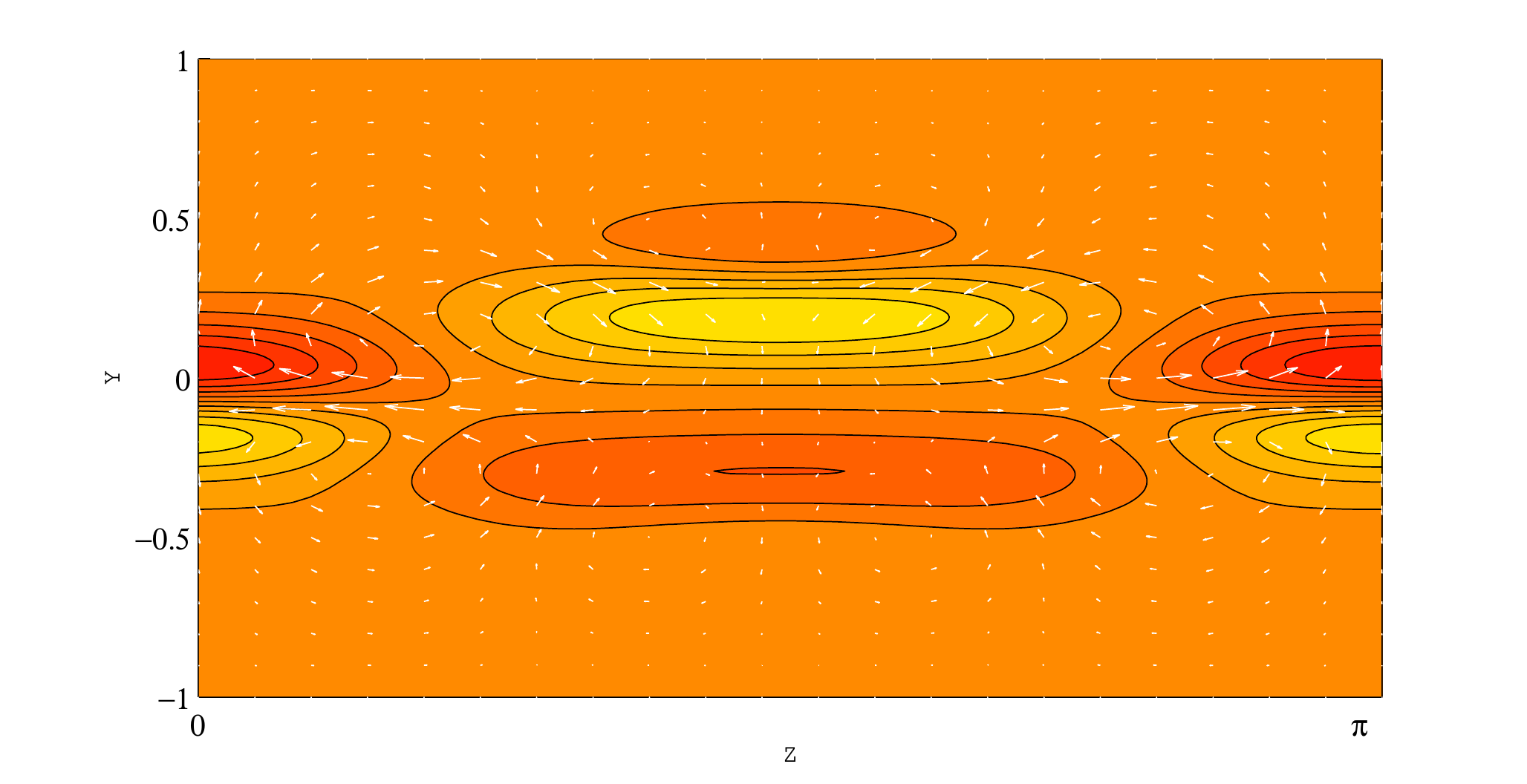} & % Ri1e6_E5e6_seed_t0
\adjincludegraphics[angle=0,height=2.5cm,width=4.5cm ,trim={{.075\width} {.029\height} {.075\width}  {.06\height}},clip]{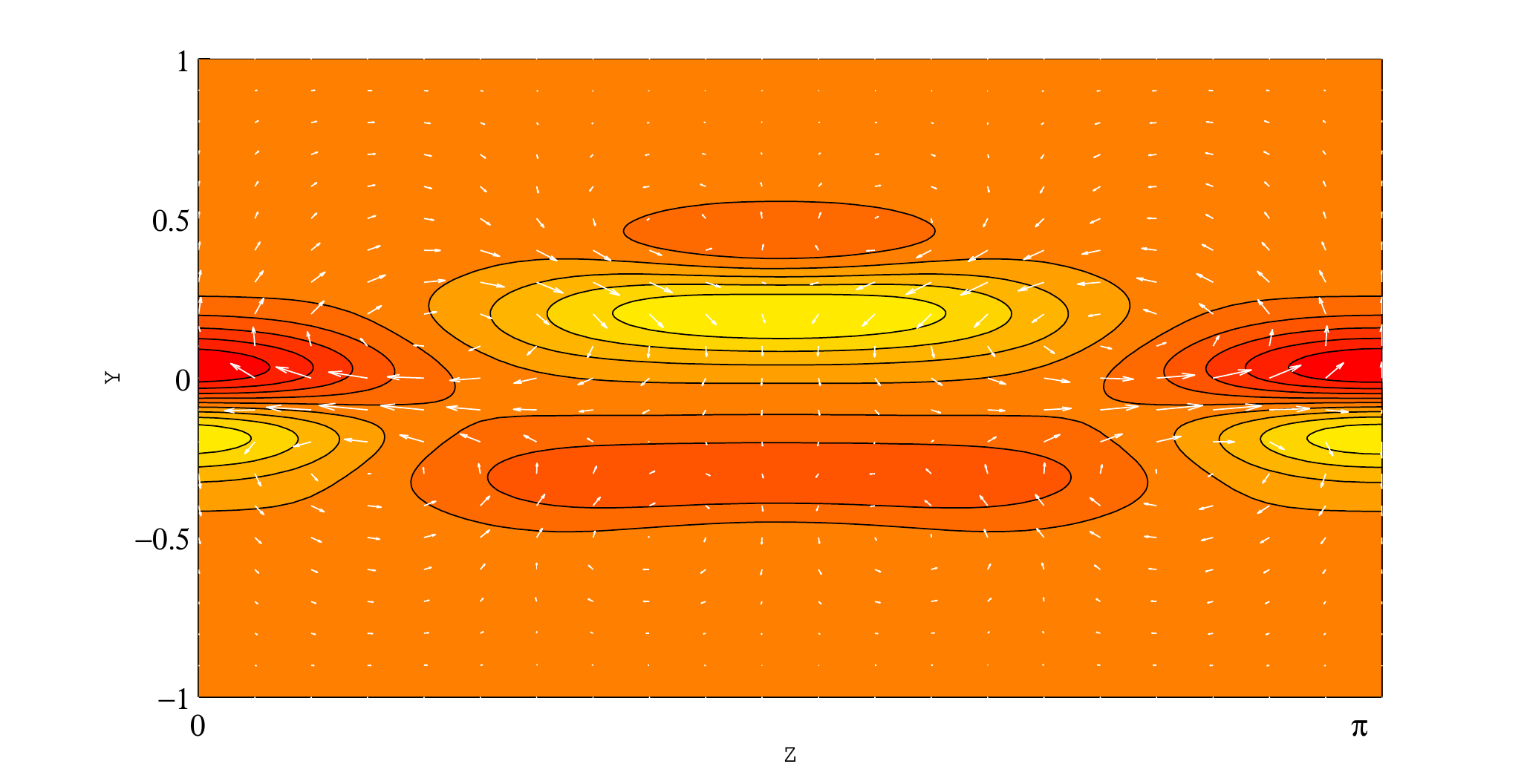}\\  % Ri01_E75e5_seed_t0
\adjincludegraphics[angle=0,height=2.5cm,width=4.5cm, trim={{.075\width} {.029\height} {.075\width}  {.06\height}},clip]{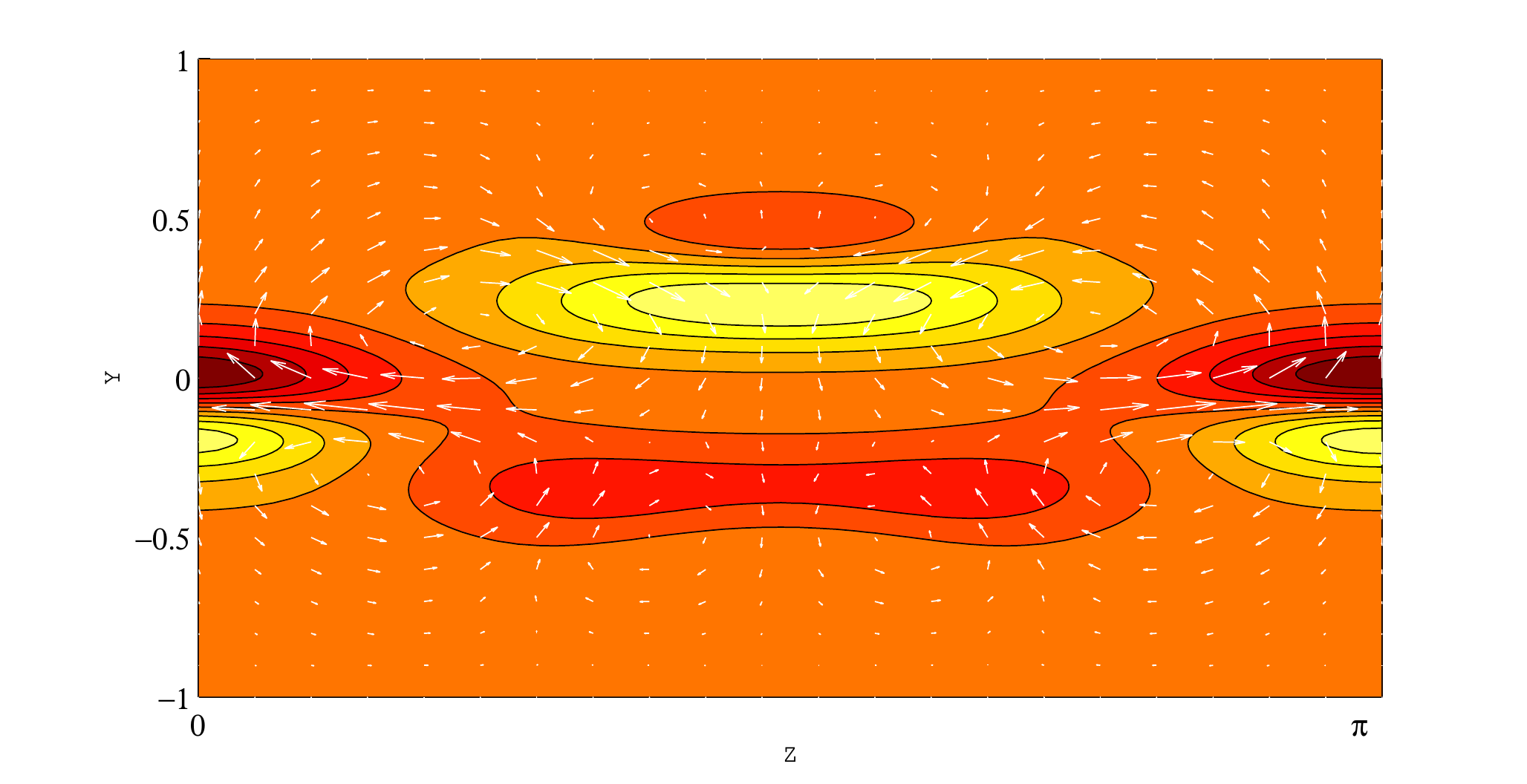}&  %Ri04_E21e4_seed_t0
\adjincludegraphics[angle=0,height=2.5cm,width=4.5cm, trim={{.075\width} {.029\height} {.075\width}  {.06\height}},clip]{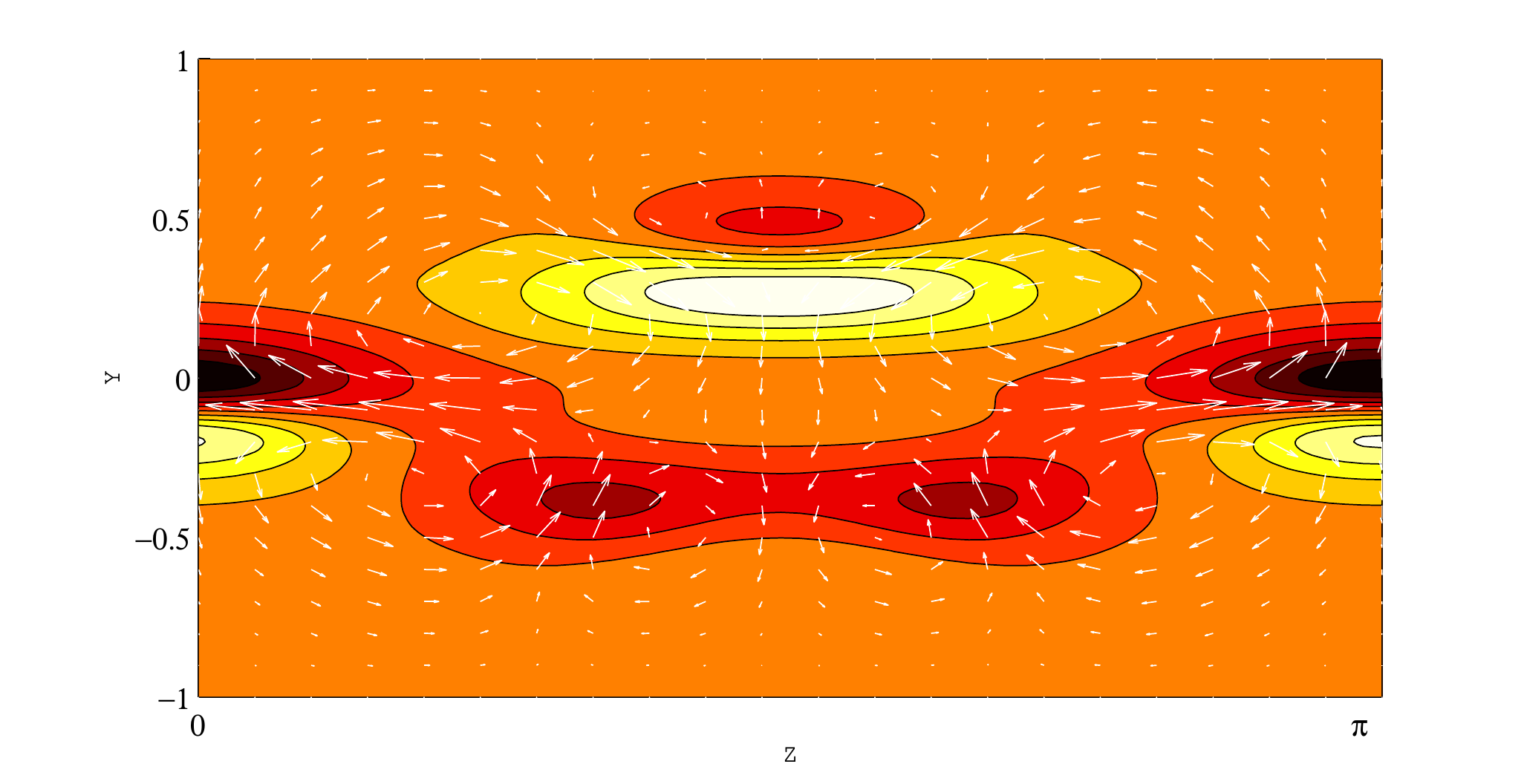} % Ri06_E4e4_seed_t0
\end{tabular}
 \caption[]
{Optimal perturbations. Contours of $yz$ cross-sections of streamwise pertubation velocity $\bu(x,y,z,t)$ (arrows indicate velocity field in plane).
Top left, $Ri_b=1 \times 10^{-6}$ at $E_0=5.0\times 10^{-5}$;
top right, $Ri_b=0.01$ at $E_0=7.5\times 10^{-5}$;
bottom left, $Ri_b=0.04$ $E_0=2.104\times 10^{-4}$;
bottom right, $Ri_b=0.06$ at $E_0=4.2\times 10^{-4}$.
 The unstratified, weakly and strongly stratified minimal seeds share the same structure. All plots use the same 9 contour levels set by the extremes of the minimal seed at $Ri_b=0.6$ (arrows rescaled as well).
 }
 \label{IC}
    \end{center}
  \end{figure}

\begin{figure}
  \centering
\adjincludegraphics[angle=0,height=4.5cm ,trim={{.04\width} {.01\height} {.04\width} {.05\height}},clip]{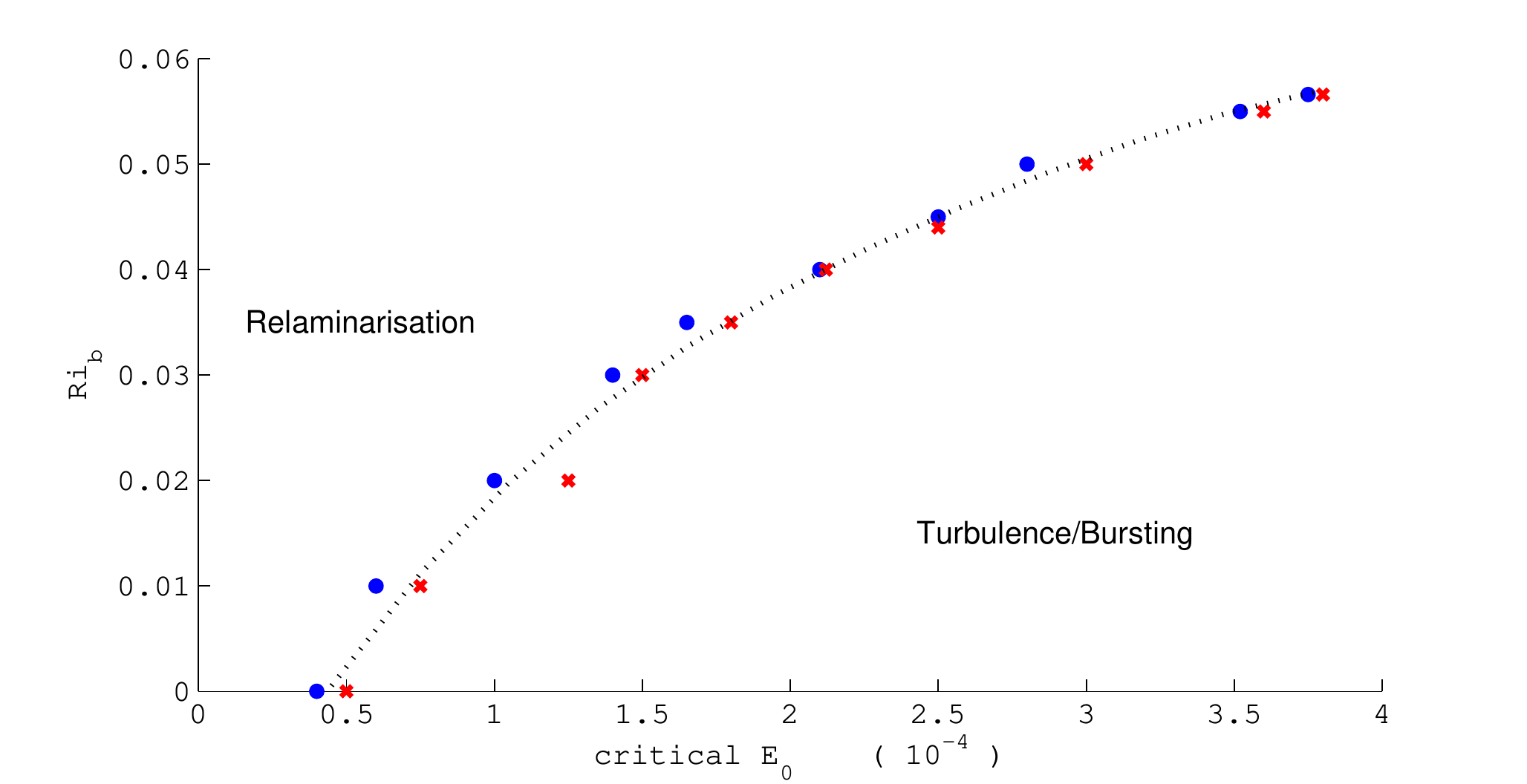}%    Critical_E0_RE400
  \caption{
Estimation of  the curve of $E_c$ for $Re=400$  and $T=50$  in box $2\pi \times 2 \times \pi$. Turbulent events are 
marked as red crosses and relaminarisation as solid blue cdots. No discernible energy threshold was observed at $Ri_b=0.06$ when  EQ7 no longer exists.}
  \label{Curve_critical_E0}
\end{figure}

The conclusion of this subsection is then that when the turbulent attractor disappears, one still can see `bursting' which is caused by the flow  trajectory being repelled out from the vicinity of the linearly-sheared base state by the unstable manifold of the edge state. This dominant (first) feature of the transition process therefore survives well after the suppression of turbulence by stable stratification and relies only on the edge state continuing to exist.  To further confirm this picture, the minimal seed to trigger transition or latterly this bursting (as $Ri_b$ increases) remains essentially the same across all $Ri_b$ studied including $Ri_b=10^{-6}$ which is essentially the unstratified case of \cite{Rabin12}: see figure \ref{IC}. However, what does change is the evolution of the minimal seed once it has experienced the burst in energy growth. Finally we collect together in figure \ref{Curve_critical_E0} the data collected on $E_c$ - the minimal energy to reach `above' the edge -  as a function of $Ri_b$ (the curve stops at $Ri_b=0.06$ since it is difficult to identify an edge beyond this point).

%
% Fig 17
%

%
% Fig 18          
%

%
% Fig 19   Thesis  fig 3.18     RRK 12-12-16
%

%
% Fig 22      GMRES OUTPUT RI=0.06 Re=400, 500
% 
%\begin{figure}[htp]  % 
%\begin{center}   
%\adjincludegraphics[angle=0,height=4cm ,trim={{.06\width} {.02\height} {.08\width} {.058\height}},clip]
%{./high_RI/GMRES_SOLC_RI06.eps} % 
%     \caption[]{Residual as a function of iteration $n$ of the Newton-GMRES rootfinding method. 
%The state at $t=80$  from  the evolution of the $E_0=4.2 \times 10^{-4}$ optimal (see figure \ref{BURST_RI06_RE400}) was used
%as initial guess for two different Reynolds numbers $Re$. The search stalls  after 45 iterations for %$Re=400$ (blue diamonds) but it converges in 22 steps
%for $Re=500$ (red circles).}
%\label{gmres_RI06}
%\end{center}
%\end{figure}

%
% Fig 20      i.c.s across  Ri_b for Re=400
%   

%
%  Fig 21      CRITICAL CURVE
%

  %____________________________________________________________________________________________________________________________
  %
  \section{Discussion \label{Disc}}

% what we have achieved

In this paper, we have demonstrated that an optimisation technique, in which the energy growth of a finite-amplitude disturbance to a known solution is maximised, can be used to generate flow fields  subsequently convergeable via a Newton-GMRES algorithm to another `nearby' solution of the Navier-Stokes equations.   That this may be possible has been noticed before in the particular case of an edge state to which a flow initiated by the minimal seed will get infinitesimally close during its evolution \cite{Pringle12, Rabin12, Kerswell14}.  Now this has been  confirmed  in section \ref{burst}. What was not clear before, however, was whether the technique could be used to find unstable solutions more generally located in phase space in the absence of an edge. The rediscovery of both steady and travelling wave snake solutions in section \ref{wide} demonstrates that the technique can also work in this situation too. 

The technique has then been used to probe very low $Re$ pCf with negative results adding further
 weight to the view that the linear-shear solution is indeed the global attractor in pCf up to $Re=127.7$. Finally, a `bursting' phenomenon was investigated in stably-stratified pCf and found to be produced by the unstable manifold of the edge state: when the edge state vanished on increasing the stratification so did the bursting. The key here is that the optimisation technique allowed the bursting to be found if it existed as the stratification was varied.
 
 Finally, the steady spanwise-localised snake solution has been been found to connect, via a global 3D state, to the 2D rolls of the Rayleigh-Benard problem. In doing so, a particularly simple delocalisation process has been found where the spanwise tails of the snake gradually reduce their spatial decay rate until this vanishes at the global linear instability threshold whereupon the state is then global. This would seem a very generic phenomenon where a  localised state moves from a region of subcriticality to one of supercriticality or vice versa.

% implications/questions

In terms of further work, the most obvious question is whether the optimisation technique can be used to find periodic orbits in which the energy varies in time. So far, only constant energy solutions (equilibria and travelling waves) have been sought and the identification of an energy plateau during the optimal's evolution has been central to indicate a `close approach'. In principle, a mildly fluctuating plateau should also be recognisable and provided the period is not too long, convergence should still be feasible.
Another issue is whether the approach can provide any insight into phase space around a linearly {\em unstable} solution. Here the answer surely depends on the time scale $ \tau$ of the linear instability. Typically the growth rates of linear instabilities in shear flows are much smaller than the typical instantaneous growth rates of energy growth optimals (e.g. in plane Poiseuille flow, the transient energy  growth is at least an order of magnitude more than that from the linear instability over O(100) advective times). This then suggests that the optimisation technique will simply ignore the linear instability for $T \ll \tau$ in preference to more potent mechanisms which  give better, albeit transient, energy growth. One such could be a nearby stable manifold of another solution as explored here.

As a final comment, it's worth emphasizing that the principles which make the optimisation approach work here in a fluid mechanical context hold true also for any dynamical system where solution multiplicity is suspected. What makes shear flows so interesting, of course, is this is now understood to be the generic situation where not only are there multiple solutions but these are typically unstable yet instrumental in determining the fluid dynamics of the flow (e.g. the bursting of section \ref{burst}). This optimisation approach then looks to be  a valuable new addition to a theoretician's  toolbox.

%
%____________________________________________________________________________________________________________________________

\begin{acknowledgments}
DO would like to thanks CONACYT for the award of a scholarship which has supported his PhD studies and the availability of free HPC time on the University of Bristol's BlueCrystal supercomputer. Both DO and RRK would like to thank Tom Eaves for sharing his nonlinear optimal perturbation routine (here converted to one focussed on total energy rather than total dissipation), John Taylor for sharing his time-stepping code `Diablo' and Dan Lucas for help parallelizing the GMRES algorithm used. We also are grateful for encouragement from the rest of the EPSRC-funded  `MUST' team at Cambridge received during the course of this work.
\end{acknowledgments}

\appendix
\section*{Appendix: Rayleigh Benard Convection}

In the normal Rayleigh-Benard set-up (e.g. \cite{Drazin} equations (8.6)-(8.8) with nonlinearities reinstated\,) the fully nonlinear equations for the total flow field $\mathbf{U}$ and temperature field $\Theta$ are
\begin{align}
\frac{\partial\, \mathbf{ U} }{\partial \tau} +\mathbf{U} \cdot \nabla \mathbf{U} &= -\nabla P + Ra\, Pr \, \Theta \,\hat{y} + Pr \nabla^2 \mathbf{ U} \label{RB_1} \\
\frac{\partial \, \Theta }{\partial \, \tau} +\mathbf{U}\cdot \nabla \Theta &=  \nabla^2 \Theta \label{RB_2}\\
\nabla \cdot \mathbf{U} &=0
\end{align}
subject to boundary conditions
\begin{equation}
 u(x,\pm 1,z,t))=\pm 1 \quad \& \quad \theta(x,\pm 1,z,t)= \mp 1.
\end{equation}
Stratified plane Couette flow (as described by equations (\ref{NS_1})-(\ref{NS_3})\, ) is retrieved under the following transformation
\begin{equation}
t = (Re Pr) \tau, 
\quad 
\mathbf{u} = \mathbf{U}/(RePr), 
\quad 
\rho = \Theta,
\quad             
p = P/(Re Pr)^2
\end{equation}              
so that $Ra = -Ri_b \, Re^2 \, Pr$. The critical $Ra$ for linear instability is $Ra_{crit}:=1708/16$ which is unchanged by introducing a unidirectional shear (e.g. \cite{Kelly77}) (the factor of $1/16$ is  because the half-channel width has been used to non-dimensionalize the system).

\end{document}